%% file: main.tex
\documentclass[twocolumn,prd,nofootinbib,superscriptaddress,longbibliography]{revtex4-1}

\usepackage{graphicx}
\usepackage[caption=false]{subfig}
\usepackage{bm}
\usepackage{amssymb,amsmath}
\usepackage{mathrsfs} 
\usepackage{latexsym}
\usepackage{mathtools} 
\usepackage{color}
\usepackage[normalem]{ulem} 
\usepackage{dcolumn}
\usepackage[colorlinks=true,citecolor=blue,urlcolor=blue]{hyperref}
\usepackage[usenames,dvipsnames]{xcolor}
\usepackage{booktabs}

\newcommand{\Austin}{\affiliation{Weinberg Institute, University of Texas at Austin, Austin, TX 78712, USA}}
\newcommand{\Torino}{\affiliation{Istituto Nazionale di Fisica Nucleare - Sezione di Torino, Torino, Italy}}

\newcommand{\etenhz}{e_\mathrm{10}}
\newcommand{\flo}{f_\mathrm{low}}
\newcommand{\fhi}{f_\mathrm{high}}

\begin{document}

\title{Misinterpreting Spin Precession as Orbital Eccentricity in Gravitational-Wave Signals}
\author{Snehal Tibrewal} \Austin
\author{Aaron Zimmerman} \Austin
\author{Jacob Lange} \Torino\Austin
\author{Deirdre Shoemaker} \Austin
\date{\today}

\begin{abstract}
    The increasing scope and breadth of gravitational wave detectors is providing the opportunity to explore new parameters in gravitational-wave astronomy. Eccentricity and spin-precession are two key observables to infer the origin of a gravitational wave (GW) source. The interpretation of GW source parameters can be plagued by degeneracy, such as the well-known degeneracy between mass and spin.  As the field has explored new parameters, questions have been raised about possible degeneracies between eccentricity and spin-precession. Although some state-of-the-art models now include these effects individually, models that incorporate spin-precession and eccentricity are only in their infancy.  Until models faithfully cover the complete parameter space of compact binary coalescence,  our ability to correctly measure the source parameters and infer the formation of the binary is compromised. Here, we present a study of the distinguishability of these two key parameters in heavy binary black hole systems. We study merger-ringdown dominated systems with detector-frame total masses ranging between $200-300M_{\odot}$ and mass ratio ranging between $1-3$. Our work finds that there is indeed a degeneracy between eccentricity and spin-precession; however, it is a highly localized effect. We find that the misidentified eccentricity estimates get worse as the detector-frame total mass increases from $200M_{\odot}$ to $300M_{\odot}$, corresponding to progressively shorter signals in the detector band. Additionally, this misidentification is highly sensitive to the inclination angle of the source system. We provide quantifiable estimates of the potency of this degeneracy in addition to identifying some of the regions of parameter space where this degeneracy exists.
\end{abstract}

\maketitle

\section{Introduction}
\label{sec:intro}

This past year marked the tenth anniversary of the first detection of gravitational waves (GW) from the merger of a binary black hole (BBH). 
The loud chirp of GW150914 \cite{LIGOScientific:2016aoc} was a landmark discovery that propelled the community into various streams of scientific investigation. Since that first discovery, the The LIGO-Virgo-Kagra (LVK) collaboration has reported 218 events in the fourth gravitational-wave catalog~\cite{LIGOScientific:2018mvr,LIGOScientific:2020ibl,LIGOScientific:2021usb,KAGRA:2021vkt,LIGOScientific:2025slb}. As the number of observed events increases, the likelihood of encountering systems with non-generic or exotic features, such as orbital eccentricity and spin-induced precession  also increases~\cite{Hoy:2024wkc, Gumbel1958_EVT}.  

These non-generic features, in turn, offer insight into one of the key open questions in the field: the identification of BBH formation mechanisms; How and when the binary formed, and in what environment. 
Binary properties like eccentricity or spins tilted below the orbital plane can indicate that the binary underwent dynamical interactions at some point in its lifetime, e.g.~\cite{Samsing:2017xmd, Rodriguez:2016vmx}. 
Thus, accurate and reliable measurement of these features is of great importance. 
Occasionally, the GWs observed by the detectors will capture only the final cycles of a binary system's inspiral and merger. 
That can give rise to the challenge of using limited information, sometimes only a handful of cycles in the sensitive band, to reconstruct as much of the binary's history as possible.

The GW  observations have revealed a potential population of massive BHs~\cite{LIGOScientific:2025pvj}, with examples like 
GW190521 having an estimated total  mass of $\sim150\,M_\odot$~\cite{LIGOScientific:2020iuh,LIGOScientific:2020ufj} 
and GW231123 with  $\sim240\,M_\odot$~\cite{LIGOScientific:2025slb,LIGOScientific:2025rsn}. 
These massive signals have few GW cycles in band, only  $\sim4$--5, making robust interpretation challenging~\cite{Fumagalli:2024gko}. 
A small number of cycles poses a challenge in placing constraints on the source properties, particularly parameters such as eccentricity and in-plane spin components. 
The initial LVK analysis of GW190521 included quasi-circular precessing templates~\cite{LIGOScientific:2020iuh,LIGOScientific:2020ufj}, 
and found mild evidence for the imprint of precession; however, it did not include any with eccentricity. 
Since then, multiple studies have reanalyzed the event; however, its origin remains ambiguous. 
Work by various groups has shown that the signal could have been emitted by an eccentric source \cite{Romero-Shaw:2020thy,Gayathri2022}, a precessing source \cite{Miller:2023ncs}, or more exotic compact objects such as boson stars \cite{CalderonBustillo:2020fyi} or cosmic strings~\cite{Cuceu:2025fzi}. 
These competing propositions highlight the challenge of positively inferring eccentricity from short signals, and the possibility that precession and eccentricity can be misidentifed with limited models.

\begin{figure}[tb]
    \includegraphics[width=0.5\textwidth]{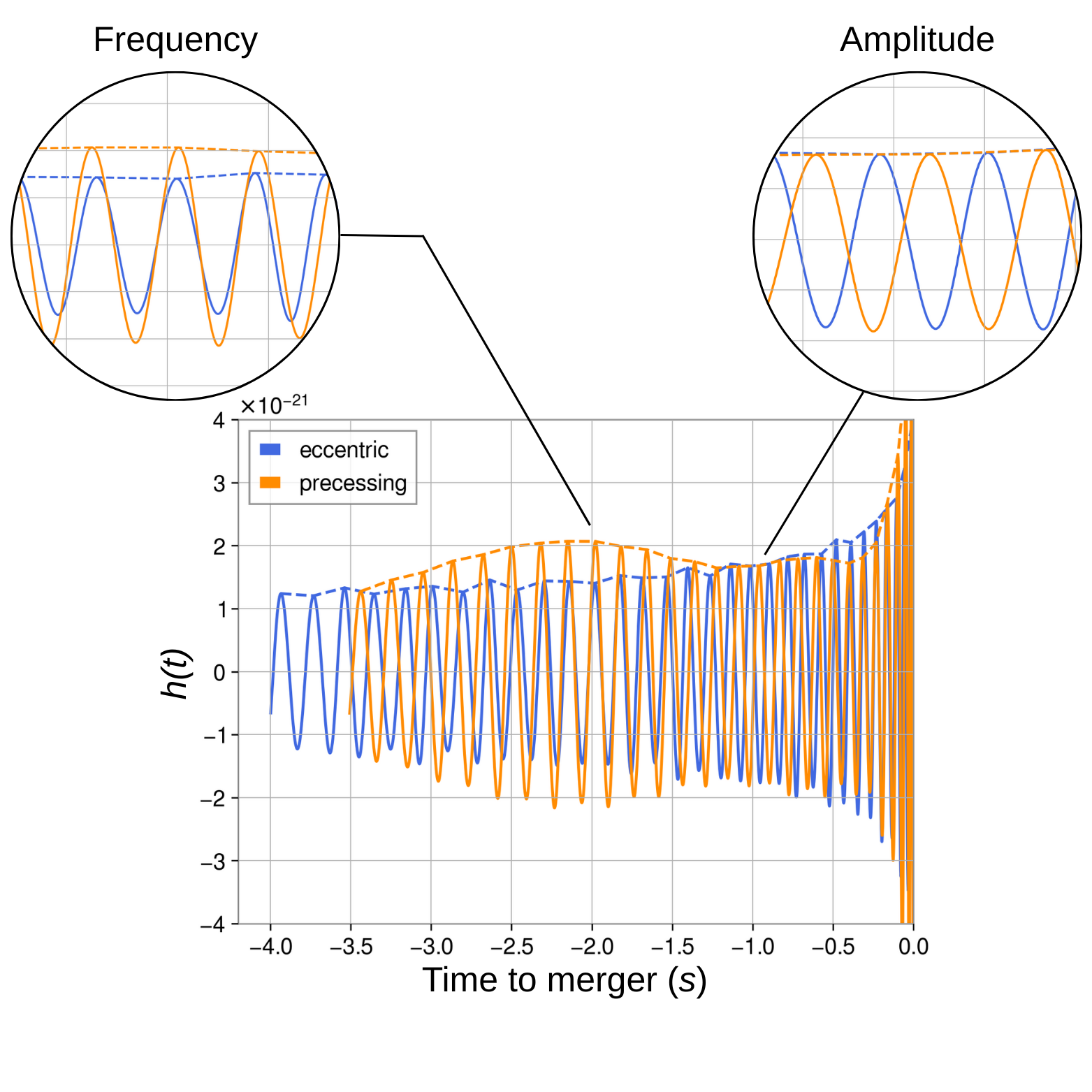}
    \caption{Overlay of an eccentric aligned-spin waveform (orange) and a quasi-circular precessing waveform (blue), illustrating qualitatively similar waveform features. }
    \label{fig:amp_freq}
\end{figure}

Eccentricity and spin-precession are relatively well-studied parameters individually. 
Both these effects cause a 
modulation in amplitude and frequency in the GW signal. 
However, these effects generally operate on different timescales. 
While eccentricity produces modulations at an orbital timescale, precession can cause modulations on longer timescales, spanning many orbits. 
Superficially, these distinct timescales can make the waveforms produced by systems displaying eccentric or precessing dynamics 
appear starkly different. 
Nevertheless, as illustrated in Fig.~\ref{fig:amp_freq}, there exists regions of parameter space in which their modulations overlap in amplitude and phase in a way that can make them qualitatively similar and potentially degenerate.
Romero-Shaw et al.~\cite{Romero-Shaw:2022fbf} reported that a sufficient number of cycles in-band provides enough information to distinguish between these effects; however, this is not the case for a GW190521-like signal, i.e., signals with few cycles in-band. 
Contrary to expectations, Xu et al.~\cite{Xu2023} focused on heavier systems but reported no correlation between eccentricity and precession in the specific numerical relativity (NR) systems they studied.
More recently, Divyajyoti et al.~\cite{Divyajyoti:2025cwq} expanded this picture by exploring parameter biases for eccentric systems with and without precessional effects when analyzed with waveform models with incomplete physics. They find that for systems exhibiting both these effects simultaneously, an eccentric aligned spin model is preferred over a quasi-circular precessing model. This highlights how waveform systematics can mimic an underlying physical degeneracy. Further investigating these effects, Gupta et al.~\cite{Gupta:2025paz} show posteriors from two distinct binary systems where eccentricity is misidentified as spin-precession. Their results point to the existence of the degeneracy uni-directionally.  
However, the overall prevalence, parameter dependence, and strength of this degeneracy in the broader parameter space remain largely unclear.

Meanwhile, the astrophysical importance of measuring eccentricity has driven developments in the modeling of GWs from eccentric systems, as well as assessing the impact of mismodeling on PE.
A number of state-of-the-art models exist that treat precession and eccentricity separately (see Sec.~\ref{sec:waveform}).
Such models show that neglecting eccentricity when present leads to biases in interpretations, e.g.~\cite{OShea:2021faf,Favata:2021vhw,Divyajyoti:2023rht,Saini:2022igm,Zeeshan:2024ovp}.
They have been used to supplement quasi-circular, precessing inferences of detected BBH GW events with additional eccentric analyses, e.g.~\cite{Romero-Shaw:2021ual,Iglesias:2022xfc,Gupte:2024jfe,Planas:2025jny,Romero-Shaw:2025vbc}, 
which in some cases suggest non-negligible eccentricities for a subset of events.
Finally, a small number of waveform models have been developed which incorporate both eccentricity and precession~\cite{Klein:2018ybm,Gamba:2024cvy,Morras:2025xfu,Morras:2025nlp}.
These are either limited to only the inspiral phase, or are relatively new models which remain computationally expensive for use in full-scale PE studies.

In this work we address the question of whether there exist regions of the parameter space where eccentricity and spin-precession mimic each other, and whether these regions can be identified or predicted in advance.
Motivated by our intuition and past results, we focus on GW signals from massive BBHs, where the signals are relatively short.
While complete, inspiral-merger-ringdown models with both eccentricity and precession would be ideal for understanding the possible degeneracy between these physical effects, here we use existing state-of-the-art models which capture only one of these at a time.
We carry out a systematic investigation, 
first performing mismatch studies to broadly identify regions of the parameter space in which precessing, quasi-circular systems might mimic eccentric systems.
Subsequently, we conduct full Bayesian PE to verify whether true statistical degeneracies arise in these regions. 
In doing so, to our knowledge we identify the first controlled examples of quasi-circular, precessing signals which are misidentified as eccentric when analyzed with eccentric models.
These examples confirm the intuition gained from the differing conclusions drawn from analyzing real events, which are complicated by the presence of real detector noise.
Our work also underscores the sensitivity of this degeneracy to variations in other binary parameters.

This paper is organized as follows. We describe the waveform model and associated conventions in Sec.~\ref{sec:waveform}. In Sec.~\ref{sec:mismatch}, we present the setup and results of the preliminary mismatch study that informs our parameter estimation section. In Sec.~\ref{sec:PE}, we detail the setup and results for PE for the identified injection set. Finally, in Sec.~\ref{sec:discussion}, we discuss the implications and robustness of our results and outline directions for future work.
Throughout this work we set $G = c = 1$, quote detector frame rather than (redshifted) source-frame masses~\cite{LIGOScientific:2025hdt}, and adopt the NR convention that the mass ratio is greater than unity, $q \geq 1$.

\section{Waveform model}
\label{sec:waveform}

A number of state-of-the-art waveform models exist with differing conventions, approximations, and physics content. This includes phenomenological models~\cite{2021PhRvD.103j4056P,2025PhRvD.111j4019C,PhysRevD.104.124027,PhysRevD.109.063012,2025arXiv250702604H,PhysRevD.105.084039,PhysRevD.105.084040}, effective-one-body (EOB) models from the SEOB family~\cite{PhysRevD.95.044028,PhysRevD.98.084028,PhysRevD.102.044055,PhysRevD.101.124040,PhysRevD.104.124087,2023PhRvD.108l4035P} and the TEOB family~\cite{2020PhRvD.102b4077N,PhysRevD.104.104045,PhysRevD.103.024014,PhysRevD.106.024020,PhysRevD.110.024031}, and the NR surrogate models~\cite{2019PhRvR...1c3015V,2019PhRvD..99f4045V}. Many of these model families also have there own associated eccentric model~\cite{2025arXiv250313062D,Gamboa2024,2022PhRvD.105d4035R,2020PhRvD.101j1501C,2021PhRvD.103f4022I}. For this work, we use SEOBNRv5 waveforms that are based on the EOB formalism. 
EOB models approximate the full inspiral–merger–ringdown regime. We adopt SEOBNRv5EHM \cite{Gamboa2024} as our non-precessing eccentric model and SEOBNRv5PHM \cite{Ramos-Buades:2023ehm,2023PhRvD.108l4035P} as our quasi-circular precessing model. 
Using waveforms from the same model family minimizes potential model-related systematics when comparing the two. 
Both models support higher-order modes. 
The modes used for SEOBNRv5EHM are \{(2,±2), (2,±1), (3,±3), (3,±2), (4,±4), (4,±3)\}, and for SEOBNRv5PHM are \{(2,±2), (2,±1), (3,±3), (3,±2), (4,±4), (4,±3), (5,±5)\} in the co-precessing frame. 
SEOBNRv5EHM allows for explicit sampling in relativistic anomaly $\zeta$ at the reference frequency, an essential parameter in the relativistic definition of eccentricity.

\section{Mismatch study}
\label{sec:mismatch}

\begin{figure*}[tb]
    \includegraphics[width=1\textwidth]{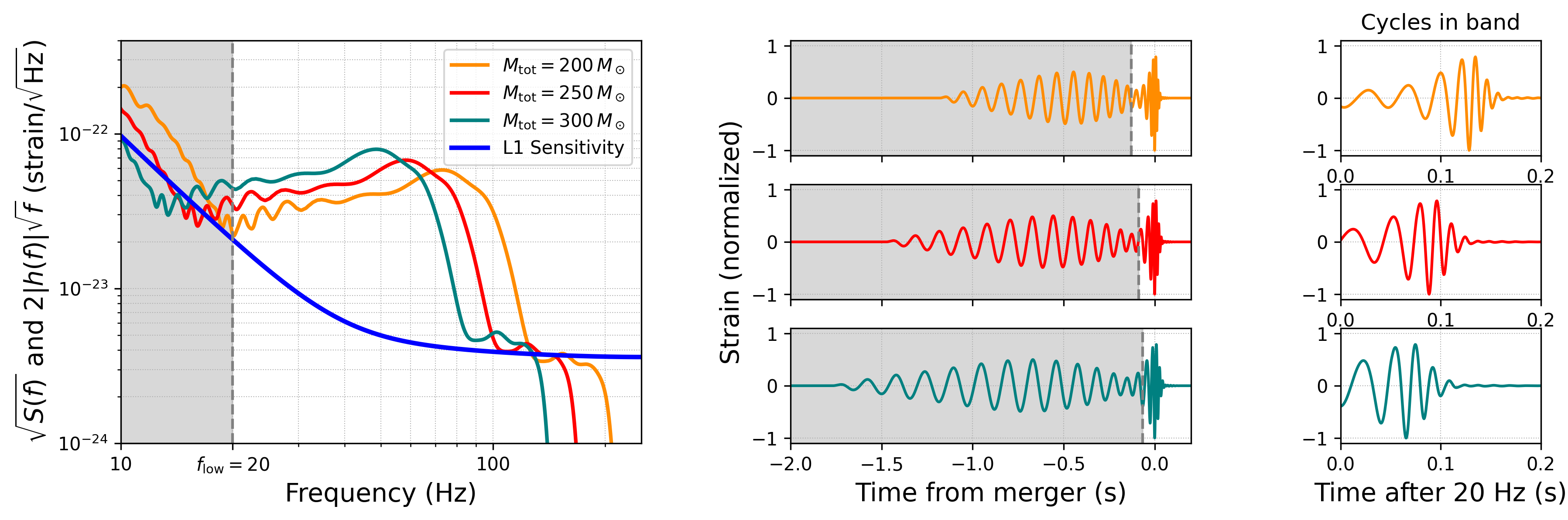}
    \caption{Example signals from precessing systems with $q=1$ and total masses $200\,M_\odot$, $250\,M_\odot$ and $300\,M_\odot$.
    These same signals are used as injections for studying eccentric inferences across total mass in Sec.~\ref{sec:PE}.
    {\it Left:} The signals in frequency domain, as compared to the amplitude spectral density used in this study. 
    {\it Center:} The signals in time domain, with the shaded portions outside the sensitive band of the detectors.
    {\it Right:} The effective in-band cycles for a detector sensitivity beginning at $20$ Hz, highlighting the reduction in cycles with increasing total mass and the growing dominance of the merger–ringdown regime.}
    \label{fig:FD_TD_short}
\end{figure*}

In order to identify regions of parameter space where precession can be misidentified as eccentricity, we would ideally scan over a wide range of precessing systems, injecting them as synthetic signals, and carrying out PE using eccentric waveform models.
However, full Bayesian PE is computationally expensive, especially for state-of-the-art eccentric models that include higher mode content.
For this reason, we first carry out a mismatch study 
to isolate the regions of parameter space where such misidentification is most likely. Specifically, we compute the mismatch 
between quasi-circular precessing signals and eccentric non-precessing signals, varying parameters that control the number of waveform cycles in band, the degree of precession, and the orbital eccentricity of the binaries. 
We perform this suite of mismatch comparisons first with non-spinning eccentric systems, and then eccentric systems with aligned spins fixed at $\chi_{1,z}=\chi_{2,z}=0.5$.
For the latter, we consider precessing systems both without and with the same aligned-spin components, allowing us to isolate the effect of differences in aligned spin between the two systems.
From this study, we identify selection criteria that highlight a promising region of parameter space where precession may be misidentified as eccentricity. 
We then use this region to inform injection choices for a smaller number of full PE analyses in Sec.~\ref{sec:PE}.
We find that a combination of aligned spin and nonzero eccentricity is required to get a small mismatch with a precessing binary with in-plane spins. In many cases, including the case where the precessing system has some part of the spin aligned with orbital momentum, zero eccentricity provides the best match between the non-precessing and precessing signals.

\subsection{Parameter choices}
\label{sec:mismatchA}

Mismatches~\cite{1999PhRvD..60b2002O} are conventionally used to quantify the similarity between two waveforms, for example when calibrating analytical models against post-Newtonian or  NR waveforms~\cite{Boyle:2007ft,Hannam:2007ik,Buonanno:2009zt,Ajith:2007qp}. 
In this work, we use the mismatch between waveforms generated by physically distinct systems to assess the similarity between eccentric and precessing signals. 
This serves as a relatively inexpensive diagnostic for the expected behavior of full PE. 
This is a suboptimal approach, as we can only maximize over a few parameters while keeping several extrinsic and intrinsic parameters fixed.

The mismatch is defined as:
\begin{equation}
\mathcal{MM}=1-\underset{\Delta t, \Delta \phi, \Delta \psi}{\max}\frac{\left< h_{p}|h_{e}\right>}{\sqrt{\left< h_{p}|h_{p}\right>\left< h_{e}|h_{e}\right>}} \,,
\label{eqn:mm_eqn}
\end{equation}
where $h_{p}$ is the precessing signal, $h_{e}$ is the eccentric signal and $\langle\cdot|\cdot\rangle$ is the standard noise-weighted inner product (Appx.~\ref{sec:AnalysisConventions}).
For the mismatch we maximize the inner product over the difference in coalescence time $\Delta t$, coalescence phase $\Delta \phi$, and polarization angle $\Delta \psi$ between the signals. 
We fix the same total mass $M_\mathrm{tot}$ and mass ratio $q$ for both signals. In all cases we fix the inclination to $\pi/3$ radians, a value intermediate between the special face-on ($\iota=0$) and edge-on ($\iota=\pi/2$) cases. This inclination value also coincides with the peak of the expected observed distribution of inclinations~\cite{Schutz:2011tw}.
Meanwhile, as the two models support different physical effects, our choices of spins and eccentricity parameters necessarily differ.

The expansive dimensional space of the precessing, non-eccentric and eccentric, non-precessing models require us to reduce the dimensionality of the problem. So we restrict the precessing configurations to have spins along a single axis, setting $\chi_{1,x} = \chi_{2,x}$ with varying magnitude across cases. 
For the first two mismatch suites, all other spin components of the precessing configurations are set to zero. In an additional suite, we set $\chi_{1,z}=\chi_{2,z}=0.5$ for both the eccentric and precessing systems, while retaining the varying in-plane spins of the precessing systems.
For eccentric systems, we vary the eccentricity at the reference frequency of $f_\mathrm{ref} = 10$ Hz, $\etenhz$ and set the relativistic anomaly at the reference frequency to $\zeta = 0.5$ radians to select a generic orbital phase at this moment.
We also allow for equal, aligned spins for eccentric signals, as described below.

\begin{figure}[tb]
    \includegraphics[width=0.45\textwidth]{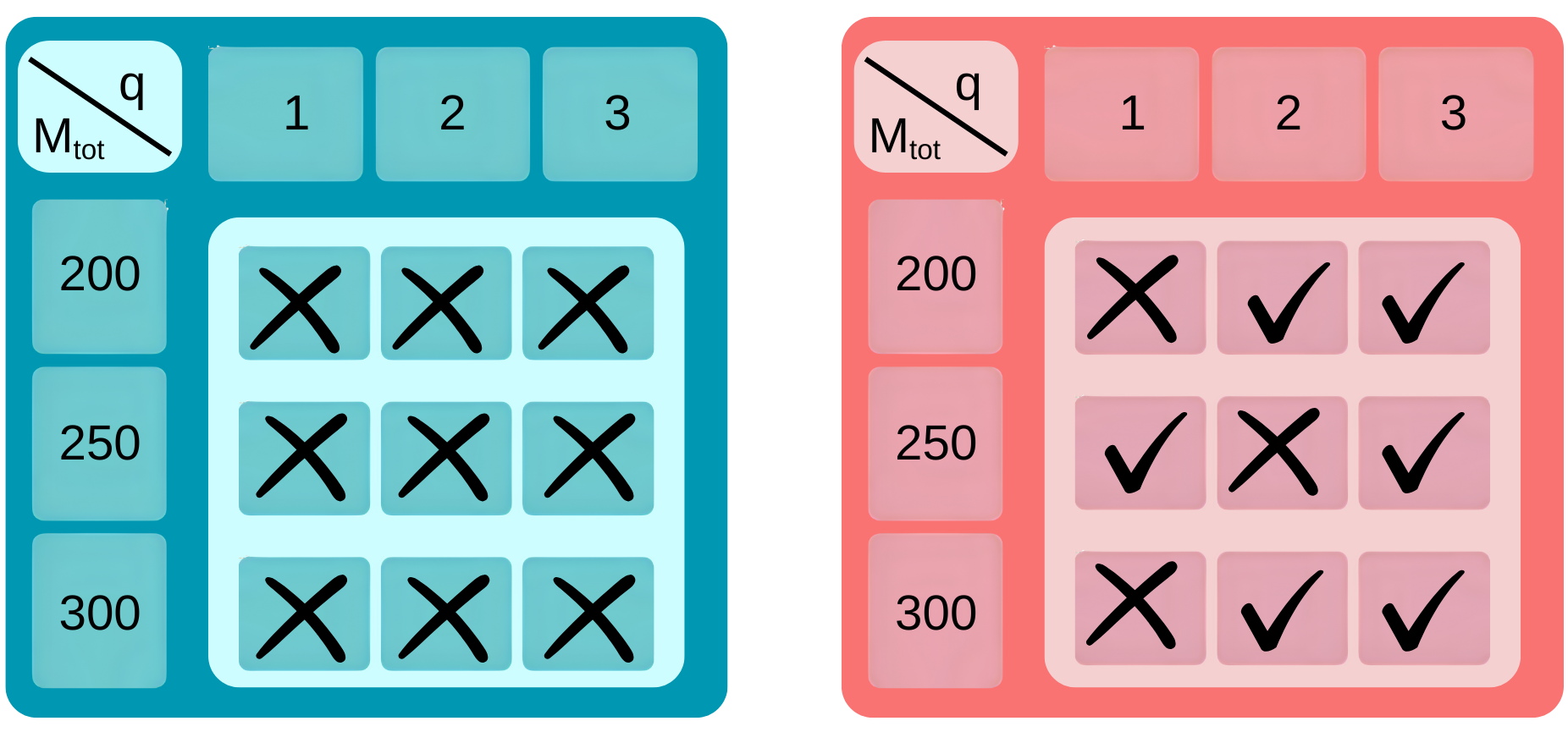}
    \caption{Configuration of the mismatch suites. The grids show mismatches between precessing systems with in-plane spins and non-spinning eccentric systems (\textit{left}), and aligned-spin eccentric systems with $\chi_{1,z}=\chi_{2,z}=0.5$ (\textit{right}). An additional suite, which can be represented by the schematic on the \textit{left}, compares aligned-spin eccentric systems with precessing systems having the same aligned-spin components in addition to their in-plane spins. All suites span across the same total mass and mass ratio values.} The total mass $M_{\mathrm{tot}}$ is given in units of $M_{\odot}$, and the mass ratio is defined as $q=m_{1}/m_{2}$, where $m_{1}\geq m_{2}$. The check symbol indicates the regions of parameter space where an eccentric system i.e. $\etenhz > 0.05$ was found to be the best match to a precessing system.
    \label{fig:mm_config}
\end{figure}

\begin{figure*}[tb]

\subfloat[$q=2$, $M_{\mathrm{tot}}=200M_\odot$]{
    \includegraphics[width=0.32\textwidth]{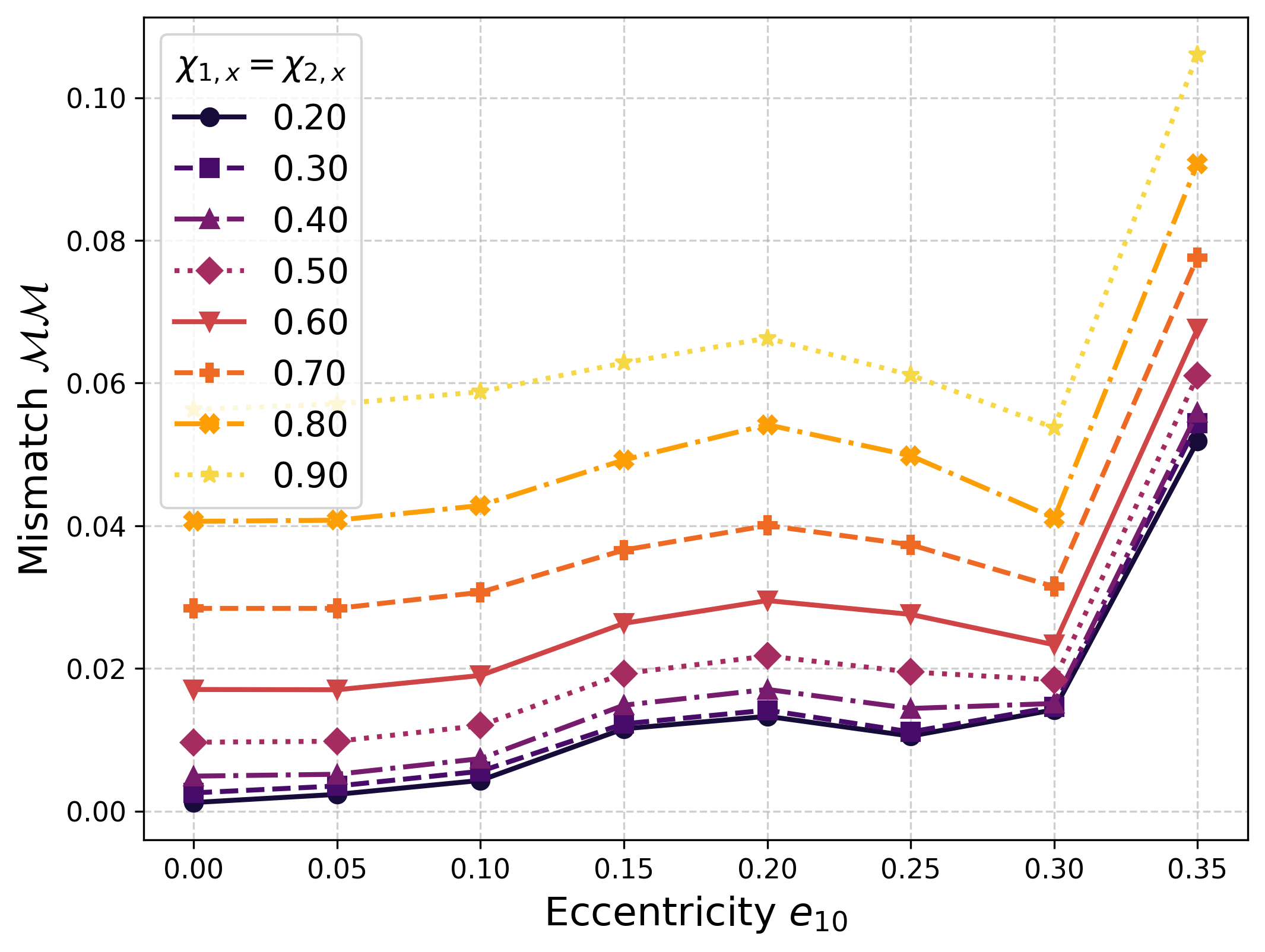}
}
\hfill
\subfloat[$q=2$, $M_{\mathrm{tot}}=200M_\odot$]{
    \includegraphics[width=0.32\textwidth]{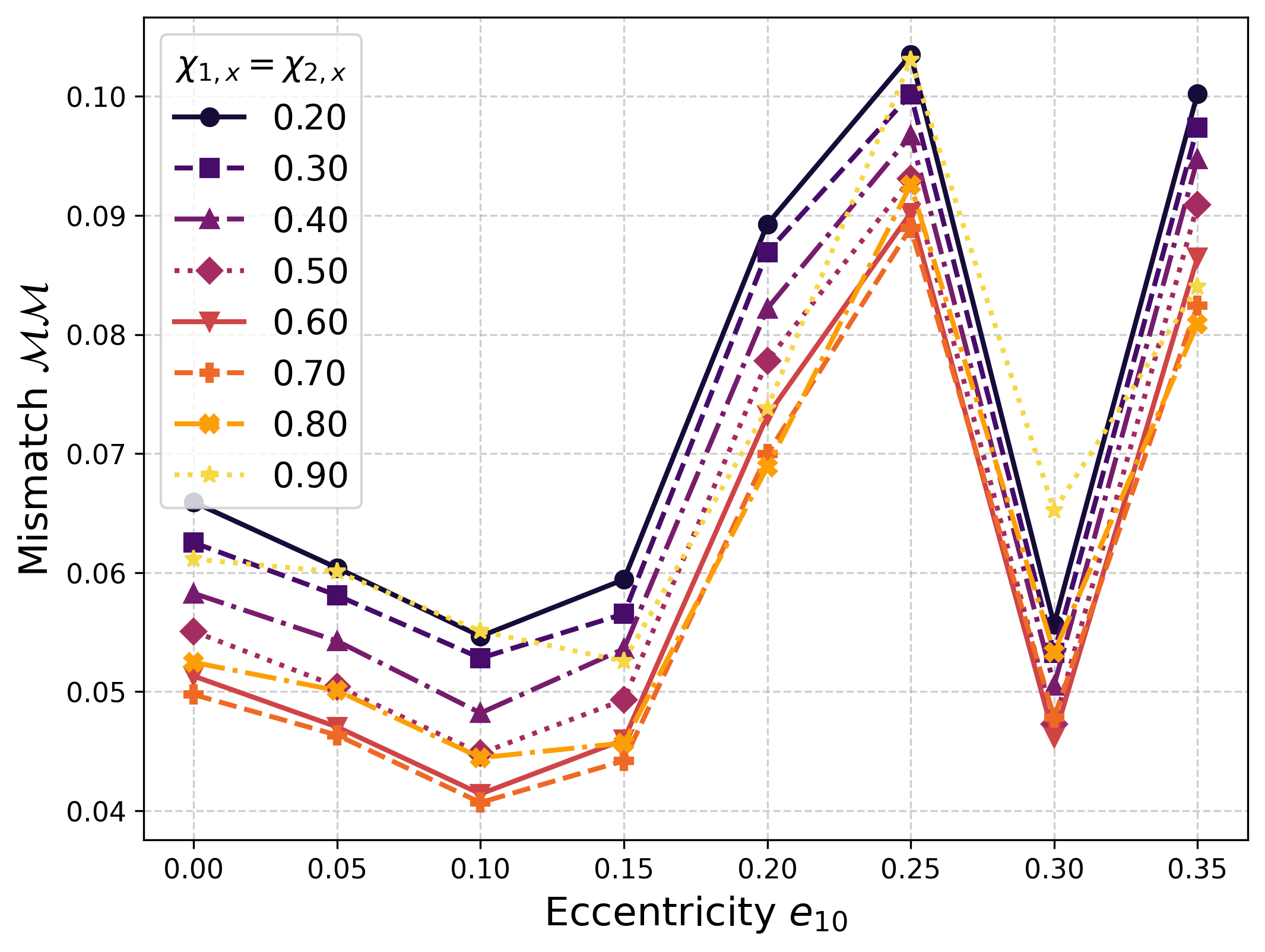}
}
\hfill
\subfloat[$q=2$, $M_{\mathrm{tot}}=200M_\odot$]{
    \includegraphics[width=0.32\textwidth]{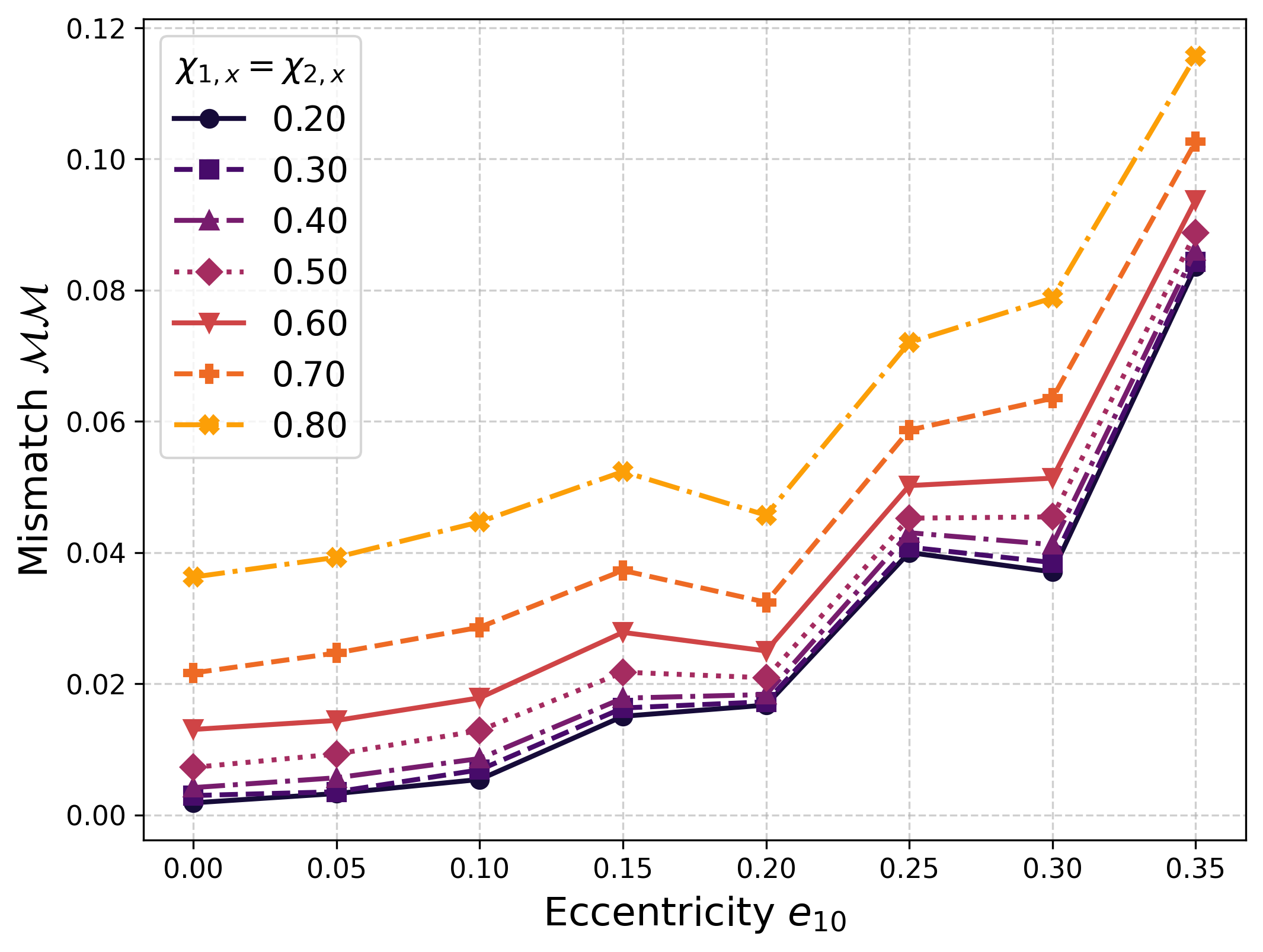}
}

\vspace{0.3cm}

\subfloat[$q=1$, $M_{\mathrm{tot}}=250M_\odot$]{
    \includegraphics[width=0.32\textwidth]{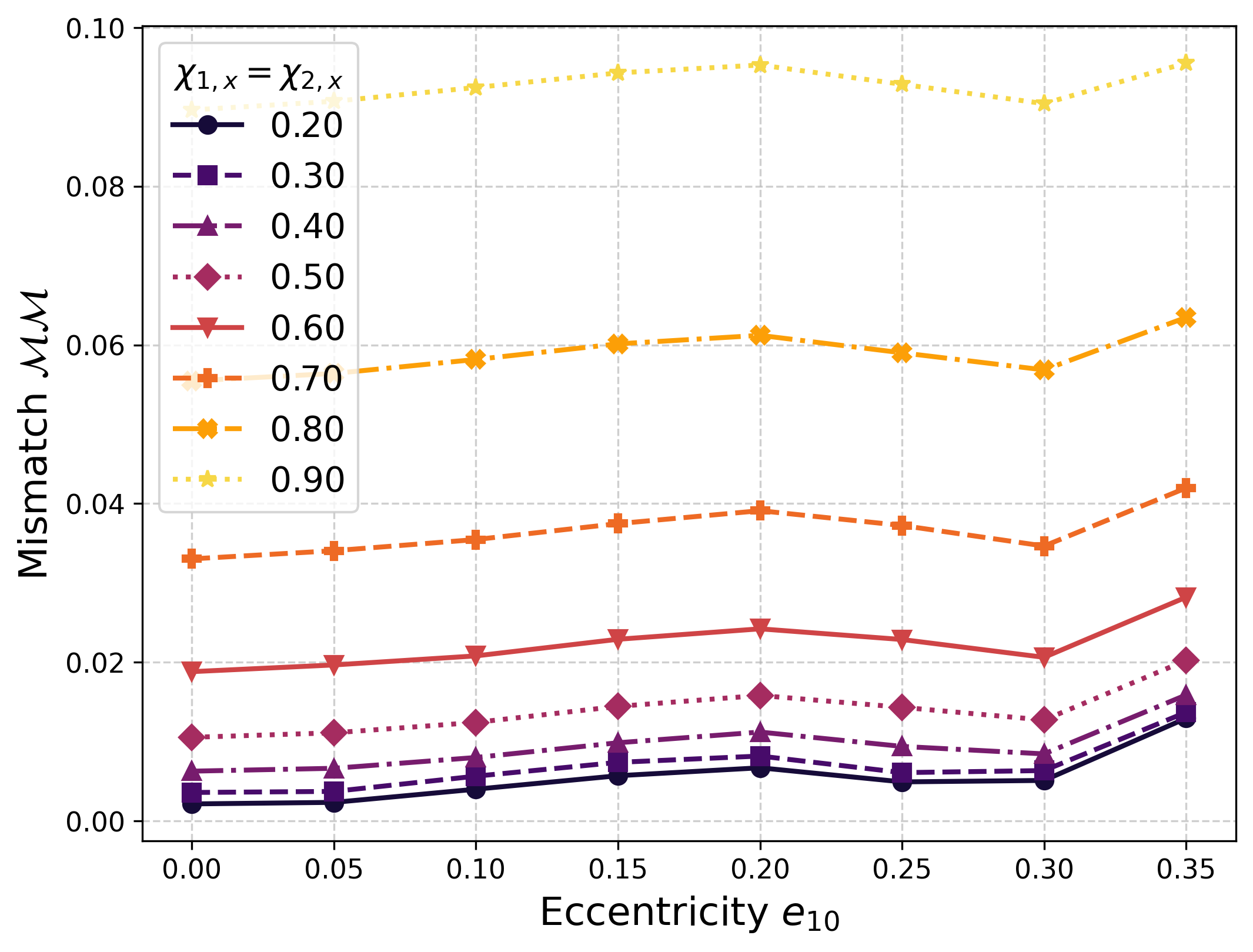}
}
\hfill
\subfloat[$q=1$, $M_{\mathrm{tot}}=250M_\odot$]{
    \includegraphics[width=0.32\textwidth]{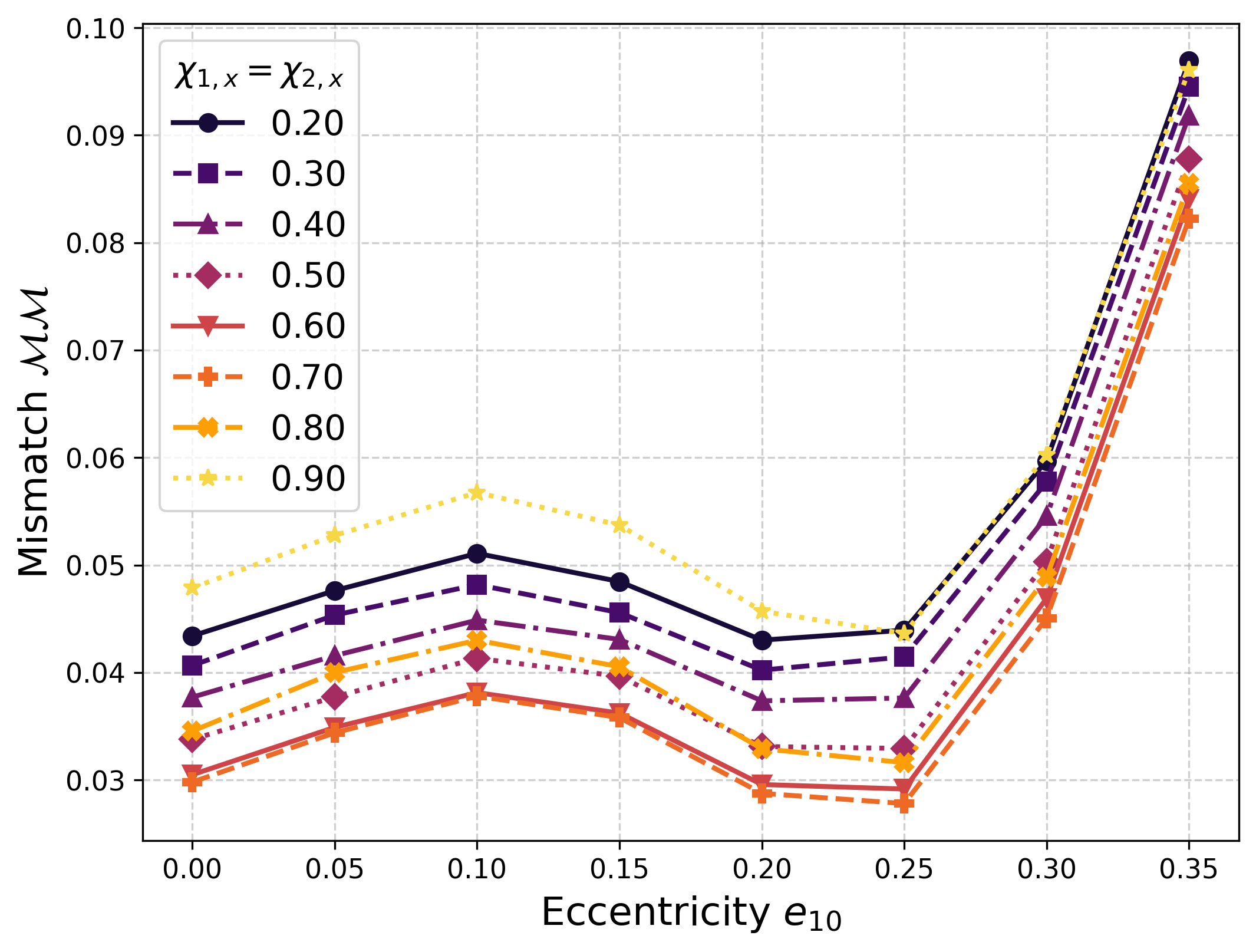}
}
\hfill
\subfloat[$q=1$, $M_{\mathrm{tot}}=250M_\odot$]{
    \includegraphics[width=0.32\textwidth]{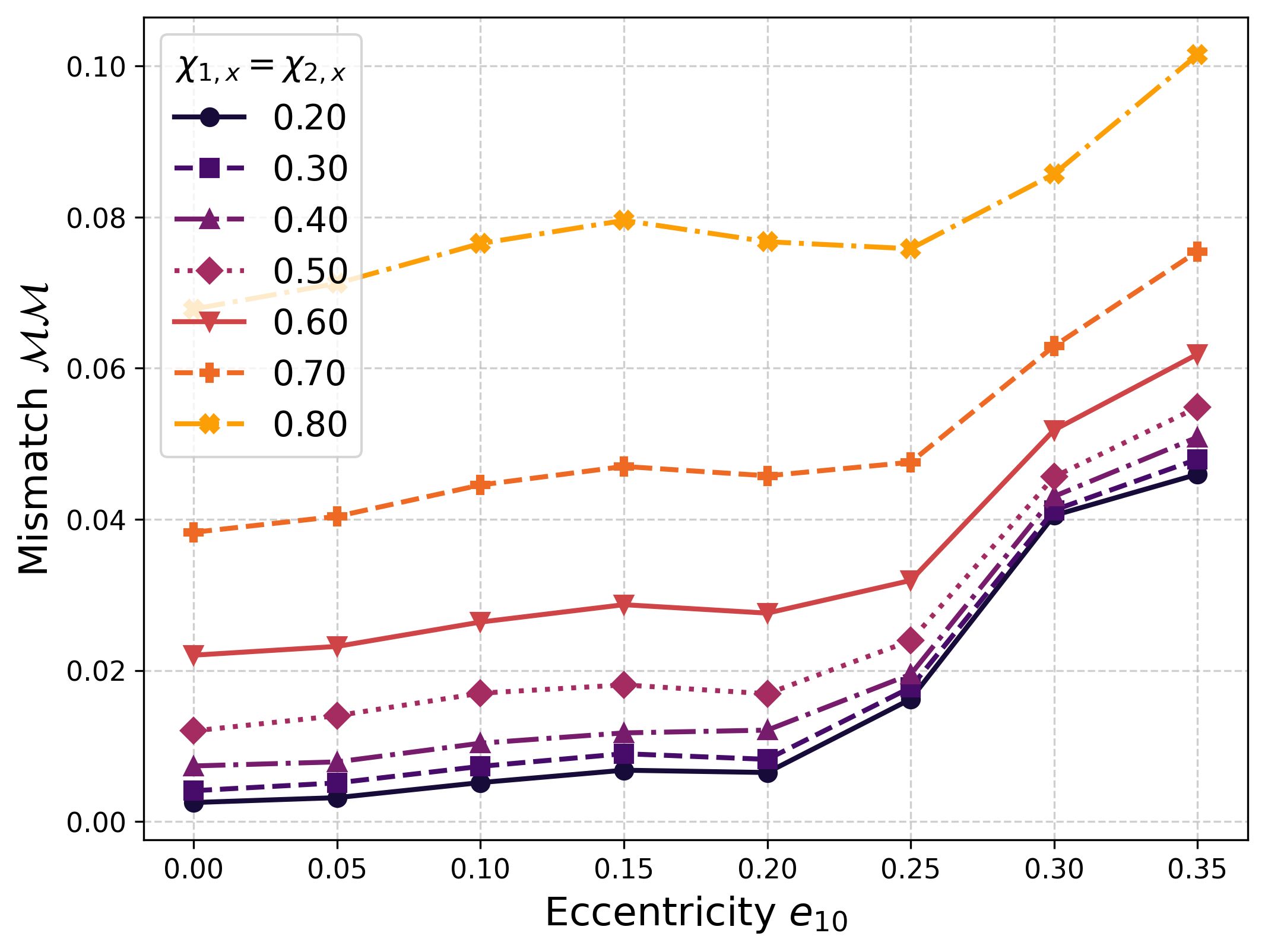}
}

\caption{This figure presents some results from the mismatch study. The plots show mismatch as a function of eccentricity for the various values of in-plane spins of the precessing system. We show 2 of the 9 ($M_{\text{tot}},\:q$) combinations here: $M_{\text{tot}}=200M_{\odot},\:q=2$ (top panel) and $M_{\text{tot}}=250M_{\odot},\:q=1$ (bottom panel). The three columns correspond to the different suite of mismatches, in-plane precessing systems versus A) non-spinning eccentric systems (\textit{left}) and B) aligned-spin eccentric systems (\textit{middle}), and C) precessing systems (with both in-plane and aligned spins) versus aligned spin eccentric systems (\textit{right}). In this additional suite (C), both the eccentric and precessing systems have the same aligned spin value. Note that we limit the in-plane spin in this case to $\chi_{1,x}= \chi_{2,x}=0.8$ as going above that pushes the spin magnitude values beyond unity, implying an unphysical system.}
\label{fig:contour_grid}
\end{figure*}

The number of cycles in band is a significant factor in identifying eccentricity~\cite{Romero-Shaw:2022fbf}. 
For a given total mass, the mass ratio and number of cycles are inversely related~\cite{Cutler:1994ys} (see Fig.~\ref{fig:FD_TD_short}); therefore we expect that misidentification of eccentricity is more likely in short duration signals. 
As a result, we focus on high total mass systems having $M_{\text{tot}}\in[200M_{\odot}, 300M_{\odot}]$ (detector-frame), while varying the mass ratios $q\in[1,3]$ to control the number of cycles in band. 

We describe the configuration setup of mismatches in Fig.~\ref{fig:mm_config}. 
The first two suites correspond to mismatch computations between precessing systems having only in-plane spins and A) non-spinning eccentric systems, $\chi_{1,z}=\chi_{2,z}=0$, or B) aligned-spin eccentric systems with $\chi_{1,z}=\chi_{2,z}=0.5$. We additionally consider a third suite in which both the eccentric and precessing systems have $\chi_{1,z}=\chi_{2,z}=0.5$, while the precessing systems retain their varying in-plane spins. For each suite, we perform mismatches for nine unique combinations of total mass and mass ratio. For each combination, we compute mismatches over a grid $\etenhz\in[0,0.35]$ and $\chi_{1,x}=\chi_{2,x}\in[0.2,0.9]$, except for the third suite in which case we limit our mismatches to $\chi_{1,x}=\chi_{2,x}\in[0.2,0.8]$ as going beyond this range results in unphysical spins.
Further technical details of the analysis settings in this study are in Appx.~\ref{sec:AnalysisConventions}.

\subsection{Mismatch study results}
\label{sec:mismatchB}

In Fig.~\ref{fig:contour_grid}, we present two of the nine unique eccentric versus precessing system mismatches for the three suites:
\begin{itemize}
    \item[A)] non-spinning eccentric vs in-plane spin precession (left)
    \item[B)] aligned-spin eccentric vs in-plane spin precession (middle)
    \item[C)] aligned-spin eccentric vs precession with both in-plane and aligned spin (right)
\end{itemize}

A complete set of mismatch plots as a function of eccentricity is provided in Appendix \ref{sec:apx_mm}.
The key takeaway from these results is that $<25\%$ of the configurations we analyze reveal regions of parameter space where a precessing system is best matched by an eccentric system. 
Even in those cases, the corresponding mismatches are typically comparable in magnitude to their quasi-circular counterparts. 
This points to the limited prevalence of eccentricity misidentification.

For the non-spinning eccentric suite of mismatches, the mismatch magnitude for a given precessing spin value increases monotonically as the precession increases. 
In contrast, the aligned spin eccentric suite of mismatches display a non-uniform oscillatory behavior, with different ($M_{\text{tot}},\:q$) combinations showing distinct orderings of the mismatch curves. 
In addition to that, the mismatches vary strongly for a given precessing system with in-plane spins only, across the eccentricity range. 
When the same aligned-spin components are included in both the eccentric and precessing systems, this non-uniform oscillatory behavior is reduced and the mismatch varies more smoothly with eccentricity. This indicates that the oscillations primarily arise from the difference in aligned-spin content between the two systems, rather than from the eccentricity--precession comparison alone. Nevertheless, for precessing systems, a minimum mismatch can occur at a nonzero eccentricity when the aligned-spin components differ.
It remains to be seen if there is a strong local minimum for the mismatch with a system with nonzero eccentricity and some nonzero spin, which could cause a misidentification of eccentricity in PE.

These results highlight that waveform overlaps can vary in possibly unexpected ways when 
complex physical dynamics are involved, making degeneracy regions highly localized and harder to predict. 
Another noticeable feature from the mismatch study is that the dependence of mismatch on eccentricity follows similar qualitative trends across all in-plane spin values for precessing systems with the only variance coming from mismatch magnitude. 
Notably, none of the non-spinning eccentric systems matches any of the precessing system better than the corresponding quasi-circular case. 
This indicates that non-spinning eccentric systems are unlikely to be confused with precessing systems, even for relatively short signals. 
Similarly, precessing signals with the same contributions from spins aligned to the orbital plane as the eccentric systems, are also unlikely to be confused with eccentric systems.
Indications of possible degeneracies arise only when precessing systems have a different aligned-spin component as compared to the eccentric system.

The rough oscillation of the mismatches in case (B) raises the natural question of whether minimizing additional parameters would reveal trends with less variation. 
We explored minimizing mismatch over relativistic anomaly (in addition to $\Delta t$, $\Delta \phi$, $\Delta \psi$) for a particular combination ($M_{\text{tot}}=250M_{\odot},\:q=1$), and found no clear pattern in the resulting mismatches.

To quantify the results from this study and isolate precessing configurations most probable to exhibit misidentification of eccentricity, we identified the following criteria. 
A precessing system with in-plane spins only was shortlisted 
\begin{enumerate}
    \item  if its mismatch with an aligned-spin eccentric system $\etenhz >0.05)$ is lower than its mismatch with an aligned-spin quasi-circular system $(\etenhz \leq 0.05)$, and 
    \item if that mismatch is the lowest within its corresponding mismatch grids, and  
    \item  if that mismatch is also lower than the mismatch of the corresponding precessing systems with any non-spinning eccentric system.
\end{enumerate}
If a precessing system satisfied all these conditions, it was selected as a candidate injected signal for our PE study. The additional matched-aligned-spin suite (C) was used to investigate the origin of the oscillatory behavior and does not alter the injection-selection procedure described above. 
These criteria resulted in one injection only (Inj~\#5). 
To collect additional cases, we relaxed the criteria by bypassing the second condition, which yielded all additional injections, culminating into a total of eight precessing injections set to be analyzed with the non-precessing, eccentric waveform model.

\begin{table}[tb]
    
    \begin{tabular}{ccccccc}
    \hline
        \toprule
        \textbf{Inj~\#} & $q (\frac{m_{1}}{m_{2}})$ & $M_{\mathrm{tot}} (M_\odot)$ & $M_{\mathrm{chirp}} (M_\odot)$ & $\chi_{\mathrm{eff}}$ & $\chi_{p}$ & $D_{L} (Mpc)$ \\
        \midrule
        \hline
        1 & 1.00 & 200 & 87.06 & 0.0 
        & $0.700$ & $2034.43$ \\

        2 & 2.00 & 200 & 81.12 & 0.0 
        & $0.385$ & $270.17$ \\

        3 & 3.00 & 200 & 73.25 & 0.0 
        & $0.192$ & $269.00$ \\

        4 & 3.00 & 200 & 73.25 & 0.0 
        & $0.269$ & $272.75$ \\

        5 & 1.00 & 250 & 108.82 & 0.0 
        & $0.700$ & $269.71$ \\

        6 & 3.00 & 250 & 91.57 & 0.0 
        & $0.308$ & $271.16$ \\

        7 & 2.00 & 300 & 121.67 & 0.0 
        & $0.385$ & $273.05$ \\

        8 & 3.00 & 300 & 109.88 & 0.0  
        & $0.308$ & $271.93$ \\
        \bottomrule
        \hline
    \end{tabular}
    \caption{Precessing injections. \newline\textit{Note:} Columns with no unit mentioned indicate unitless quantities.}
\label{table:inj_tab}
\end{table}

\begin{table}[tb]

\begin{tabular}{|l|l|}
\toprule
\hline
\textbf{Parameter} & \textbf{Value} \\
\midrule
\hline
SNR ($\rho$) & 30.0 \\
Inclination ($\iota$) & $\pi/3$ \\
Right ascension ($\alpha$) & $0.52359878\,\mathrm{rad}$ \\
Declination ($\delta$) & $1.0471976\,\mathrm{rad}$ \\
Polarization angle ($\psi$) & $0.0\,\mathrm{rad}$ \\
Coalescence phase ($\phi$) & $0.0\,\mathrm{rad}$ \\
\bottomrule
\hline
\end{tabular}
\caption{Fixed parameters across all injections}
\label{table:fixed_param}
\end{table}

\section{Parameter estimation study}
\label{sec:PE}

In this section, we present the results of our PE study on the eight precessing systems (summarized in Table \ref{table:inj_tab}) injected using SEOBNRv5PHM with $l_\mathrm{max}=4$ modes and recovered using the eccentric model SEOBNRv5EHM also with $l_\mathrm{max}=4$ modes. 
The aim of this analysis is to assess if PE recovers apparent evidence for eccentricity when the true signal is governed by spin-induced precession, and thereby identify degeneracies between the two effects. 
One of the eight precessing systems, Inj~\#5, displays substantial evidence for nonzero eccentricity, even though the injected signal was quasi-circular. 
This is an equal-mass binary with a total mass of $250 M_{\odot}$ and $\chi_{p}=0.7$. 
The eccentricity posterior for this particular system peaks at $\etenhz=0.16^{+0.05}_{-0.06}$. 
All other injections yield eccentricity posteriors that either peak at zero or exhibit a broad, plateau-like morphology that lacks evidence to support presence of eccentricity.
Interestingly, Inj~\#5 is also the only system shortlisted by the stringent criteria described in Sec.~\ref{sec:mismatchB}, which was eventually relaxed to obtain the additional systems. 
In the following subsections, we discuss the results in detail.

\subsection{PE methodology}
\label{sec:PEMethods}

We use Bayesian inference on injected signals using the parameter inference pipeline {\tt RIFT} 0.0.15.9-918-g39ce2f0a~\cite{Pankow:2015cra, Lange:2018pyp, Wofford:2022ykb, Wagner:2025bih,PhysRevD.96.104041}, specifically commit
\texttt{39ce2f0ac01bfabc68db5ab5efd01cde9cda5d66}. 
{\tt RIFT} utilizes a grid-based likelihood calculation approach that aims to converge to the true value over the course of multiple iterations.  
This formulation allows for parallelization and is therefore substantially faster than standard MCMC or nested sampling based PE pipelines. 
This is especially useful when using computationally expensive waveforms such as EOB models. 
Another distinguishing feature of {\tt RIFT} is that it marginalizes over extrinsic parameters and computes their posteriors in a separate stage. 
This accelerates the likelihood calculations for intrinsic parameters most relevant to our degeneracy study. 

For each system in Table~\ref{table:inj_tab}, we inject it into the three-detector network of Hanford, Livingston, and Virgo at their design sensitivity~\cite{LIGO-T2200043} and assuming a zero-noise realization.
The extrinsic parameters are given in Table~\ref{table:fixed_param}, and the distance is adjusted to ensure that all injections have a signal to noise ratio (SNR) of $\rho = 30$.

We use broad and agnostic priors on the intrinsic and extrinsic parameters for our eccentric signal model, as described in Appx.~\ref{sec:ParamConventions}.
Since SEOBNRv5EHM is an aligned-spin model, we adopt the aligned-spin prior on the component of the spin along the orbital angular momentum~\cite{Lange:2018pyp}, which tends to favor small spins. 
We set the prior on eccentricity $\etenhz$ and relativistic anomaly $\zeta$ to be uniform with $\etenhz \in [0,0.35]$ and $\zeta \in [0,2\pi)$ at the reference frequency.

\subsection{Results of injection-recovery}

\begin{figure}[tb]
\includegraphics[width=\columnwidth]{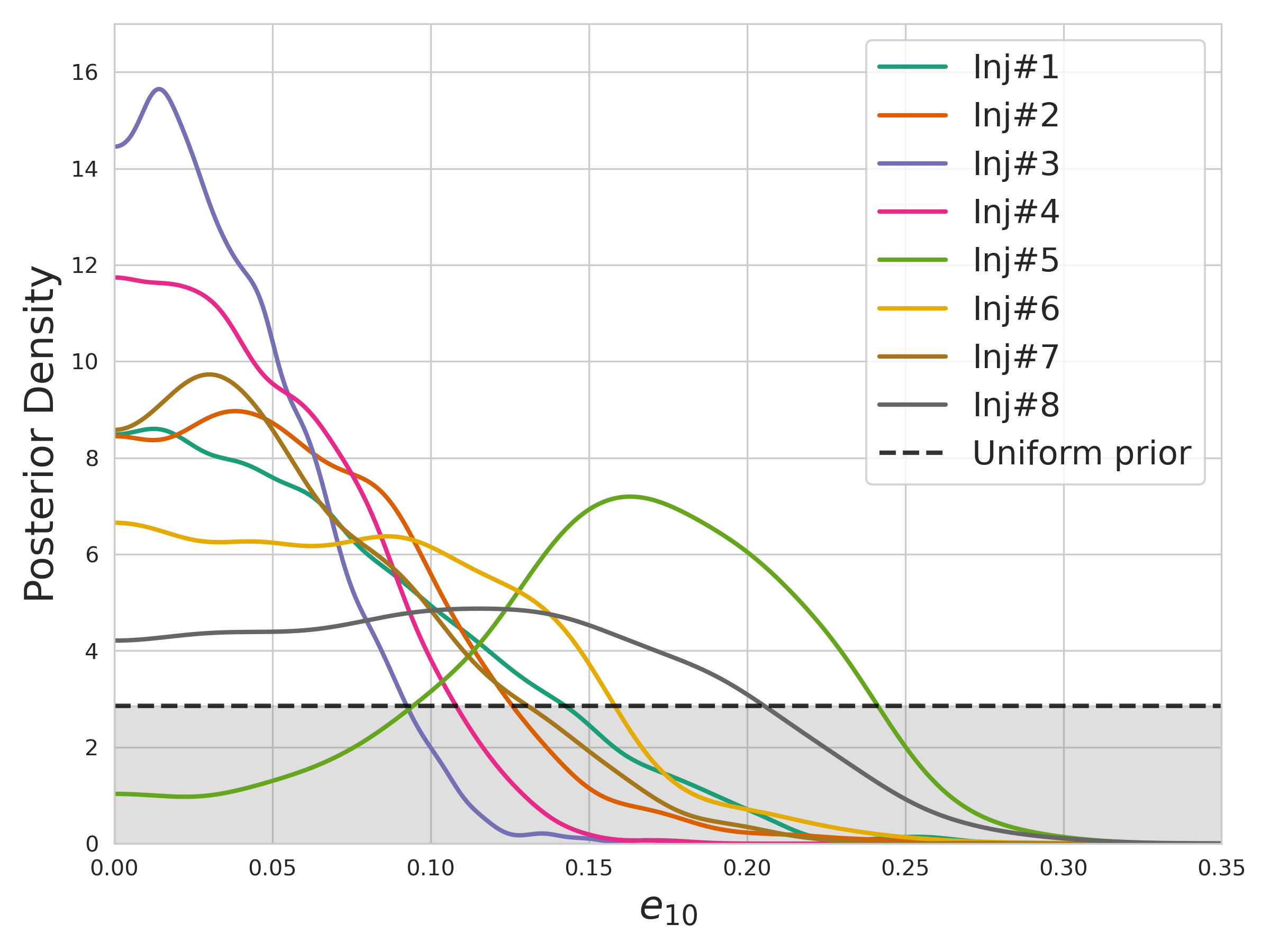}
\caption{This plot shows the recovered marginalized posterior KDEs for $\etenhz$ for each of our precessing injections.}
\label{fig:EccentricPosteriors}
\end{figure}

In Fig.~\ref{fig:EccentricPosteriors} we see that for the majority of our injections, the recovery favors zero eccentricity. 
There's only a singular case where the marginalized posterior $p(\etenhz \mid d)$ clearly peaks away from $\etenhz = 0$. 
Inj~\#5 in Table~\ref{table:inj_tab} is our most significant case of misidentification. It is an equal-mass, precessing system of total mass $250\,M_\odot$ with $\chi_{1,x} = \chi_{2,x} = 0.7$, resulting in $\chi_p=0.7$ and $\chi_\mathrm{eff}=0$. 
We can use these marginalized posteriors to evaluate the Bayes factor  $\mathcal{B}_{E/C}$ comparing the hypothesis of nonzero eccentricity to the quasi-circular limit.
For this we invert the
Savage-Dickey density ratio~\cite{Dickey:1971, Verdinelli:1995}, giving 
\begin{equation}
\mathcal{B}_{E/C} = \frac{\pi(e=0)}{p(e=0 \mid d)} .
\end{equation}
Generally, $\mathcal{B}>1$ indicates support for the hypothesis.
To further add to the robustness of this metric, we quantify the uncertainty in the Savage-Dickey density ratio using bootstrap resampling. 
For each posterior, we draw samples with replacement, recompute the boundary-corrected density estimate at $(\etenhz=0)$, and repeat this procedure $1000$ times. 
We report the resulting median Bayes factor and the corresponding $68\%$ credible interval.
The posterior for Inj~\#5 provides $\mathcal{B}_{E/C} = 2.55^{+0.28}_{-0.27}$ which shows mild support for eccentricity.

\begin{figure}[tb]
    \includegraphics[width=0.45\textwidth]{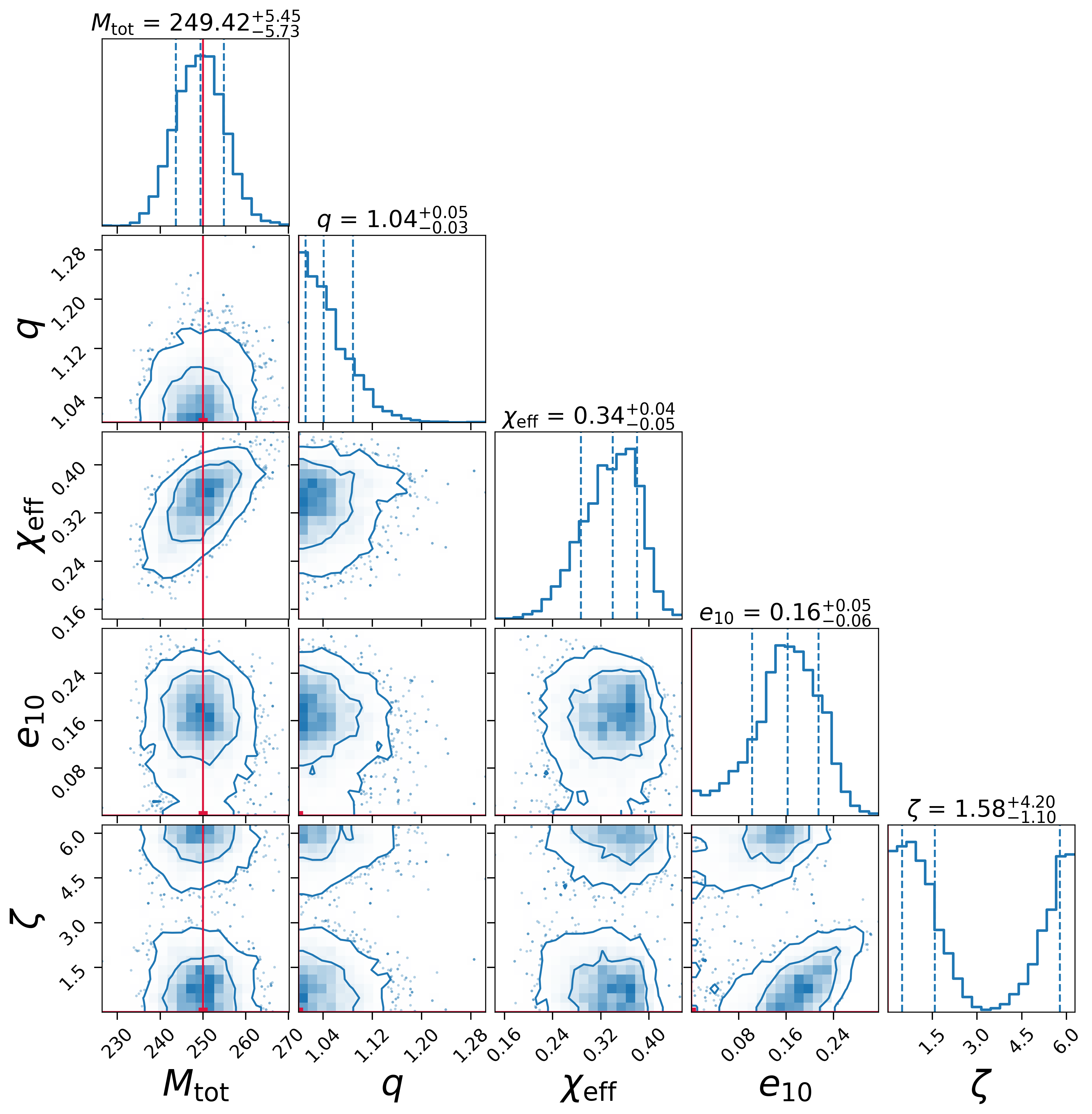}
    \caption{Corner plot for Inj\#5 with 2D and 1D posteriors for total mass $M_\mathrm{tot}$, mass ratio $q$ (where $q=m_{1}/m_{2}$), effective spin parameter $\chi_\mathrm{eff}$, eccentricity $\etenhz$ and relativistic anomaly $\zeta$. All posteriors are reported at the reference frequency, $f_\mathrm{ref}=10$ Hz.}
    \label{fig:fig5}
\end{figure}

The PE results for Inj~\#5 can be found in Fig.~\ref{fig:fig5}. 
When recovering this injection with the eccentric model, we find that the injected masses and the aligned-spin component of the secondary ($\chi_{2,z} = 0$) lie within the 90\% credible intervals; however, the primary aligned-spin component $\chi_{1,z}$ is overestimated, leading to a significant bias in $\chi_\mathrm{eff}$. 
Most importantly, we find that the eccentricity posterior peaks at $\etenhz=0.16^{+0.05}_{-0.06}$. 
Since the signal was injected with zero eccentricity, this is a strong indication of the statistical degeneracy between eccentricity and precession in this part of the parameter space. 
The posterior for relativistic anomaly $\zeta$ is unimodal and peaks at $0$ (note that it is a periodic variable).

Apart from Inj~\#5, we observe more ambiguous evidence for the presence of eccentricity in only two other cases. 
These are Injs~\#6 and \#8 in Tab.~\ref{table:inj_tab}. 
These injections result in plateaued eccentricity posteriors that span $0.0\lesssim \etenhz \lesssim 0.2$. 
Although the posteriors exhibit a broad plateau, computation of Bayes factors reveals $\mathcal{B}_{E/C} = 0.67^{+0.04}_{-0.04}$ for Inj~\#8 and $\mathcal{B}_{E/C} = 0.44^{+0.02}_{-0.02}$ for Inj~\#6. 
A $\mathcal{B}_{E/C}<1$ implies lack of evidence for the hypothesis, i.e.~the evidence from these posteriors indicate very mild preference for a quasi-circular signal.

\begin{figure}[tb]
    \includegraphics[width=0.45\textwidth]{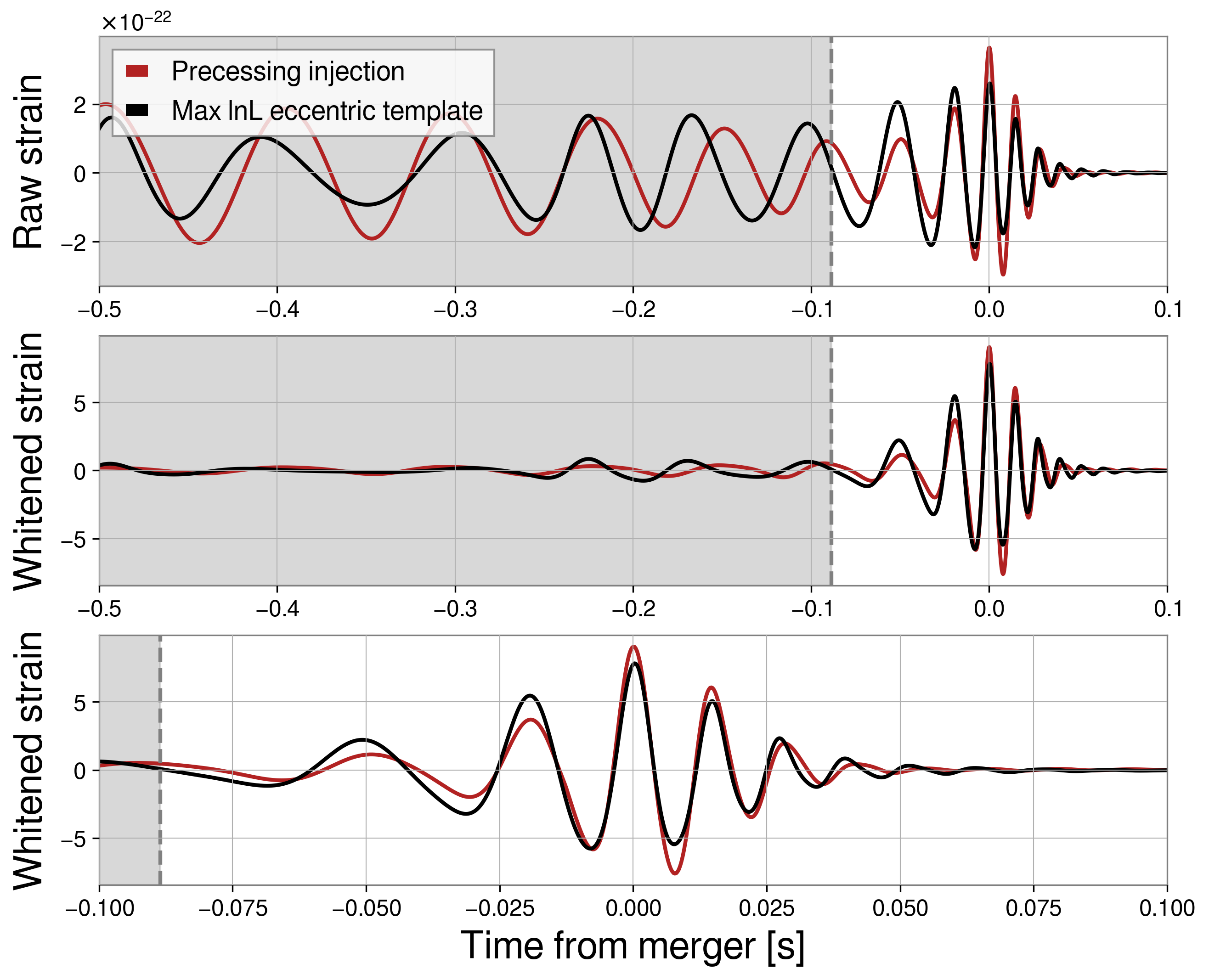}
    \caption{Inj~\#5 (in red) plotted alongside the highest likelihood point computed (in black). The gray dotted line marks the beginning of the detector sensitivity band at 20 Hz. {\it Top and middle:} Zoom out, showing that last $0.5$ s of the raw and whitened signals respectively.
    {\it Bottom:} Zoom in on the region above $20$ Hz for the whitened strains.}
    \label{fig:WaveformRec}
\end{figure}

Finally, to qualitatively understand the misidentification of the eccentricity, in Fig.~\ref{fig:WaveformRec} we show the waveform used for Inj~\#5 and the reconstructed waveform for the highest likelihood point from our posterior distribution. 
We see that the injected and recovered signals differ significantly over long time scales. 
However, it is only the last 4-6 cycles that contribute majorly to the likelihood, and the signals have significant overlap during these final cycles. 

\subsection{Follow-up investigations}

To further understand trends surrounding the degeneracy, we conducted two additional studies. 
For both, we focus on the precessing system that resulted in the most significant misidentification (Inj~\#5 in Table~\ref{table:inj_tab}) and surrounding regions in the parameter space. 
For the first study, we generated two more injections by varying total mass such that we have a set of three injections with $M_{\text{tot}}\in\{200M_{\odot}, 250M_{\odot}, 300M_{\odot}\}$. 
Our aim was to study the exact same physical system, scaled only by mass, such that we only vary the number of cycles in band. 
For this reason, we injected these three systems with the remaining parameters identical to Inj~\#5, but matching the parameters that vary over the coalescence at a varying reference frequency such that $M_\mathrm{tot} \, f_\mathrm{ref} = 250\,M_{\odot}\times 10\,{\rm Hz}$ is fixed. 
We compare the three recovered posteriors.

\begin{figure}[tb]
    \includegraphics[width=0.45\textwidth]{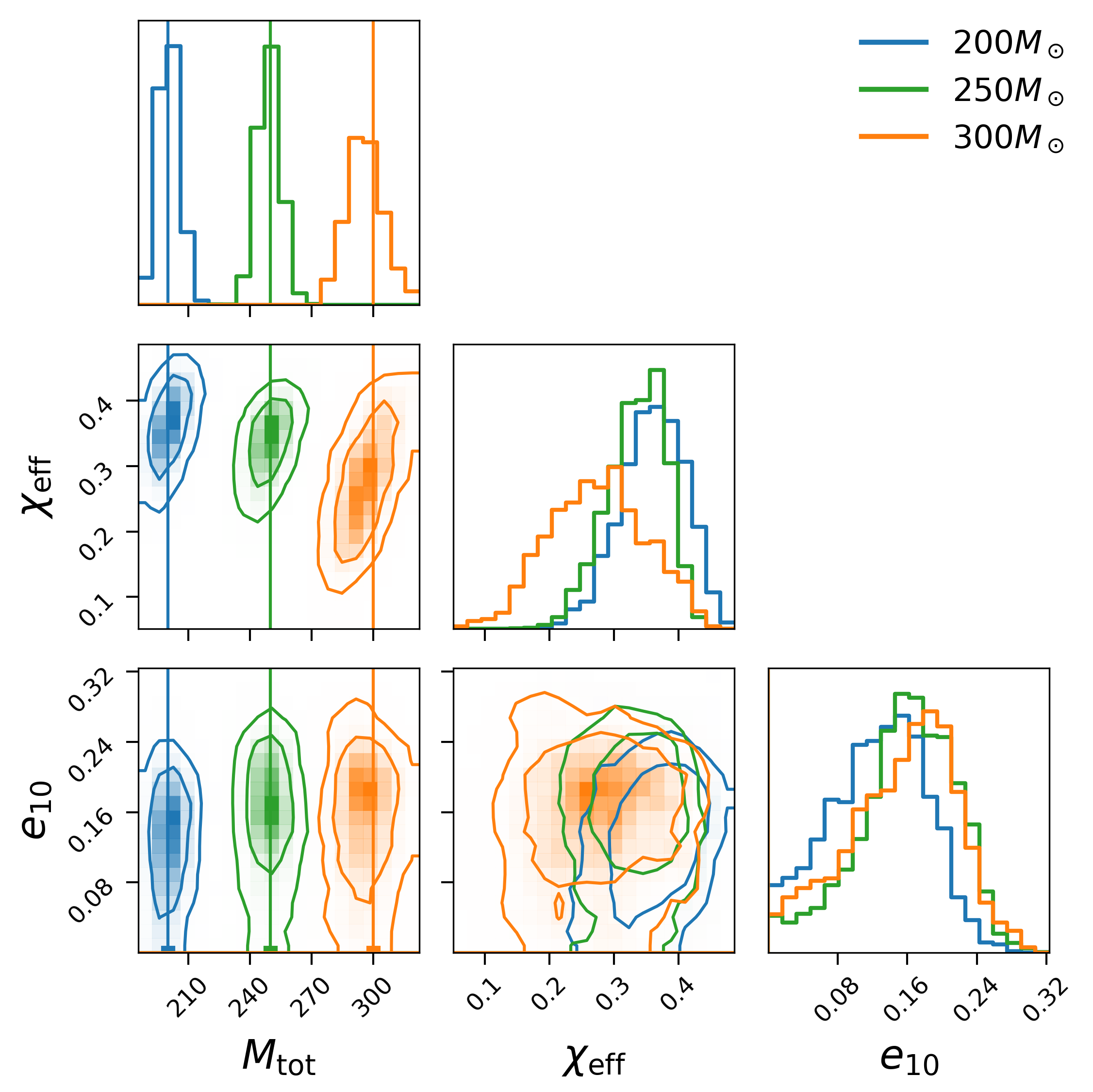}
    \caption{Comparison across masses: This corner plot shows the 1D marginal distribution as well as the 2D 50\% and 90\% credible intervals of the resulting posteriors from injections of Inj~\#5 from Table~\ref{table:inj_tab} with two other Inj~\#5-like signals with varying $M_{\text{tot}} \, f_\mathrm{ref}$. The blue, green, and orange distributions represent the posteriors from the $\{200M_{\odot}, 250M_{\odot}, 300M_{\odot}\}$ respectively. These results indicate an inverse scaling between bias in eccentricity (i.e.~misidentifying eccentricity) and the number of in-band cycles.}
    \label{fig:MassStudy}
\end{figure}

As can be seen in Fig.~\ref{fig:MassStudy}, as the total mass increases, the eccentricity posteriors peak at marginally higher values of eccentricity. 
With $200\,M_{\odot}$, $250\,M_{\odot}$ and $300\,M_{\odot}$ corresponding to 7, 6 and 5 cycles in-band respectively, the degree of the misidentified eccentricity appears to scale inversely to the number of cycles in-band. 
We also observe that all three cases report a biased posterior $\chi_\mathrm{eff}$ (recall that the injected $\chi_\mathrm{eff}$ vanishes). 
The bias seemingly decreases as total mass (and inferred eccentricity) increases. 
This hints at a mild inverse correlation between $\chi_\mathrm{eff}$ and inferred eccentricity. 
Further investigation is needed to explore these relationships.

\begin{figure}[tb]
    \includegraphics[width=0.45\textwidth]{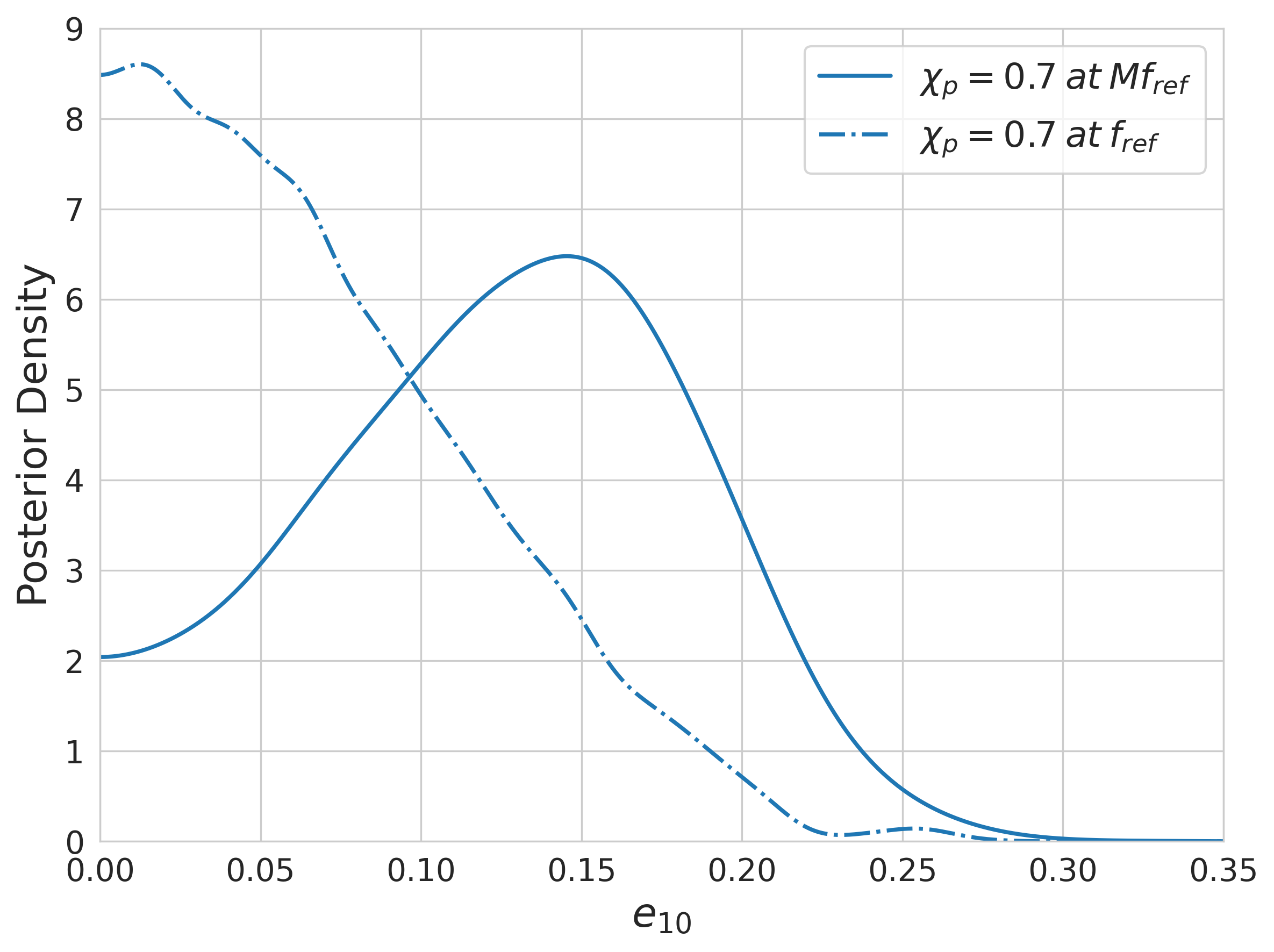}
    \caption{Comparing different precessional dynamics for the same $q=1$, $200M_{\odot}$ system. Here we inject the same $\chi_{p}$ by fixing $\chi_{1,x}=\chi_{2,x} = 0.7$ at different reference frequencies, which induces differing precessional effects.
    }
    \label{fig:M200_diffSpins}
\end{figure}

In contrast, Fig.~\ref{fig:M200_diffSpins} shows that varying only the details of the precessional dynamics while keeping every other parameter fixed results in the posterior distribution consistent with zero eccentricity. 
For this comparison we create an additional injection, repeating our $M_\mathrm{tot} = 200\, M_\odot$ case from the mass comparison but instead setting the injection reference frequency to $f_\mathrm{ref} = 10$ Hz (as with our baseline Inj~\#5 with $250\, M_\odot$).
This emphasizes the sensitivity of statistical degeneracy of these systems across parameter space.


\begin{figure}[tb]
    \includegraphics[width=0.45\textwidth]{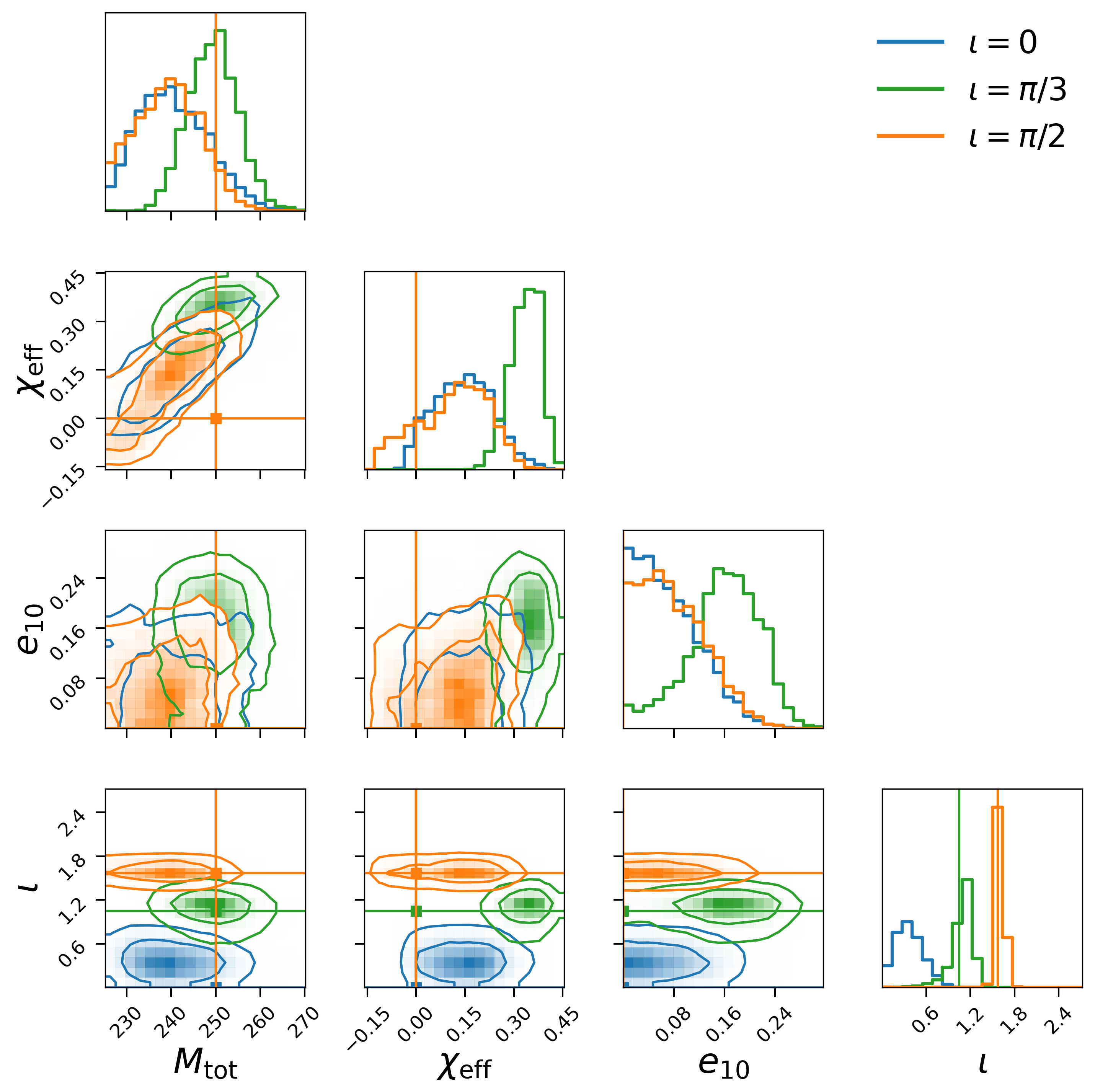}
    \caption{Comparison across inclinations: Similar to Fig.~\ref{fig:MassStudy}, this corner plot shows the 1D marginal distribution as well as the 2D 50\% and 90\% credible intervals of the resulting posteriors from injections of Inj~\#5 from Table~\ref{table:inj_tab} with two other Inj~\#5-like signals with varying inclination. The blue, green, and orange distributions represent the posteriors from $0,\pi/3,\pi/2$ respectively. This highlights the high localization in parameter space that can produce biases in different parameters including eccentricity.}
    \label{fig:InclStudy}
\end{figure}

For the second study, we generated two more systems by varying the inclination angle while keeping all other parameters the same. 
The inclination affects the relative strength of higher modes to the fundamental quadrupolar emission, which can induce modulations in the signal. 
This motivated our exploration of the sensitivity of the misidentification of eccentricity to the degree of inclination. 
Fig.~\ref{fig:InclStudy} displays a comparative corner plot for the inferred parameters for the system of Inj~\#5 while varying the inclination.
Interestingly, deviation from the inclination angle in either direction results in a correctly inferred eccentricity $\etenhz \sim 0$, i.e.~no misidentification, while increasing the bias in other parameters like the total mass and inclination. 
This further demonstrates that the degeneracy between eccentricity and precession effects can be highly localized in parameter space.

\section{Discussion}
\label{sec:discussion}

Our results clearly show that spin-precession can be misidentified as eccentricity, but only within specific regions of the parameter space. Out of the eight precessing, quasi-circular injections we targeted, only one demonstrates a significant bias towards nonzero eccentricity when recovered with an eccentric, aligned spin model. 
This confirms that the eccentricity-precession degeneracy is not generic for all short signals but instead emerges under specific parameter permutations for masses, spins, and inclination angles.

Here we summarize the key takeaways from our study:
\begin{itemize}
    \item \textbf{Eccentricity and spin-precession can be degenerate but only in highly localized regions of the parameter space}, depending on combinations of total mass, mass ratio, and strong in-plane spin components that mimic eccentric modulation over the final few observable cycles.
    \item \textbf{Mismatch-based diagnostics proved effective in pinpointing these regions}, with the one configuration flagged by our selection criteria corresponding to the only degenerate system uncovered by PE.
    \item \textbf{Non-spinning eccentric signals do not mimic spin-precession waveform morphology}, showing that the degeneracy is driven by a combination of eccentricity \textit{and} nonzero aligned spins.
    \item \textbf{Precessing systems where part of the spin is aligned with the orbital angular momentum do not mimic eccentric waveform morphology with the same aligned-spin component}, showing that precession alone does not mimic eccentricity. A special configuration of in-plane spins and aligned spins is required to trigger this effect.
    \item \textbf{For degenerate systems, the eccentricity posteriors peak at marginally higher values as the signal becomes shorter i.e.~more massive}. This implies a mild increase in the misidentified eccentricity with the decrease of number of cycles in-band which underscores the need for accurate waveform modeling in that regime. 
    \item \textbf{This eccentricity-precession degeneracy is highly sensitive to the angle of inclination}. We find that the misidentification disappears for inclinations less than or greater than our injected value. Further exploration is needed to understand the effect of inclination angle on the degeneracy. 
\end{itemize}
The physical picture that emerges from this work is largely consistent with prior work. 
In the high total mass regime, the modulations induced by spin-precessional dynamics can start to mimic the eccentricity modulations--especially in the last few cycles captured by the detectors. 
The distinctive features that can otherwise distinguish these effects have a stronger presence in the inspiral phase of the waveform. 
For the massive systems we look at, it is the merger-ringdown that predominantly contributes to the signal-to-noise ratio. 
These effects are captured by our mismatch-based diagnostic step, successfully identifying the precessing system with similar waveform morphology as an aligned spin, eccentric system. In contrast, none of the non-spinning eccentric system produce a good match with the precessing systems. 

At the same time, our investigations find that the degeneracy remains sensitive to a multitude of parameters simultaneously. 
Relatively small changes in parameters like in-plane spins or inclination of the signal waveform can amplify or dampen the degeneracy entirely. 
This highlights how subtle changes in the waveform morphology are tuned to the multi-dimensional parameter space. 
Subsequently, the degeneracy is likely sensitive to systematic differences in waveform models. 
For the high total mass systems we studied, merger-ringdown play an influential role on the signal and so any inaccuracies in modeling of this domain is enhanced. 
Note that because the SEOBNRv5EHM model is calibrated with quasi-circular NR simulations, there is an upper limit to $\etenhz$ beyond which significant eccentricity is present during merger, and so the model may lose accuracy.
At the masses explored in this paper, the upper limit is near the largest values of $\etenhz$ where our posteriors have support, and so we expect our results are robust to this modeling limitation, but it would be interesting to explore the same systems with future models that account for the impact of eccentricity on merger-ringdown.

This work has direct implications for the analysis of short duration signals like GW190521 and GW231123.
With only the last few cycles in band, the PE for these events can be susceptible to biases introduced due to lack of complete physics in waveform models. 
Based on the results of this work, caution should be exercised when interpreting results for such systems from analyses which neglect the effects of eccentricity or orbital precession.

This work points to multiple directions for better understanding and mitigating the possible misidentification of eccentricity.
The sensitivity of this degeneracy to the morphology of the signal, for example through varying inclination and precessional dynamics, means that a wider (and computationally expensive) exploration of parameter space is required to better understand it.
Ultimately, self-consistent models which incorporate both eccentricity and spin-precession are required to fully account for such degeneracies and measure the properties of astrophysical BBHs.
Complete inspiral-merger-ringdown models with such effects are only just becoming available~\cite{Gamba:2024cvy}.
The accurate development of these and future models relies heavily on the availability of NR waveforms that capture these physical effects.
Therefore, the expansion of NR catalogs especially with higher eccentricities, higher precession, and well resolved higher modes are needed to increase robustness as well as expand the domain of validity for waveform models. 

Overall, our study provides the first explicit and controlled example of GW signals from massive BBHs whose spin-precession can be misidentified as orbital eccentricity.
As waveform models continue to improve, and as more heavy binaries are detected, understanding and mitigating this degeneracy will be critical for confidently measuring eccentricity in such systems.
A clear measurement of eccentricity in a BBH would rule out the standard isolated evolutionary channel for that system, deepening our understanding of the origin of GW sources and the astrophysical processes that create them.

\section*{Acknowledgments}
We would like to thank Aasim Z.~Jan, Isobel Romero-Shaw, Charlie Hoy, Sylvia Biscoveanu, Lucy Thomas and Lorenzo Piccari for their helpful comments. The posterior samples for this work can be found at \url{https://doi.org/10.5281/zenodo.17886746}.
ST, JL, and DS were supported by NSF grant PHY-2207780 at the University of Texas at Austin.
AZ was supported by PHY-2308833.
JL is grateful for the support from the Italian Ministry of University and Research (MUR) via the PRIN 2022ZHYFA2, \textit{GRavitational wavEform models for coalescing compAct binaries with eccenTricity} (GREAT) while at INFN-Torino.
The authors are grateful for computational resources provided by the LIGO Lab and supported by NSF Grants PHY-0757058 and PHY-0823459.
This work has preprint numbers UT-WI-37-2025 and LIGO-P2500767. 

\appendix
\section{Parameter conventions}
\label{sec:ParamConventions}

We mostly use standard conventions in GW astronomy, e.g.~\cite{LIGOScientific:2025hdt}.
Two important exceptions are that we adopt a mass ratio $q \geq 1$ as commonly used in the NR community, and we report detector frame masses in all cases, without undoing the effects of cosmic redshifting of the signals.
This is appropriate given that our focus is on signals with a small number of cycles in the detector's sensitive frequency band, and it is the detector frame  (rather than true source-frame) parameters that determine the signal as observed.

The (detector-frame) component masses of the binary are denoted $m_i$, where $i$ indexes over $i \in \{1,2\}$, and we denote the more massive primary with 1 and the less massive secondary as 2.
The total mass $M_\mathrm{tot} = m_1 + m_2$, the mass ratio is $q = m_1/m_2 \geq 1$, and the chirp mass $\mathcal M_c$ is defined as
\begin{align}
    \mathcal M_c = \frac{(m_1m_2)^{3/5}}{M_\mathrm{tot}^{1/5}} \,.
\end{align}
In our mismatch and PE studies, all quantities that vary over the binary coalescence, such as spin quantities and the eccentricity, are quoted at a reference frequency of $f_\mathrm{ref} = 10$ Hz.

The dimensionless spins are $\vec \chi_i$, and we decompose them in a Cartesian coordinate system where the binary angular momentum is along the $z$ axis and the $x$ and $y$ axes span the orbital plane, with the components lying on the $x$-axis at the reference frequency.
Conventionally two combinations of the spins are reported due to their importance in the overall evolution of the signal.
The effective aligned spin~\cite{Ajith:2009bn, Santamaria:2010yb} is defined as
\begin{equation}
\chi_{\rm eff} = \frac{m_1 \vec{\chi}_1 \cdot \hat{L} + m_2 \vec{\chi}_2 \cdot \hat{L}}{m_1 + m_2} = \frac{m_1 \chi_{1,z} + m_2 \chi_{2,z} }{m_1 + m_2},
\end{equation}
where $\vec{\chi}_i$ are the dimensionless spin vectors of the two components and $\hat{L}$ is the unit vector along the orbital angular momentum. 
The effective spin is approximately conserved during inspiral~\cite{Gerosa:2015tea}, and encodes the leading spin-effects on the GW signal, at least at comparable masses.

The effective precession parameter $\chi_p$ quantifies the in-plane spin components that drive precession~\cite{Schmidt:2014iyl} and is defined as
\begin{equation}
\chi_p = \max\left( \chi_{1,\perp}, \frac{4 q + 3}{4 + 3 q} \chi_{2,\perp} \right),
\end{equation}
where $\chi_{i,\perp}$ denote the components of $\vec{\chi}_i$ perpendicular to $\hat{L}$, and again $q = m_1/m_2 \geq 1$.
For our precessing injections (both in mismatch and full PE studies) we conventionally set $\chi_\mathrm{eff} = 0$, and $\chi_{1,x} = \chi_{2,x}$ with the remaining spin components zero, such that $\chi_p = \chi_{1,x}$.

\section{Settings for mismatch and PE studies}
\label{sec:AnalysisConventions}

Our noise-weighed inner product between two signals $h_1$ and $h_2$ is the standard form for stationary Gaussian noise~\cite{LIGOScientific:2025hdt, Veitch:2014wba, Thrane:2018qnx}, and is used in both our mismatch study and in the GW likelihood used in PE.
It is represented by (see also~\cite{Thrane:2018qnx} for discrete representations)
\begin{align}
    \langle h_1 | h_2 \rangle & = 4 \, \mathrm{Re} \, \int_{\flo}^{\fhi}\frac{h_1^*(f) h_2(f)}{S_n(f)} df \,.
\end{align}
Here, $S_n(f)$ is the PSD, and $\flo$ and $\fhi$ are the lower and upper frequency cutoffs for the integral.
For the mismatch studies we use the PSD for the Advanced LIGO design~\cite{LIGO-T2200043}; for the PE studies we inject the signal consistently into a network consisting of the Hanford, Livingston, and Virgo detectors, using their design PSDs~\cite{LIGO-T2200043}.
For both the mismatch and PE studies we set $\flo = 20\, \mathrm{Hz}$.
For the mismatch study we used $\fhi = 2046$ Hz, and for PE we used $\fhi = 1024$ Hz corresponding to a sample rate of $2048$ Hz; these cutoffs are expected to be sufficient to resolve higher modes during merger for the high-mass systems we study. 

For the mismatch study we compare a precessing signal $h_p$ to eccentric signal $h_e$ in a single detector.
Many of the intrinsic parameters are degenerate in this case~\cite{Cutler:1994ys}, but for concrete application we assume both signals to be located on the sky at spherical polar coordinates $\theta = \pi/3$ and $\varphi = \pi/6$ in the detector's reference frame.
For $h_p$ we set $\phi = 0$, $\psi = 0$, with arbitrary coalescence time $t = 0$.
These same parameters for $h_e$ are then set by the maximization within the mismatch.

Bayesian inference involves evaluating the likelihood of data $d$ given at set of parameters $\boldsymbol{\theta}$ based on the signal and noise models being used. 
This likelihood is then combined with the provided priors and sampled to give the final posteriors, in our case using the RIFT software package.
Bayes' theorem is written as:
\begin{equation}
p(\boldsymbol{\theta} \mid d) 
= \frac{\mathcal{L}(d \mid \boldsymbol{\theta}) \, \pi(\boldsymbol{\theta})}
{\mathcal{Z}} \,
\end{equation}
where $p(\boldsymbol{\theta} \mid d)$ is the posterior, $\mathcal{L}(d \mid \boldsymbol{\theta})$ is the likelihood, $\pi(\boldsymbol{\theta})$ is the prior, and $\mathcal{Z}$ is the evidence (marginal likelihood) which normalizes the probability density.
We use the standard GW likelihood, e.g.~\cite{Veitch:2014wba,Thrane:2018qnx,LIGOScientific:2025yae}, which assumes a coherent signal across the detector network and which assumes independent, stationary, Gaussian noise in each detector as described by the PSD $S_n(f)$.

We use broad, relatively agnostic priors, uniform in (detector frame) component masses with specified boundaries, uniform in localization volume (assuming a Euclidean cosmology for simplicity), uniform in binary orientation, and with a flat prior on the coalescence time in a small window around the injection. The distance prior is proportional to $d_L^2$.
As discussed in Sec.~\ref{sec:PEMethods}, we adopt a prior for $\chi_{i,z}$ appropriate for isotropic spins drawn with magnitudes uniform in $\chi_i \in [0,1]$, then projected onto the $z$-axis.
This favors smaller aligned-spin values.
Our eccentricity prior is uniform, $\etenhz\in [0,0.35]$, as is our relativistic anomaly $\zeta \in [0,2\pi)$.
The ranges are given in Table~\ref{tab:Priors}.

\begin{table}[t]

\caption{Prior ranges for parameters used in the analyses.}
\label{tab:Priors}
\renewcommand{\arraystretch}{1.4}
\begin{tabular}{|c|c|c|}
\hline
\textbf{Parameter} & \textbf{Prior} & \textbf{Range} \\
\hline

$\mathcal{M}$ & Uniform in component masses &
$72 - 140\,M_\odot$ \\
\hline

$\eta$ & Uniform in component masses &
$0.07 - 0.25$ \\
\hline

$\chi_{1,z}$ & Aligned-spin $z$prior & $-0.99 - 0.99$ \\
\hline

$\chi_{2,z}$ & Aligned-spin $z$prior & $-0.05 - 0.05$ \\
\hline

$\etenhz$ & Uniform &
$0 - 0.35$ \\
\hline

$\zeta$ & Uniform &
$0 - 2\pi$ rad \\
\hline

$d_L$ & Proportional to $d_L^2$ &
$1 - 4000$ Mpc \\
\hline

$\iota$ & Uniform sine &
$0 - \pi$ rad \\
\hline

$\alpha$ & Uniform & $0 - 2\pi$ rad \\
\hline

$\delta$ & Uniform cosine & $\frac{-\pi}{2} - \frac{\pi}{2}$ rad \\
\hline

$\psi$ & Uniform & $0 - \pi$ rad \\
\hline

$\phi_c$ & Uniform & $0 - 2\pi$ rad \\
\hline

$t_c$ & Uniform & $1000000000.0\pm\Delta$ \\
\hline

\end{tabular}

\end{table}

We inject the precessing systems using the model SEOBNRv5PHM and recover using the eccentric model SEOBNRv5EHM due to the computational speedup it offered relative to injecting with the eccentric model and recovering with the precessing model. 
All injections have $f_{\mathrm{min}} = f_{\mathrm{ref}} = 10$ Hz, and are made into a zero-noise realization. 
We initiate the injection waveforms at an earlier starting frequency than the integration limit $\flo$ to ensure that all higher-order modes begin at frequencies lower than $\flo$, thereby mitigating potential systematics. 
For consistency, we fix the SNR to 30.0 for all injections by adjusting the luminosity distance, as shown in Table~\ref{table:inj_tab}. 
The coalescence time for all injections was fixed at $t_c = 1000000000.0$. 
The remaining injection parameters are specified in Table~\ref{table:inj_tab} and Table~\ref{table:fixed_param}.

\section{Further mismatch study results}
\label{sec:apx_mm}

\begin{figure*}[!ht]

\subfloat[$M_\mathrm{tot}=200M_{\odot},\; q=1$]{
    \includegraphics[width=0.32\textwidth]{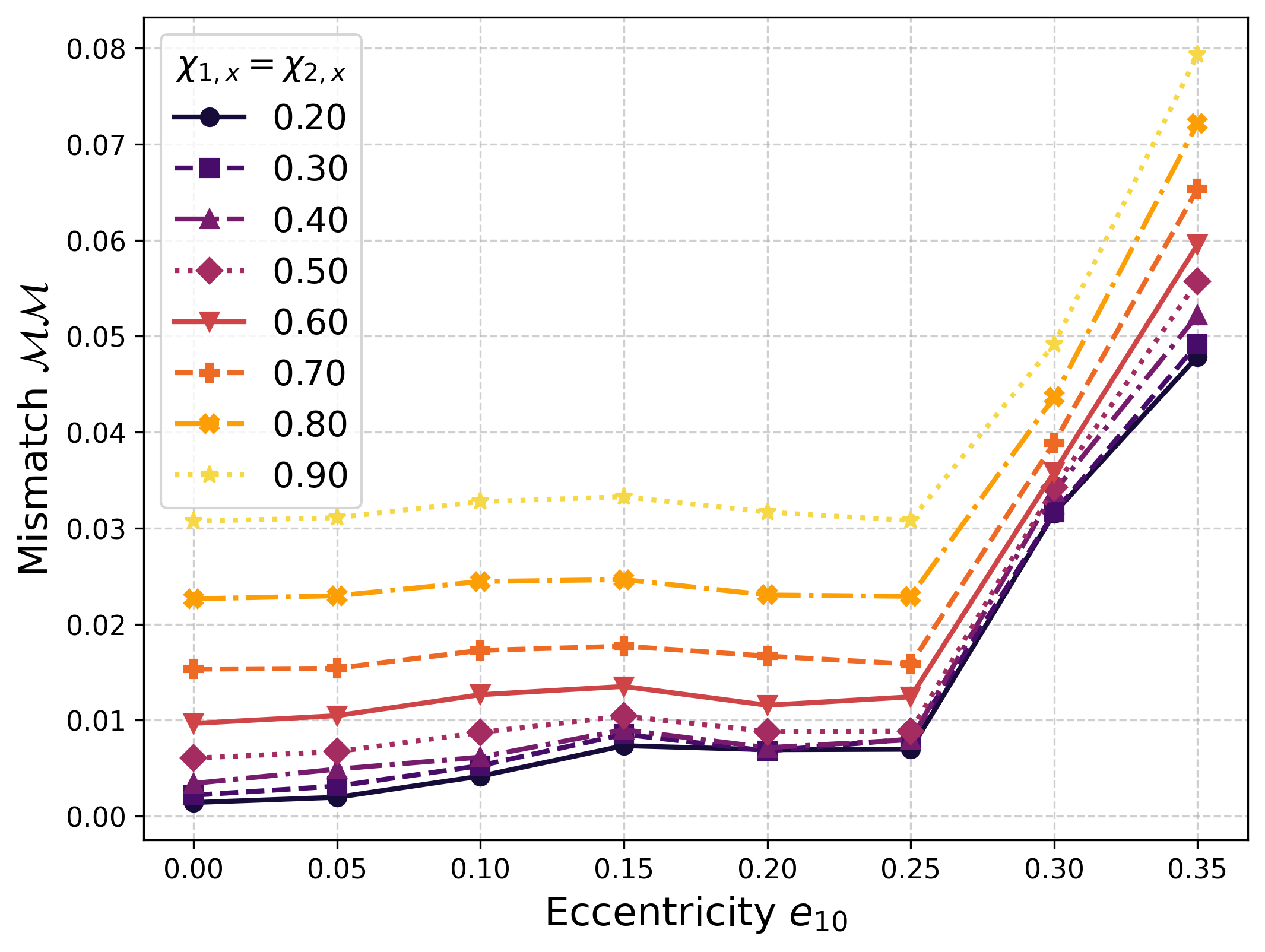}
}\hfill
\subfloat[$M_\mathrm{tot}=200M_{\odot},\; q=2$]{
    \includegraphics[width=0.32\textwidth]{Plots/MM_Appendix_final/NS/mm_q2_mtot200_NS_mm_v_ecc_final.png}
}\hfill
\subfloat[$M_\mathrm{tot}=200M_{\odot},\; q=3$]{
    \includegraphics[width=0.32\textwidth]{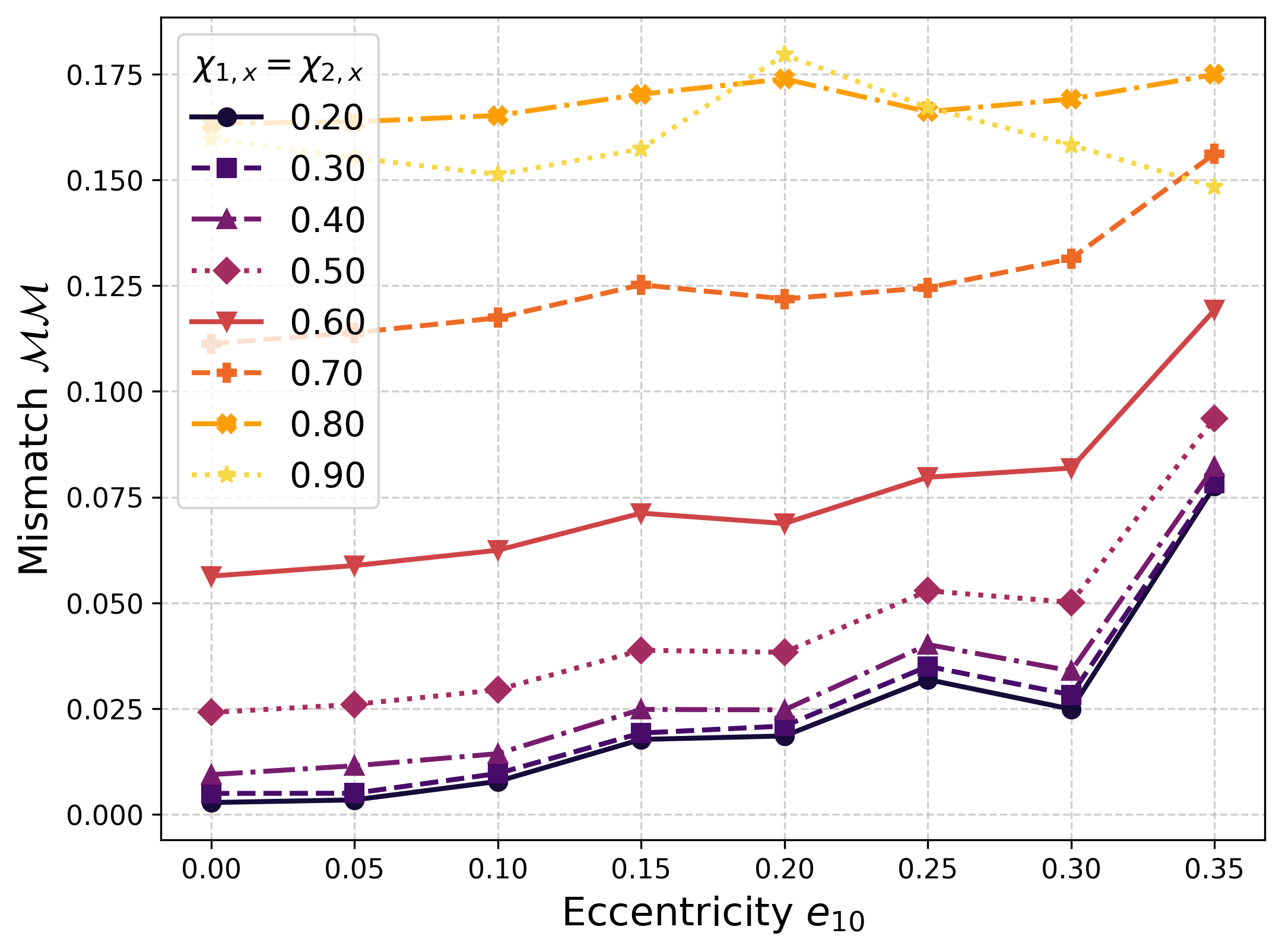}
}

\vspace{6pt}

\subfloat[$M_\mathrm{tot}=250M_{\odot},\; q=1$]{
    \includegraphics[width=0.32\textwidth]{Plots/MM_Appendix_final/NS/mm_q1_mtot250_NS_mm_v_ecc_final.png}
}\hfill
\subfloat[$M_\mathrm{tot}=250M_{\odot},\; q=2$]{
    \includegraphics[width=0.32\textwidth]{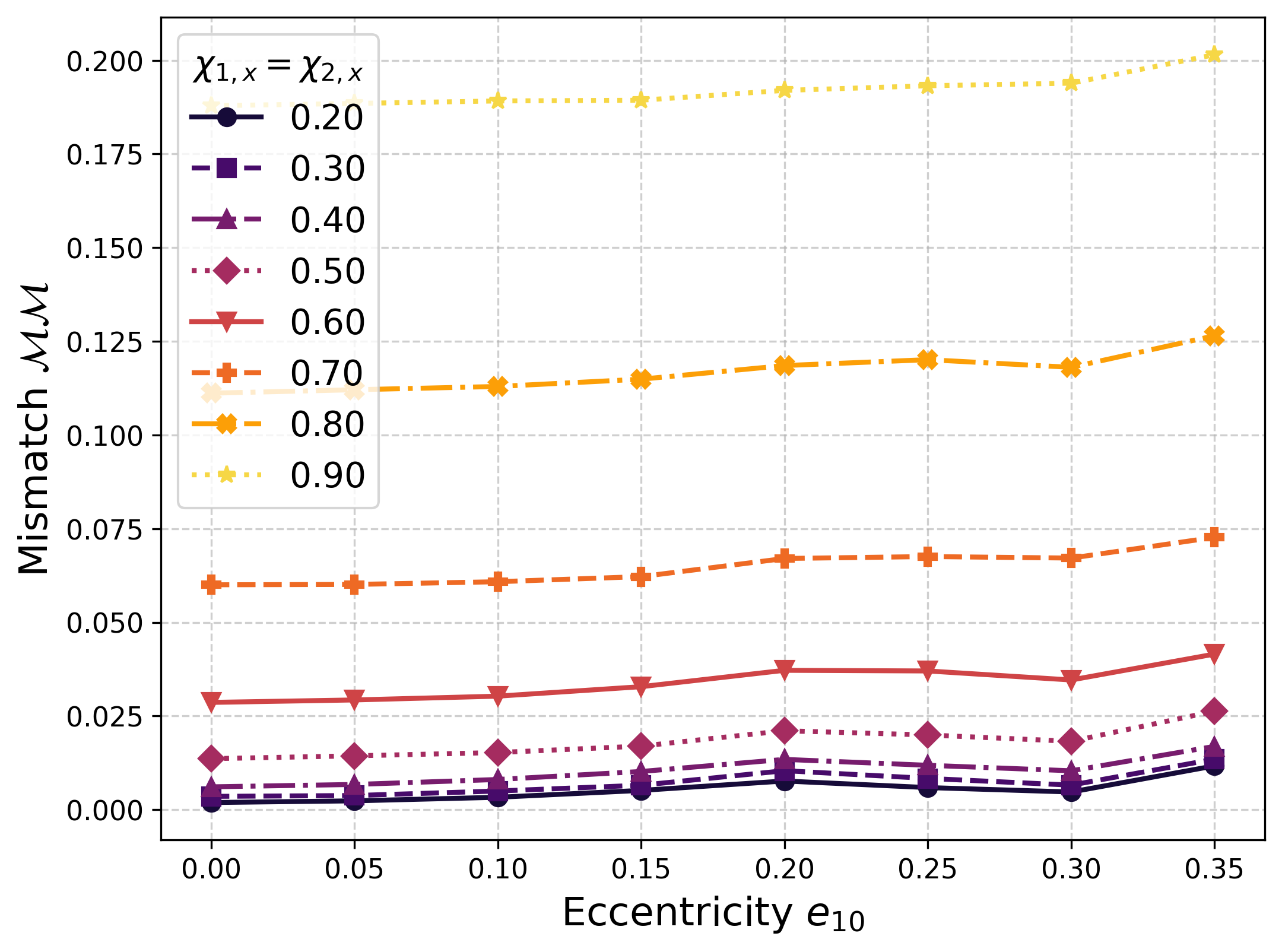}
}\hfill
\subfloat[$M_\mathrm{tot}=250M_{\odot},\; q=3$]{
    \includegraphics[width=0.32\textwidth]{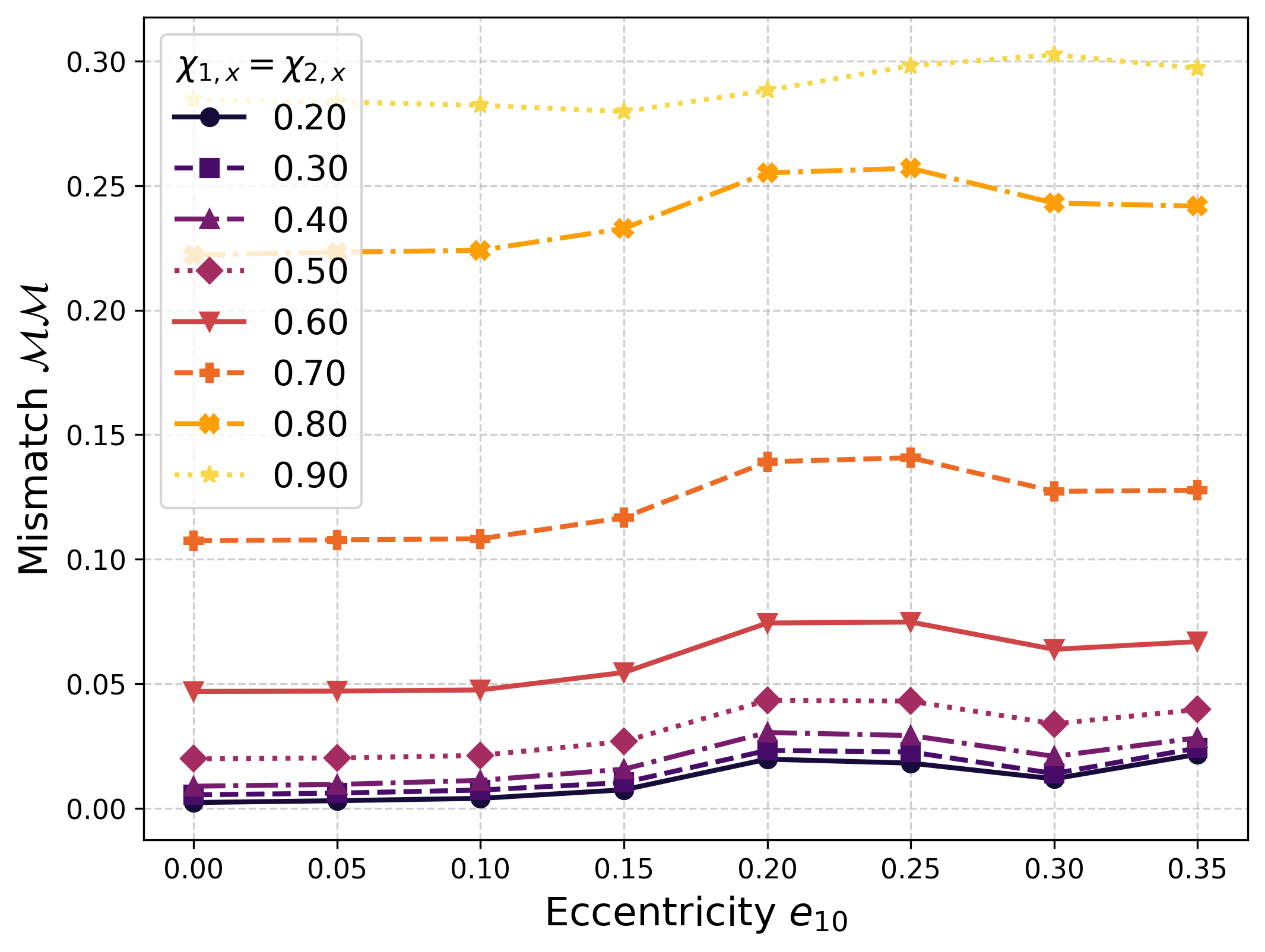}
}

\vspace{6pt}

\subfloat[$M_\mathrm{tot}=300M_{\odot},\; q=1$]{
    \includegraphics[width=0.32\textwidth]{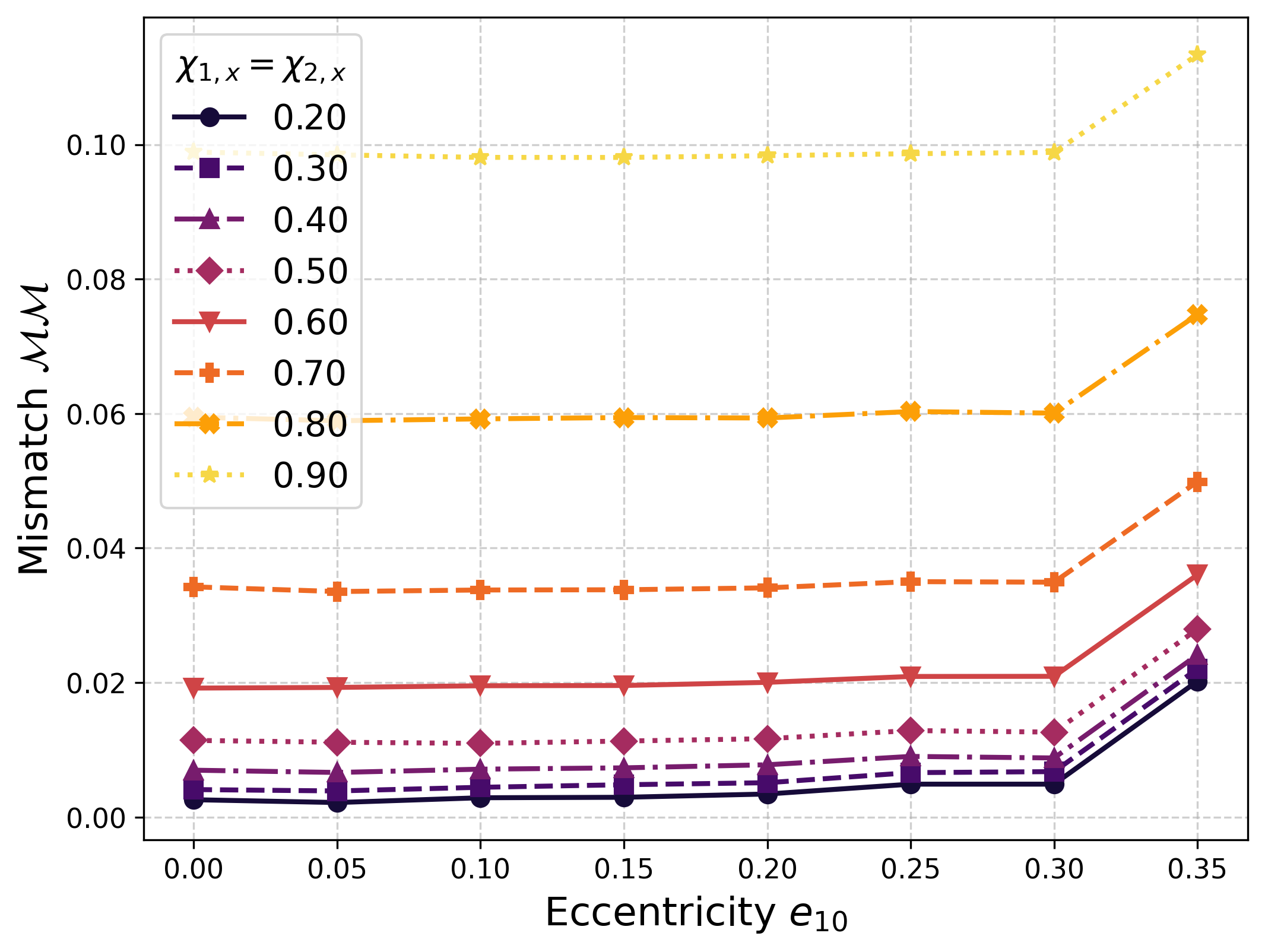}
}\hfill
\subfloat[$M_\mathrm{tot}=300M_{\odot},\; q=2$]{
    \includegraphics[width=0.32\textwidth]{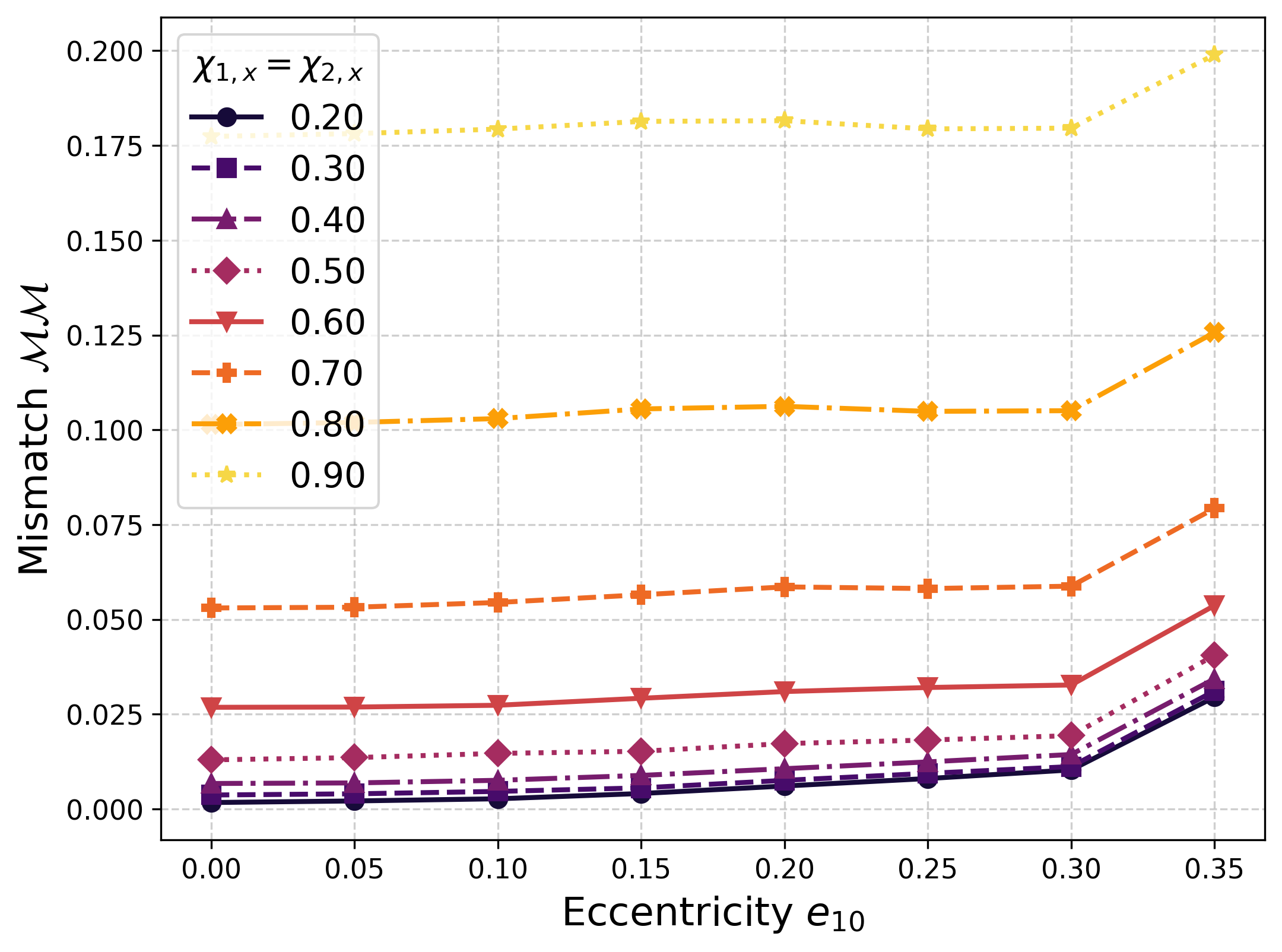}
}\hfill
\subfloat[$M_\mathrm{tot}=300M_{\odot},\; q=3$]{
    \includegraphics[width=0.32\textwidth]{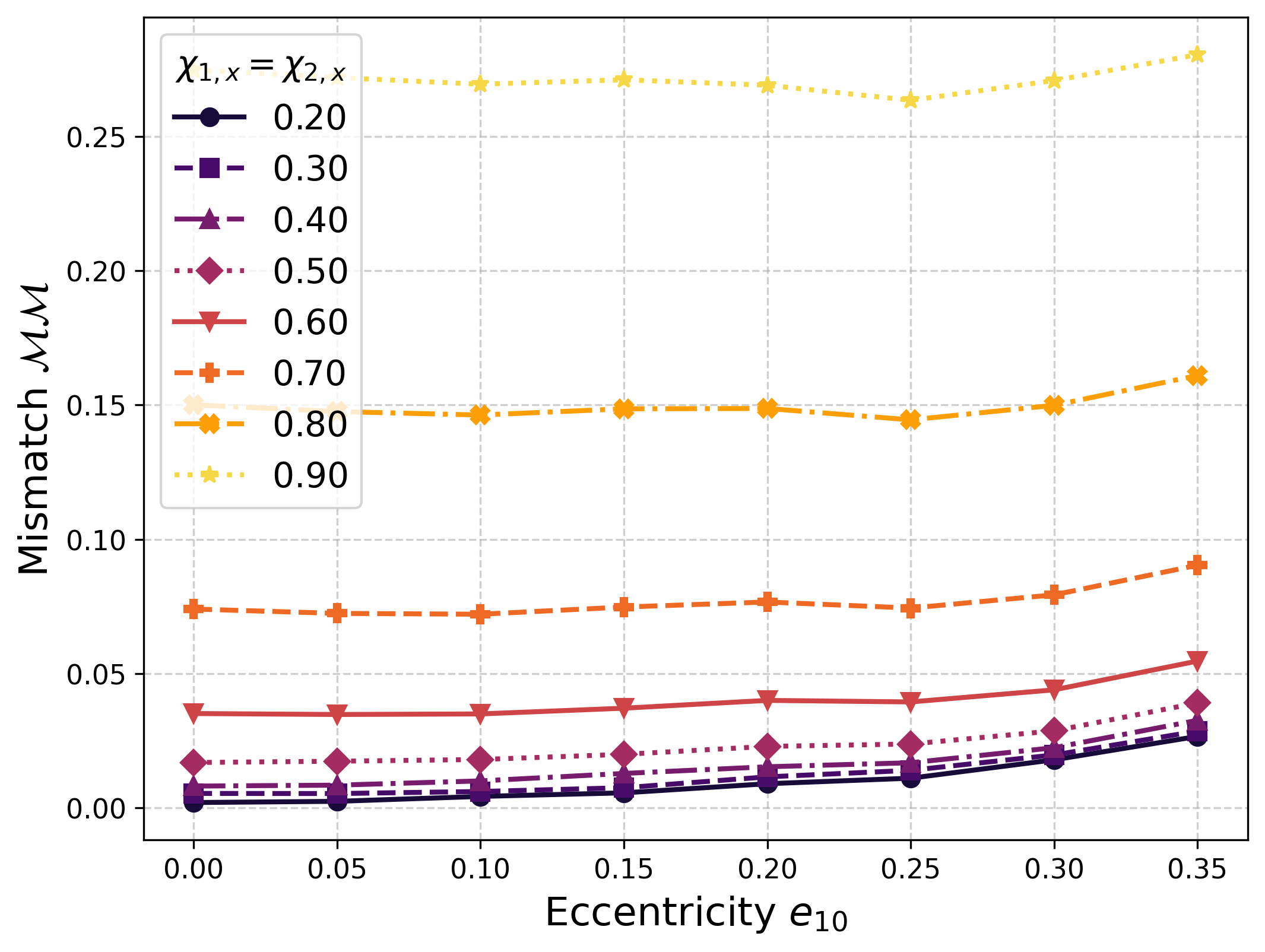}
}

\caption{Mismatches between precessing(in-plane spins only) and non-spinning eccentric systems for all 9 combinations.}
\label{fig:gridnospin}
\end{figure*}

\begin{figure*}[!ht]

\subfloat[$M_\mathrm{tot}=200M_{\odot},\; q=1$]{
    \includegraphics[width=0.32\textwidth]{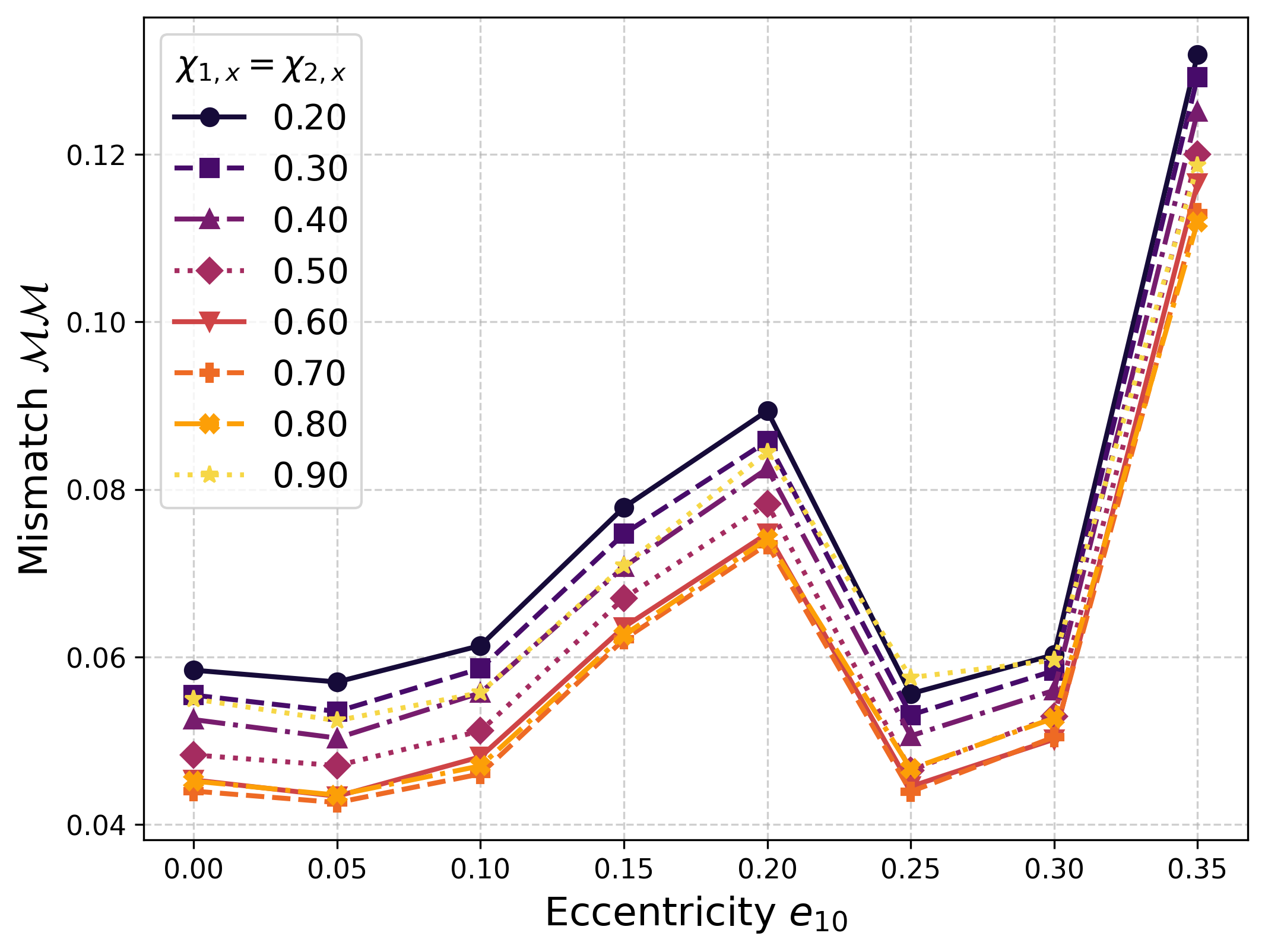}
}\hfill
\subfloat[$M_\mathrm{tot}=200M_{\odot},\; q=2$]{
    \includegraphics[width=0.32\textwidth]{Plots/MM_Appendix_final/AS/mm_q2_mtot200_AS_mm_v_ecc_final.png}
}\hfill
\subfloat[$M_\mathrm{tot}=200M_{\odot},\; q=3$]{
    \includegraphics[width=0.32\textwidth]{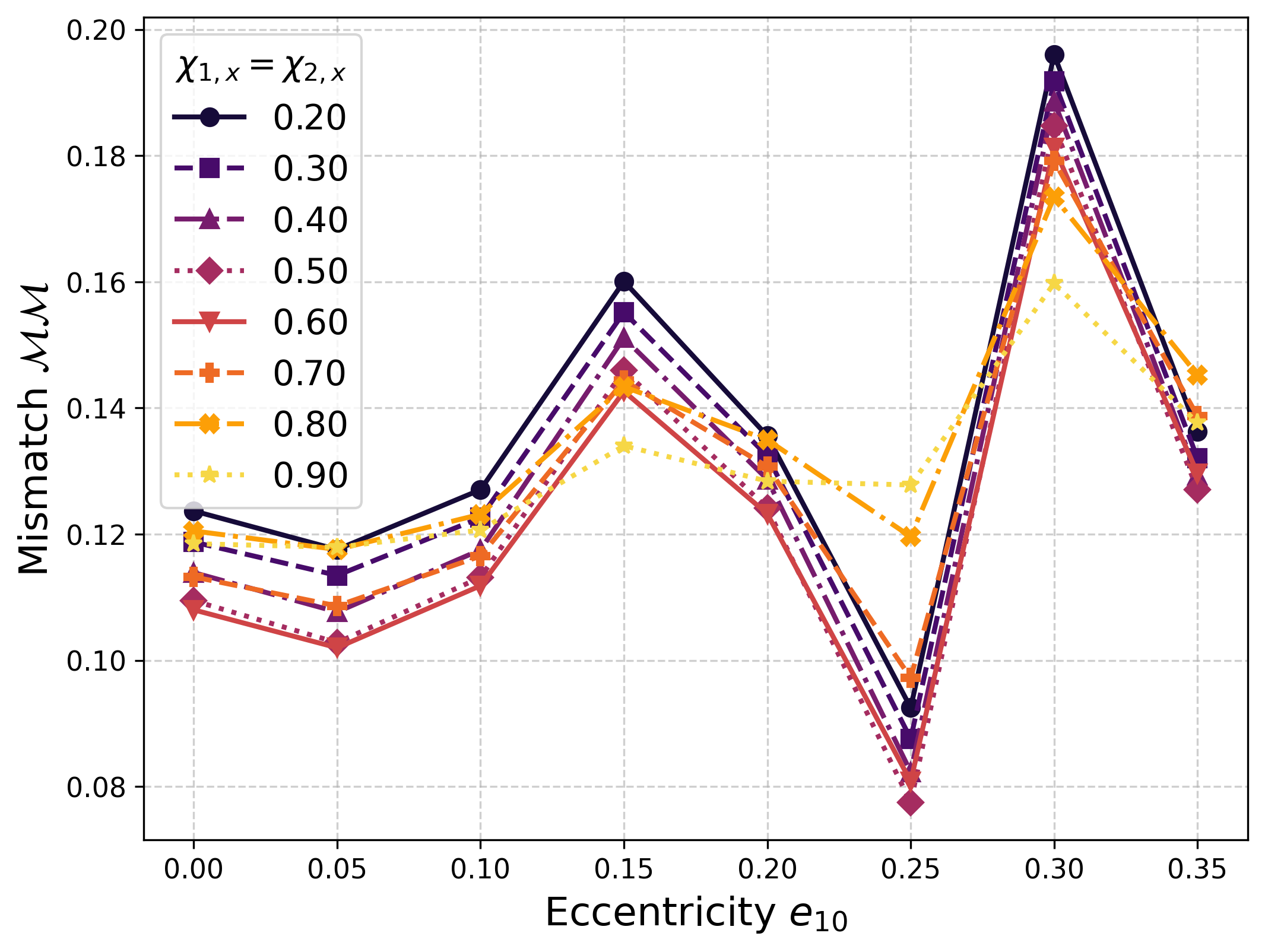}
}

\vspace{6pt}

\subfloat[$M_\mathrm{tot}=250M_{\odot},\; q=1$]{
    \includegraphics[width=0.32\textwidth]{Plots/MM_Appendix_final/AS/mm_q1_mtot250_AS_mm_v_ecc_final.png}
}\hfill
\subfloat[$M_\mathrm{tot}=250M_{\odot},\; q=2$]{
    \includegraphics[width=0.32\textwidth]{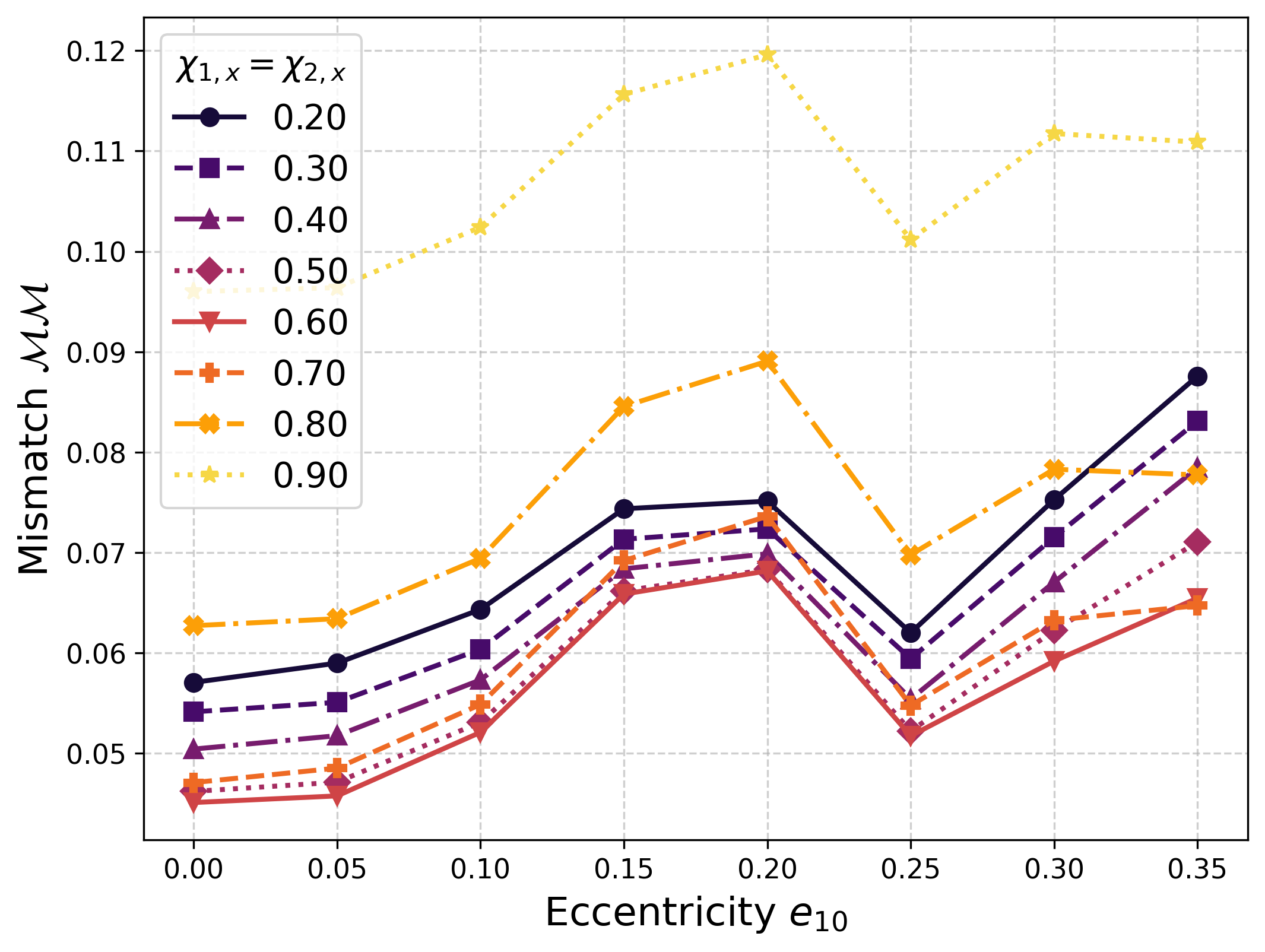}
}\hfill
\subfloat[$M_\mathrm{tot}=250M_{\odot},\; q=3$]{
    \includegraphics[width=0.32\textwidth]{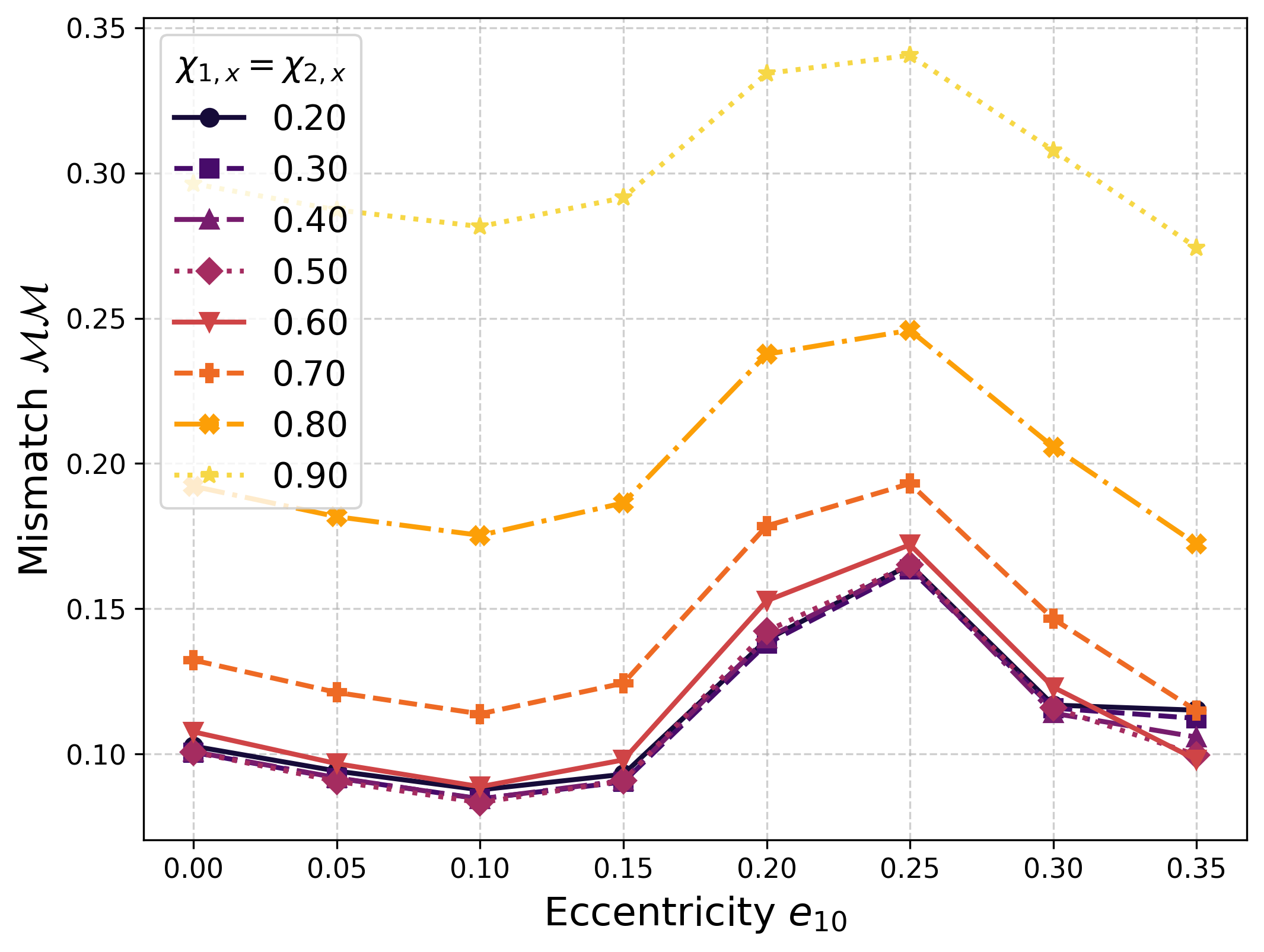}
}

\vspace{6pt}

\subfloat[$M_\mathrm{tot}=300M_{\odot},\; q=1$]{
    \includegraphics[width=0.32\textwidth]{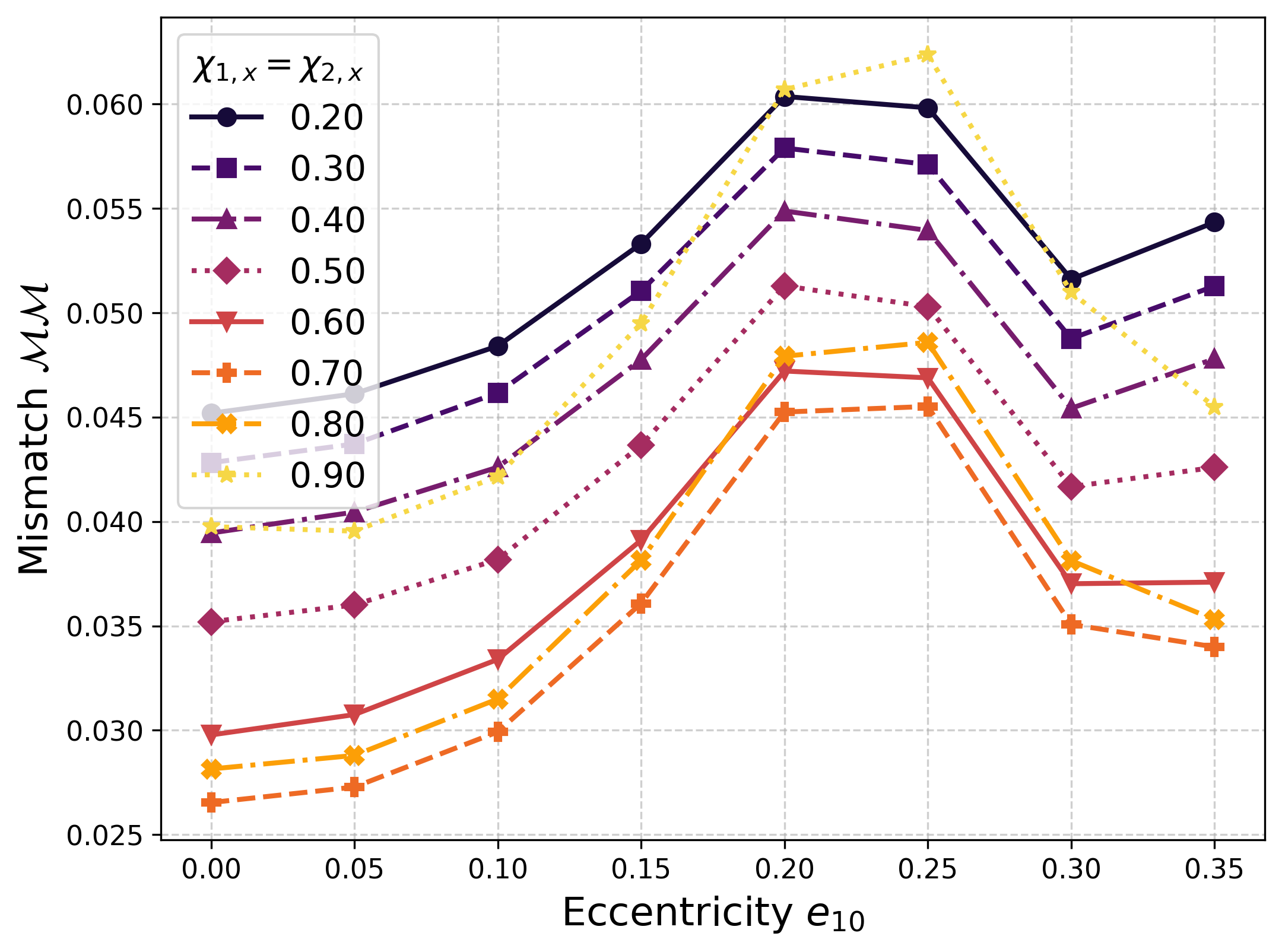}
}\hfill
\subfloat[$M_\mathrm{tot}=300M_{\odot},\; q=2$]{
    \includegraphics[width=0.32\textwidth]{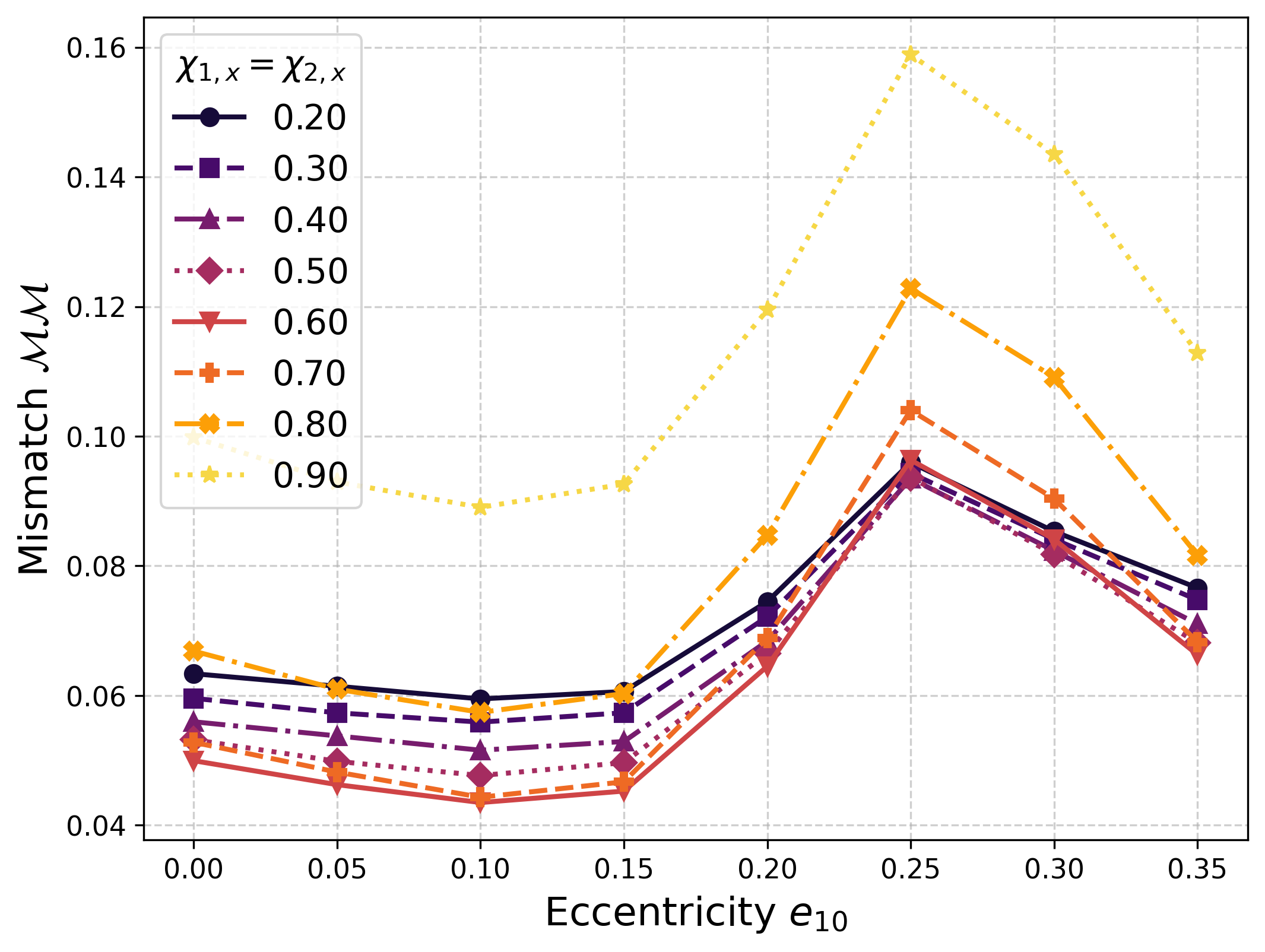}
}\hfill
\subfloat[$M_\mathrm{tot}=300M_{\odot},\; q=3$]{
    \includegraphics[width=0.32\textwidth]{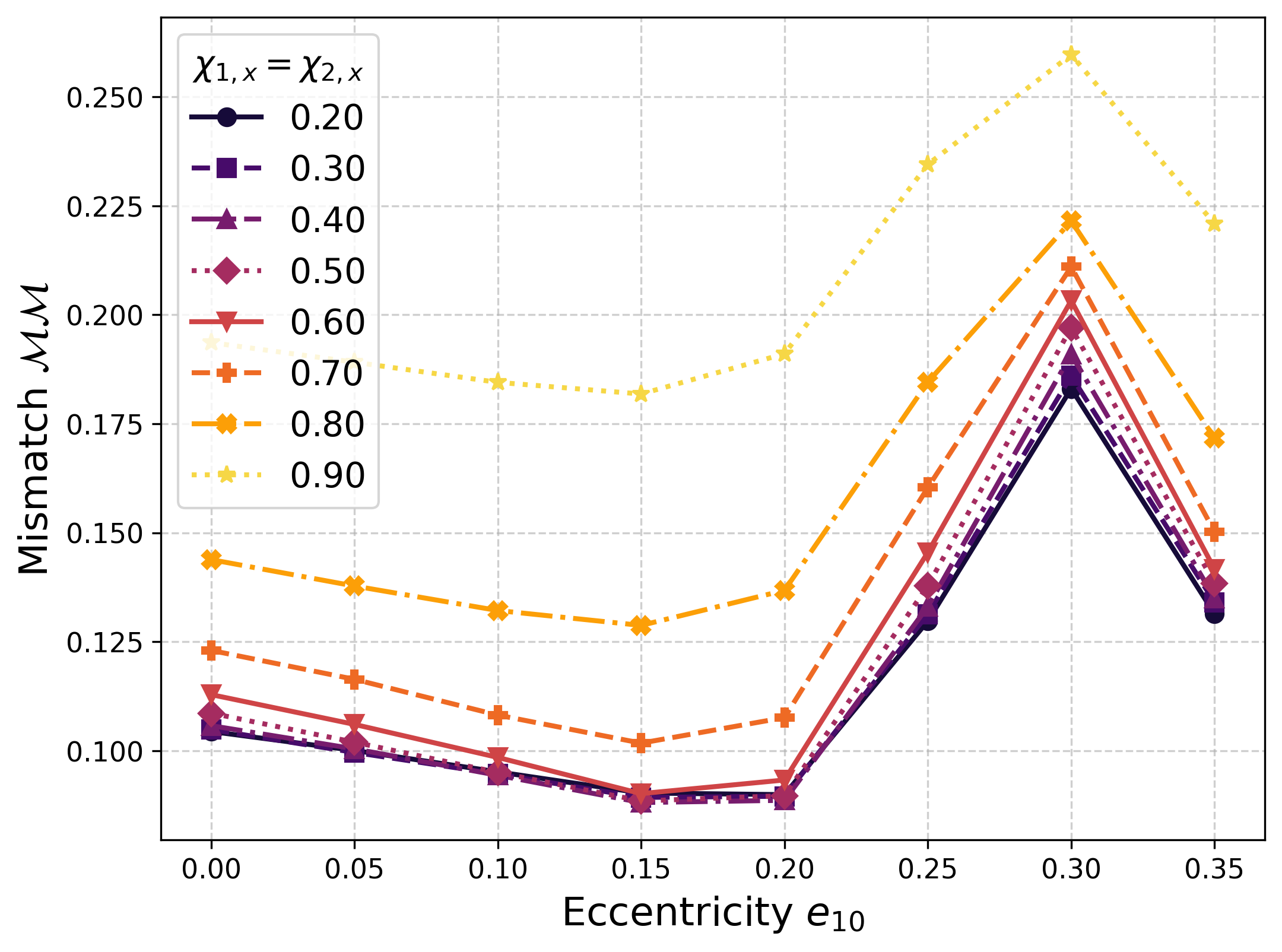}
}

\caption{Mismatches between precessing(in-plane spins only) and aligned-spin eccentric systems for all 9 combinations.}
\label{fig:gridaligned}
\end{figure*}

\begin{figure*}[!ht]

\subfloat[$M_\mathrm{tot}=200M_{\odot},\; q=1$]{
    \includegraphics[width=0.32\textwidth]{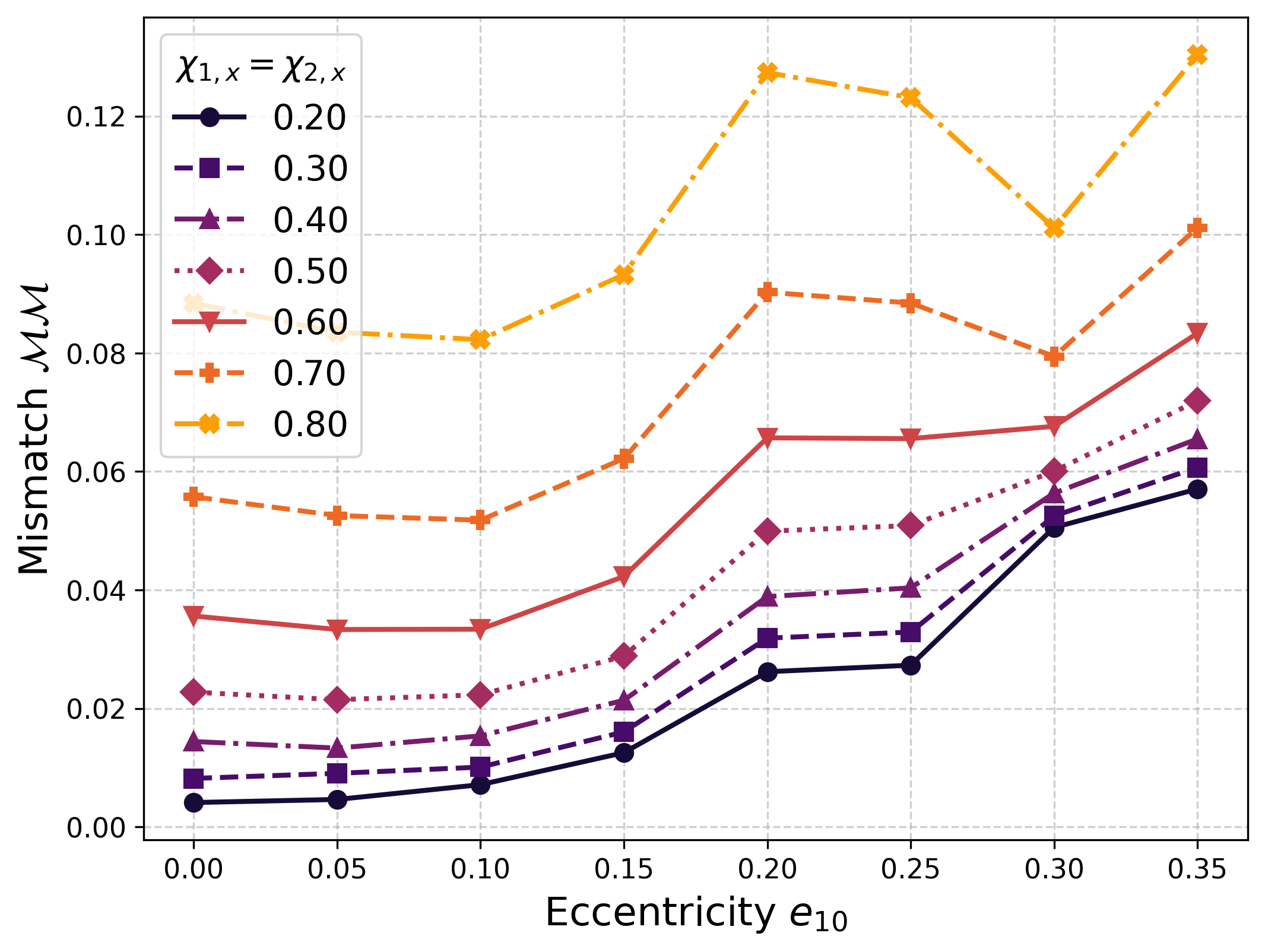}
}\hfill
\subfloat[$M_\mathrm{tot}=200M_{\odot},\; q=2$]{
    \includegraphics[width=0.32\textwidth]{Plots/MM_Appendix_final/ASPrec/mm_q2_mtot200_AS-ASPrec_mm_v_ecc_final.png}
}\hfill
\subfloat[$M_\mathrm{tot}=200M_{\odot},\; q=3$]{
    \includegraphics[width=0.32\textwidth]{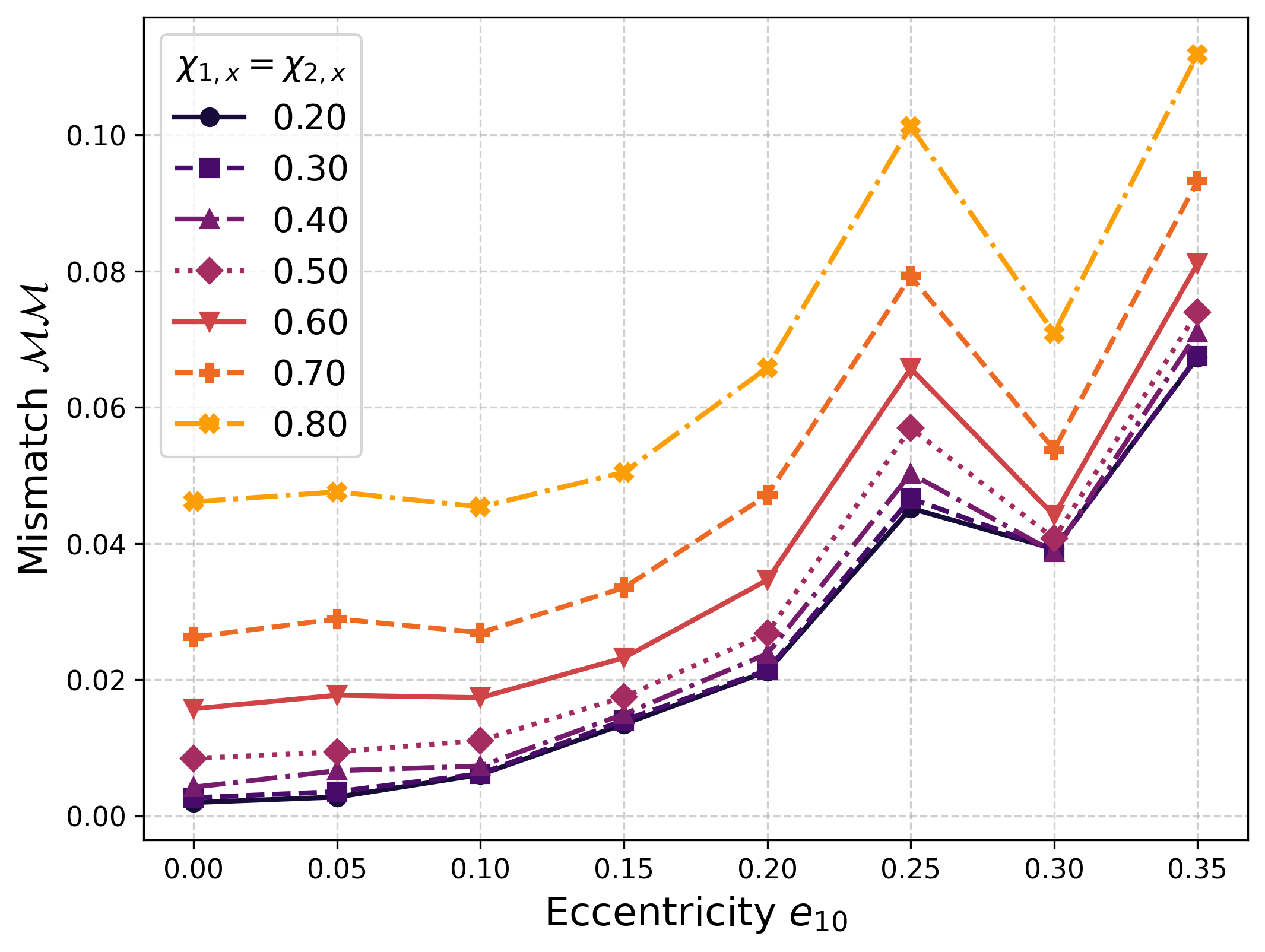}
}

\vspace{6pt}

\subfloat[$M_\mathrm{tot}=250M_{\odot},\; q=1$]{
    \includegraphics[width=0.32\textwidth]{Plots/MM_Appendix_final/ASPrec/mm_q1_mtot250_AS-ASPrec_mm_v_ecc_final.png}
}\hfill
\subfloat[$M_\mathrm{tot}=250M_{\odot},\; q=2$]{
    \includegraphics[width=0.32\textwidth]{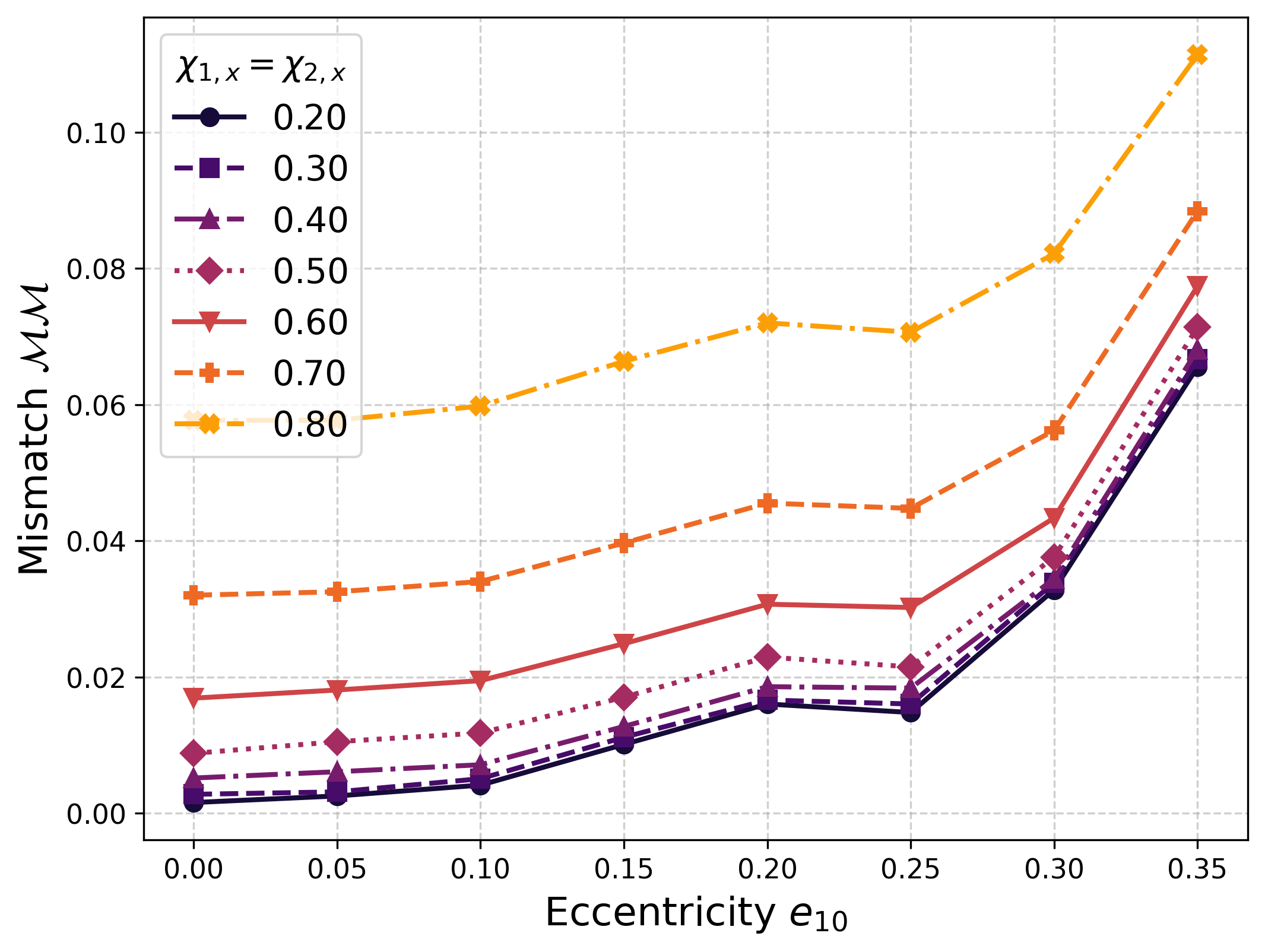}
}\hfill
\subfloat[$M_\mathrm{tot}=250M_{\odot},\; q=3$]{
    \includegraphics[width=0.32\textwidth]{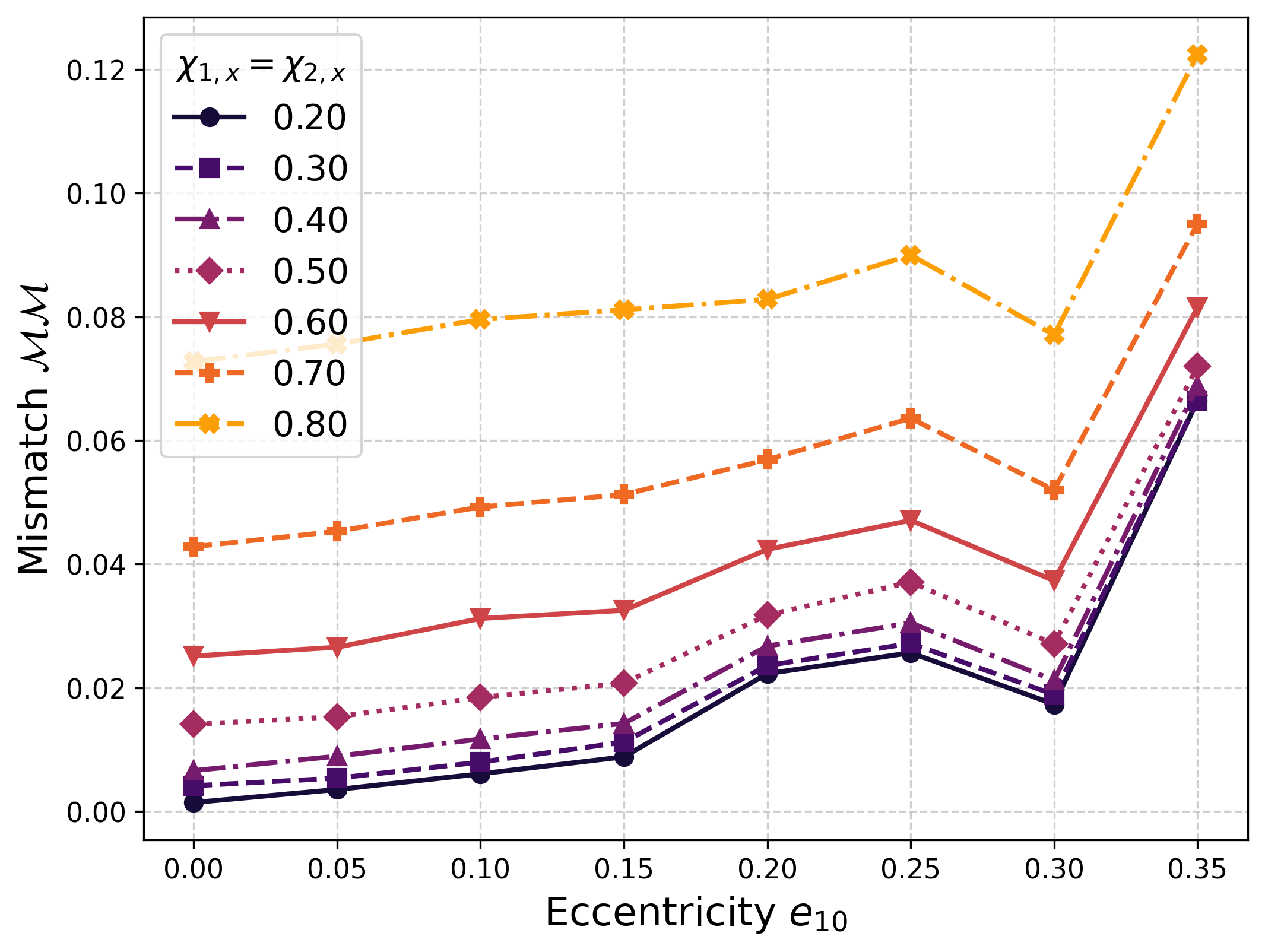}
}

\vspace{6pt}

\subfloat[$M_\mathrm{tot}=300M_{\odot},\; q=1$]{
    \includegraphics[width=0.32\textwidth]{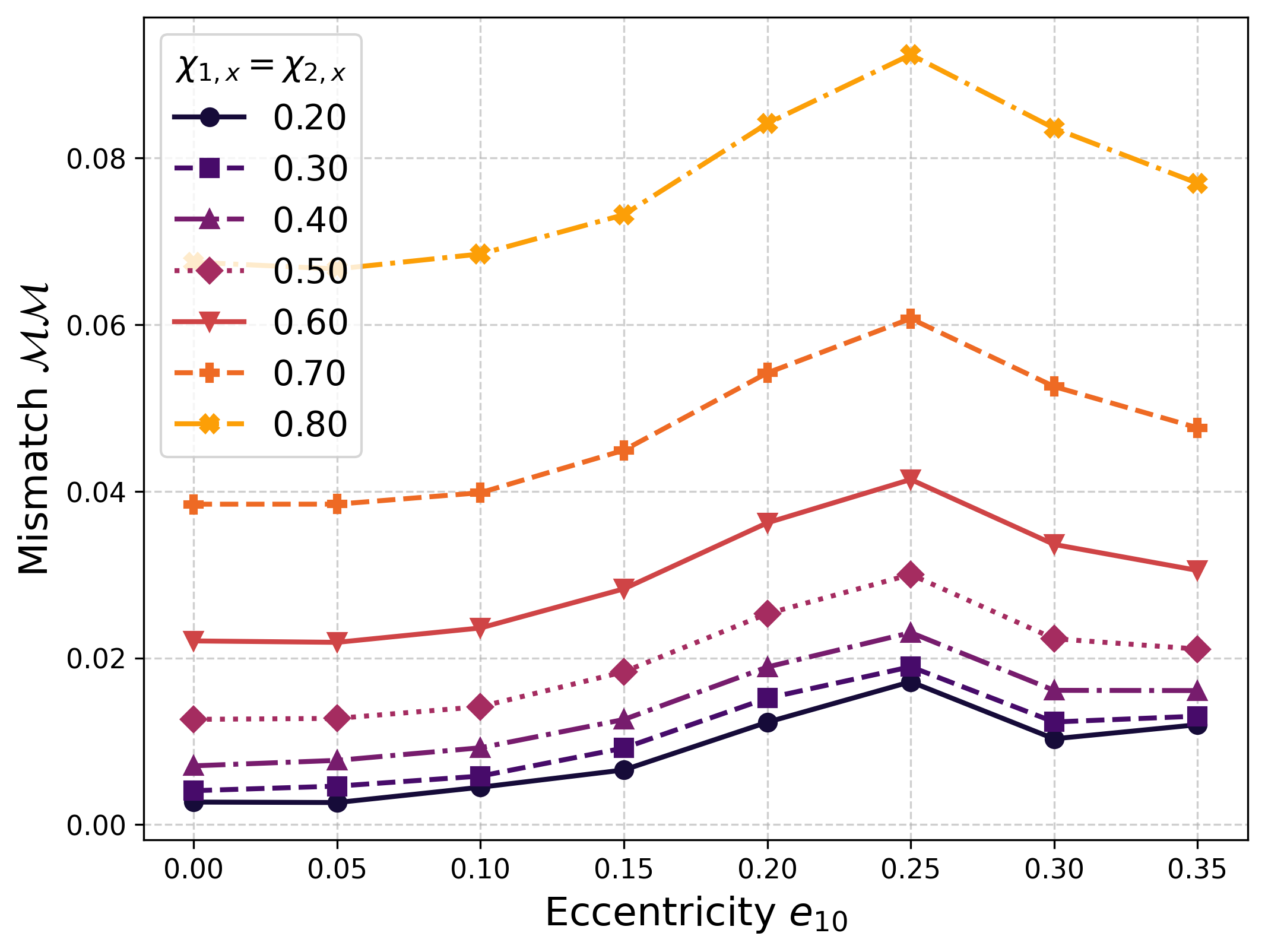}
}\hfill
\subfloat[$M_\mathrm{tot}=300M_{\odot},\; q=2$]{
    \includegraphics[width=0.32\textwidth]{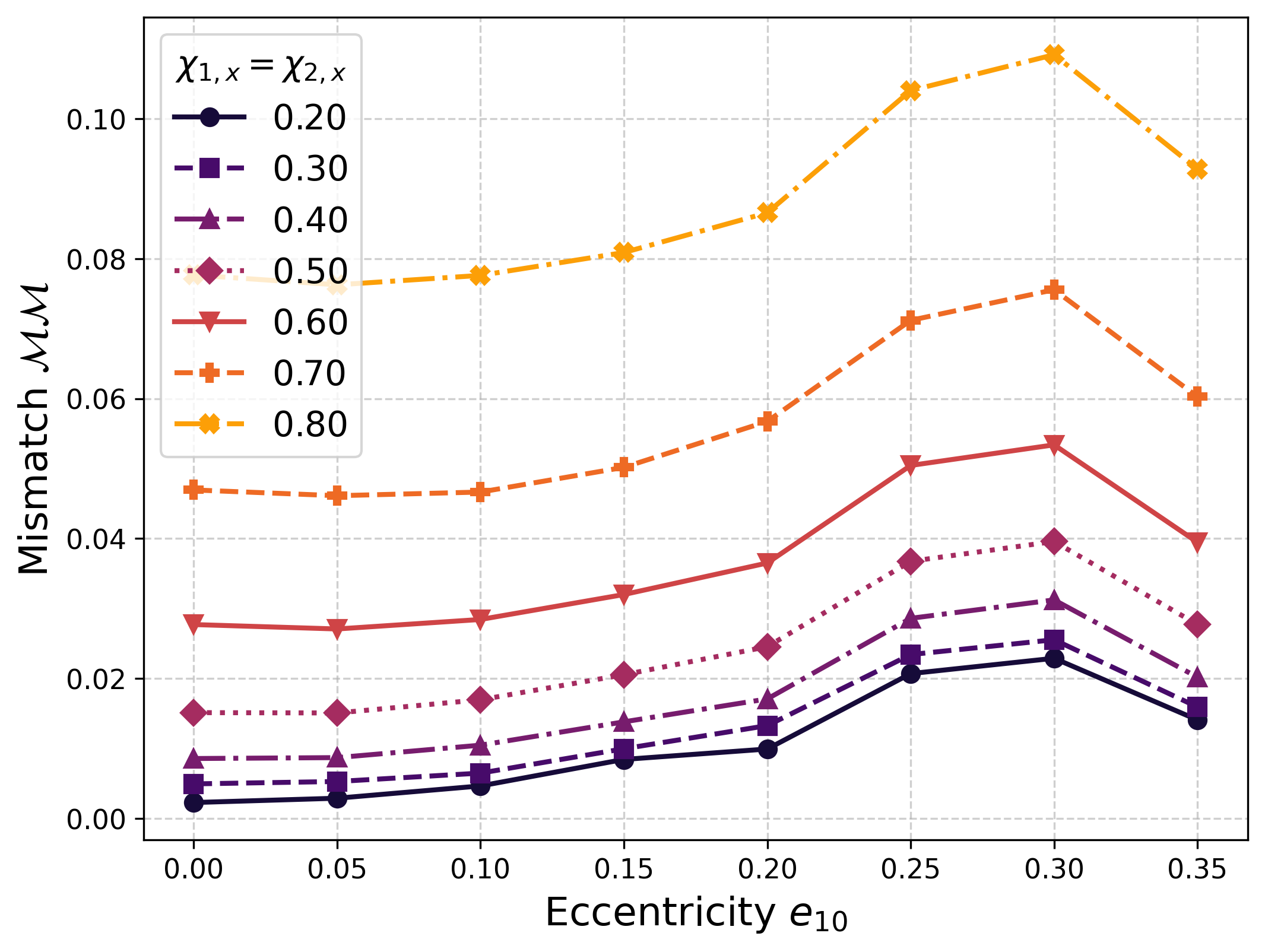}
}\hfill
\subfloat[$M_\mathrm{tot}=300M_{\odot},\; q=3$]{
    \includegraphics[width=0.32\textwidth]{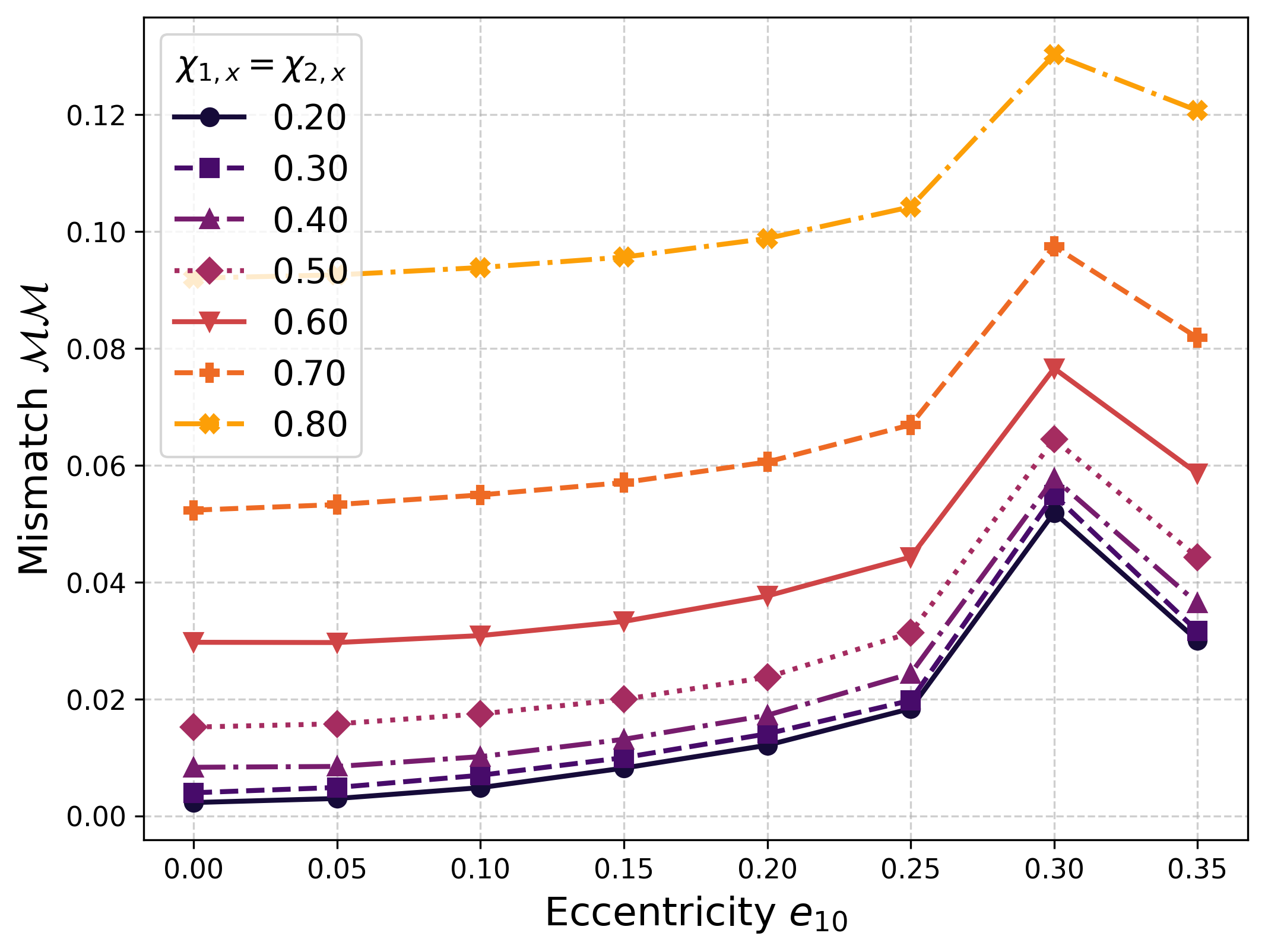}
}

\caption{Mismatches between precessing(in-plane spins+aligned spin) and aligned spin eccentric systems for all 9 combinations. The aligned spins in both precessing and eccentric system have the same magnitude in this case.}
\label{fig:gridaligned_prec}
\end{figure*}

We give further results of our mismatch study (Sec.~\ref{sec:mismatch}) in Figs.~\ref{fig:gridnospin}, ~\ref{fig:gridaligned} and~\ref{fig:gridaligned_prec}. For the case where precessing system have in-plane spins only, Fig.~\ref{fig:gridnospin} shows the mismatches when the eccentric waveform model is required to be non-spinning.
The mismatch values are relatively flat with eccentricity, and generally the mismatch increases with increasing eccentricity.
The result is that the quasi-circular waveform tends to provide the best match with the precessing waveform.
We conclude that some aligned spin is needed for misidentifying eccentricity.
For the same precession configuration as mentioned above, Fig.~\ref{fig:gridaligned} shows the case where the eccentric waveform model has aligned spins $\chi_{1,z} = \chi_{2,z} = 0.5$.
In this case the mismatch shows greater structure as a function of eccentricity, generally increasing and then decreasing in a region $\etenhz \gtrsim 0.2$.
Broadly the trends are similar with increasing spin precession, but not monotonic.
Further, in some many cases the quasi-circular limit provides the best match to the precessing system.
We conclude that misidentification of eccentricity in a precessing signal with in-plane spins only is possible at moderate eccentricities $\etenhz\sim 0.2$--$0.3$ but the high variability means that the degeneracy is sensitive to the binary parameters, for example the initial orientation of the precessing spins, the relativistic anomaly at the reference frequency, or other phases.
Finally, in Fig.~\ref{fig:gridaligned_prec} we show the mismatch results for the case where the precessing signal has in-plane spins in addition to having the same aligned spin as in the eccentric system.
One can see that the mismatches are less afflicted by the non-monotonic trends seen in Fig.~\ref{fig:gridaligned}. Albeit, the mismatch magnitude in this configuration for the degenerate case identified in this work $(M_\mathrm{tot}=250,\; q=1)$  remains higher as compared to the configuration in Fig.~\ref{fig:gridaligned}, making that particular system prone to misidentification regardless.
This reinstates the notion that precession alone is less susceptible to being misidentified as eccentricity. A specific sensitive configuration of aligned and in-plane spins are required to induce the degeneracy.

\input{main.bbl}
\end{document}

%% file: main.bbl
%

%% file: main.bbl
\begin{thebibliography}{87}%
\makeatletter
\providecommand \@ifxundefined [1]{%
 \@ifx{#1\undefined}
}%
\providecommand \@ifnum [1]{%
 \ifnum #1\expandafter \@firstoftwo
 \else \expandafter \@secondoftwo
 \fi
}%
\providecommand \@ifx [1]{%
 \ifx #1\expandafter \@firstoftwo
 \else \expandafter \@secondoftwo
 \fi
}%
\providecommand \natexlab [1]{#1}%
\providecommand \enquote  [1]{``#1''}%
\providecommand \bibnamefont  [1]{#1}%
\providecommand \bibfnamefont [1]{#1}%
\providecommand \citenamefont [1]{#1}%
\providecommand \href@noop [0]{\@secondoftwo}%
\providecommand \href [0]{\begingroup \@sanitize@url \@href}%
\providecommand \@href[1]{\@@startlink{#1}\@@href}%
\providecommand \@@href[1]{\endgroup#1\@@endlink}%
\providecommand \@sanitize@url [0]{\catcode `\\12\catcode `\$12\catcode
  `\&12\catcode `\#12\catcode `\^12\catcode `\_12\catcode `\%12\relax}%
\providecommand \@@startlink[1]{}%
\providecommand \@@endlink[0]{}%
\providecommand \url  [0]{\begingroup\@sanitize@url \@url }%
\providecommand \@url [1]{\endgroup\@href {#1}{\urlprefix }}%
\providecommand \urlprefix  [0]{URL }%
\providecommand \Eprint [0]{\href }%
\providecommand \doibase [0]{http://dx.doi.org/}%
\providecommand \selectlanguage [0]{\@gobble}%
\providecommand \bibinfo  [0]{\@secondoftwo}%
\providecommand \bibfield  [0]{\@secondoftwo}%
\providecommand \translation [1]{[#1]}%
\providecommand \BibitemOpen [0]{}%
\providecommand \bibitemStop [0]{}%
\providecommand \bibitemNoStop [0]{.\EOS\space}%
\providecommand \EOS [0]{\spacefactor3000\relax}%
\providecommand \BibitemShut  [1]{\csname bibitem#1\endcsname}%
\let\auto@bib@innerbib\@empty
\bibitem [{\citenamefont {Abbott}\ \emph {et~al.}(2016)\citenamefont {Abbott}
  \emph {et~al.}}]{LIGOScientific:2016aoc}%
  \BibitemOpen
  \bibfield  {author} {\bibinfo {author} {\bibfnamefont {B.~P.}\ \bibnamefont
  {Abbott}} \emph {et~al.} (\bibinfo {collaboration} {LIGO Scientific,
  Virgo}),\ }\bibfield  {title} {\enquote {\bibinfo {title} {{Observation of
  Gravitational Waves from a Binary Black Hole Merger}},}\ }\href {\doibase
  10.1103/PhysRevLett.116.061102} {\bibfield  {journal} {\bibinfo  {journal}
  {Phys. Rev. Lett.}\ }\textbf {\bibinfo {volume} {116}},\ \bibinfo {pages}
  {061102} (\bibinfo {year} {2016})},\ \Eprint
  {http://arxiv.org/abs/1602.03837} {arXiv:1602.03837 [gr-qc]} \BibitemShut
  {NoStop}%
\bibitem [{\citenamefont {Abbott}\ \emph {et~al.}(2019)\citenamefont {Abbott}
  \emph {et~al.}}]{LIGOScientific:2018mvr}%
  \BibitemOpen
  \bibfield  {author} {\bibinfo {author} {\bibfnamefont {B.~P.}\ \bibnamefont
  {Abbott}} \emph {et~al.} (\bibinfo {collaboration} {LIGO Scientific,
  Virgo}),\ }\bibfield  {title} {\enquote {\bibinfo {title} {{GWTC-1: A
  Gravitational-Wave Transient Catalog of Compact Binary Mergers Observed by
  LIGO and Virgo during the First and Second Observing Runs}},}\ }\href
  {\doibase 10.1103/PhysRevX.9.031040} {\bibfield  {journal} {\bibinfo
  {journal} {Phys. Rev. X}\ }\textbf {\bibinfo {volume} {9}},\ \bibinfo {pages}
  {031040} (\bibinfo {year} {2019})},\ \Eprint
  {http://arxiv.org/abs/1811.12907} {arXiv:1811.12907 [astro-ph.HE]}
  \BibitemShut {NoStop}%
\bibitem [{\citenamefont {Abbott}\ \emph {et~al.}(2021)\citenamefont {Abbott}
  \emph {et~al.}}]{LIGOScientific:2020ibl}%
  \BibitemOpen
  \bibfield  {author} {\bibinfo {author} {\bibfnamefont {R.}~\bibnamefont
  {Abbott}} \emph {et~al.} (\bibinfo {collaboration} {LIGO Scientific,
  Virgo}),\ }\bibfield  {title} {\enquote {\bibinfo {title} {{GWTC-2: Compact
  Binary Coalescences Observed by LIGO and Virgo During the First Half of the
  Third Observing Run}},}\ }\href {\doibase 10.1103/PhysRevX.11.021053}
  {\bibfield  {journal} {\bibinfo  {journal} {Phys. Rev. X}\ }\textbf {\bibinfo
  {volume} {11}},\ \bibinfo {pages} {021053} (\bibinfo {year} {2021})},\
  \Eprint {http://arxiv.org/abs/2010.14527} {arXiv:2010.14527 [gr-qc]}
  \BibitemShut {NoStop}%
\bibitem [{\citenamefont {Abbott}\ \emph {et~al.}(2024)\citenamefont {Abbott}
  \emph {et~al.}}]{LIGOScientific:2021usb}%
  \BibitemOpen
  \bibfield  {author} {\bibinfo {author} {\bibfnamefont {R.}~\bibnamefont
  {Abbott}} \emph {et~al.} (\bibinfo {collaboration} {LIGO Scientific,
  VIRGO}),\ }\bibfield  {title} {\enquote {\bibinfo {title} {{GWTC-2.1: Deep
  extended catalog of compact binary coalescences observed by LIGO and Virgo
  during the first half of the third observing run}},}\ }\href {\doibase
  10.1103/PhysRevD.109.022001} {\bibfield  {journal} {\bibinfo  {journal}
  {Phys. Rev. D}\ }\textbf {\bibinfo {volume} {109}},\ \bibinfo {pages}
  {022001} (\bibinfo {year} {2024})},\ \Eprint
  {http://arxiv.org/abs/2108.01045} {arXiv:2108.01045 [gr-qc]} \BibitemShut
  {NoStop}%
\bibitem [{\citenamefont {Abbott}\ \emph {et~al.}(2023)\citenamefont {Abbott}
  \emph {et~al.}}]{KAGRA:2021vkt}%
  \BibitemOpen
  \bibfield  {author} {\bibinfo {author} {\bibfnamefont {R.}~\bibnamefont
  {Abbott}} \emph {et~al.} (\bibinfo {collaboration} {KAGRA, VIRGO, LIGO
  Scientific}),\ }\bibfield  {title} {\enquote {\bibinfo {title} {{GWTC-3:
  Compact Binary Coalescences Observed by LIGO and Virgo during the Second Part
  of the Third Observing Run}},}\ }\href {\doibase 10.1103/PhysRevX.13.041039}
  {\bibfield  {journal} {\bibinfo  {journal} {Phys. Rev. X}\ }\textbf {\bibinfo
  {volume} {13}},\ \bibinfo {pages} {041039} (\bibinfo {year} {2023})},\
  \Eprint {http://arxiv.org/abs/2111.03606} {arXiv:2111.03606 [gr-qc]}
  \BibitemShut {NoStop}%
\bibitem [{\citenamefont {Abac}\ \emph
  {et~al.}(2025{\natexlab{a}})\citenamefont {Abac} \emph
  {et~al.}}]{LIGOScientific:2025slb}%
  \BibitemOpen
  \bibfield  {author} {\bibinfo {author} {\bibfnamefont {A.~G.}\ \bibnamefont
  {Abac}} \emph {et~al.} (\bibinfo {collaboration} {LIGO Scientific, VIRGO,
  KAGRA}),\ }\bibfield  {title} {\enquote {\bibinfo {title} {{GWTC-4.0:
  Updating the Gravitational-Wave Transient Catalog with Observations from the
  First Part of the Fourth LIGO-Virgo-KAGRA Observing Run}},}\ }\href@noop {}
  {\  (\bibinfo {year} {2025}{\natexlab{a}})},\ \Eprint
  {http://arxiv.org/abs/2508.18082} {arXiv:2508.18082 [gr-qc]} \BibitemShut
  {NoStop}%
\bibitem [{\citenamefont {Hoy}\ \emph {et~al.}(2025)\citenamefont {Hoy},
  \citenamefont {Fairhurst},\ and\ \citenamefont {Mandel}}]{Hoy:2024wkc}%
  \BibitemOpen
  \bibfield  {author} {\bibinfo {author} {\bibfnamefont {Charlie}\ \bibnamefont
  {Hoy}}, \bibinfo {author} {\bibfnamefont {Stephen}\ \bibnamefont
  {Fairhurst}}, \ and\ \bibinfo {author} {\bibfnamefont {Ilya}\ \bibnamefont
  {Mandel}},\ }\bibfield  {title} {\enquote {\bibinfo {title} {{Rarity of
  precession and higher-order multipoles in gravitational waves from merging
  binary black holes}},}\ }\href {\doibase 10.1103/PhysRevD.111.023037}
  {\bibfield  {journal} {\bibinfo  {journal} {Phys. Rev. D}\ }\textbf {\bibinfo
  {volume} {111}},\ \bibinfo {pages} {023037} (\bibinfo {year} {2025})},\
  \Eprint {http://arxiv.org/abs/2408.03410} {arXiv:2408.03410 [gr-qc]}
  \BibitemShut {NoStop}%
\bibitem [{\citenamefont {Gumbel}(1958)}]{Gumbel1958_EVT}%
  \BibitemOpen
  \bibfield  {author} {\bibinfo {author} {\bibfnamefont {Emil~Julius}\
  \bibnamefont {Gumbel}},\ }\href@noop {} {\emph {\bibinfo {title} {{Statistics
  of Extremes}}}}\ (\bibinfo  {publisher} {Columbia University Press},\
  \bibinfo {year} {1958})\BibitemShut {NoStop}%
\bibitem [{\citenamefont {Samsing}(2018)}]{Samsing:2017xmd}%
  \BibitemOpen
  \bibfield  {author} {\bibinfo {author} {\bibfnamefont {Johan}\ \bibnamefont
  {Samsing}},\ }\bibfield  {title} {\enquote {\bibinfo {title} {{Eccentric
  Black Hole Mergers Forming in Globular Clusters}},}\ }\href {\doibase
  10.1103/PhysRevD.97.103014} {\bibfield  {journal} {\bibinfo  {journal} {Phys.
  Rev. D}\ }\textbf {\bibinfo {volume} {97}},\ \bibinfo {pages} {103014}
  (\bibinfo {year} {2018})},\ \Eprint {http://arxiv.org/abs/1711.07452}
  {arXiv:1711.07452 [astro-ph.HE]} \BibitemShut {NoStop}%
\bibitem [{\citenamefont {Rodriguez}\ \emph {et~al.}(2016)\citenamefont
  {Rodriguez}, \citenamefont {Zevin}, \citenamefont {Pankow}, \citenamefont
  {Kalogera},\ and\ \citenamefont {Rasio}}]{Rodriguez:2016vmx}%
  \BibitemOpen
  \bibfield  {author} {\bibinfo {author} {\bibfnamefont {Carl~L.}\ \bibnamefont
  {Rodriguez}}, \bibinfo {author} {\bibfnamefont {Michael}\ \bibnamefont
  {Zevin}}, \bibinfo {author} {\bibfnamefont {Chris}\ \bibnamefont {Pankow}},
  \bibinfo {author} {\bibfnamefont {Vasilliki}\ \bibnamefont {Kalogera}}, \
  and\ \bibinfo {author} {\bibfnamefont {Frederic~A.}\ \bibnamefont {Rasio}},\
  }\bibfield  {title} {\enquote {\bibinfo {title} {{Illuminating Black Hole
  Binary Formation Channels with Spins in Advanced LIGO}},}\ }\href {\doibase
  10.3847/2041-8205/832/1/L2} {\bibfield  {journal} {\bibinfo  {journal}
  {Astrophys. J. Lett.}\ }\textbf {\bibinfo {volume} {832}},\ \bibinfo {pages}
  {L2} (\bibinfo {year} {2016})},\ \Eprint {http://arxiv.org/abs/1609.05916}
  {arXiv:1609.05916 [astro-ph.HE]} \BibitemShut {NoStop}%
\bibitem [{\citenamefont {Abac}\ \emph
  {et~al.}(2025{\natexlab{b}})\citenamefont {Abac} \emph
  {et~al.}}]{LIGOScientific:2025pvj}%
  \BibitemOpen
  \bibfield  {author} {\bibinfo {author} {\bibfnamefont {A.~G.}\ \bibnamefont
  {Abac}} \emph {et~al.} (\bibinfo {collaboration} {LIGO Scientific, VIRGO,
  KAGRA}),\ }\bibfield  {title} {\enquote {\bibinfo {title} {{GWTC-4.0:
  Population Properties of Merging Compact Binaries}},}\ }\href@noop {} {\
  (\bibinfo {year} {2025}{\natexlab{b}})},\ \Eprint
  {http://arxiv.org/abs/2508.18083} {arXiv:2508.18083 [astro-ph.HE]}
  \BibitemShut {NoStop}%
\bibitem [{\citenamefont {Abbott}\ \emph
  {et~al.}(2020{\natexlab{a}})\citenamefont {Abbott} \emph
  {et~al.}}]{LIGOScientific:2020iuh}%
  \BibitemOpen
  \bibfield  {author} {\bibinfo {author} {\bibfnamefont {R.}~\bibnamefont
  {Abbott}} \emph {et~al.} (\bibinfo {collaboration} {LIGO Scientific,
  Virgo}),\ }\bibfield  {title} {\enquote {\bibinfo {title} {{GW190521: A
  Binary Black Hole Merger with a Total Mass of $150 M_{\odot}$}},}\ }\href
  {\doibase 10.1103/PhysRevLett.125.101102} {\bibfield  {journal} {\bibinfo
  {journal} {Phys. Rev. Lett.}\ }\textbf {\bibinfo {volume} {125}},\ \bibinfo
  {pages} {101102} (\bibinfo {year} {2020}{\natexlab{a}})},\ \Eprint
  {http://arxiv.org/abs/2009.01075} {arXiv:2009.01075 [gr-qc]} \BibitemShut
  {NoStop}%
\bibitem [{\citenamefont {Abbott}\ \emph
  {et~al.}(2020{\natexlab{b}})\citenamefont {Abbott} \emph
  {et~al.}}]{LIGOScientific:2020ufj}%
  \BibitemOpen
  \bibfield  {author} {\bibinfo {author} {\bibfnamefont {R.}~\bibnamefont
  {Abbott}} \emph {et~al.} (\bibinfo {collaboration} {LIGO Scientific,
  Virgo}),\ }\bibfield  {title} {\enquote {\bibinfo {title} {{Properties and
  Astrophysical Implications of the 150 M$_\odot$ Binary Black Hole Merger
  GW190521}},}\ }\href {\doibase 10.3847/2041-8213/aba493} {\bibfield
  {journal} {\bibinfo  {journal} {Astrophys. J. Lett.}\ }\textbf {\bibinfo
  {volume} {900}},\ \bibinfo {pages} {L13} (\bibinfo {year}
  {2020}{\natexlab{b}})},\ \Eprint {http://arxiv.org/abs/2009.01190}
  {arXiv:2009.01190 [astro-ph.HE]} \BibitemShut {NoStop}%
\bibitem [{\citenamefont {Abac}\ \emph
  {et~al.}(2025{\natexlab{c}})\citenamefont {Abac} \emph
  {et~al.}}]{LIGOScientific:2025rsn}%
  \BibitemOpen
  \bibfield  {author} {\bibinfo {author} {\bibfnamefont {A.~G.}\ \bibnamefont
  {Abac}} \emph {et~al.} (\bibinfo {collaboration} {LIGO Scientific, VIRGO,
  KAGRA}),\ }\bibfield  {title} {\enquote {\bibinfo {title} {{GW231123: a
  Binary Black Hole Merger with Total Mass 190-265 $M_{\odot}$}},}\ }\href@noop
  {} {\  (\bibinfo {year} {2025}{\natexlab{c}})},\ \Eprint
  {http://arxiv.org/abs/2507.08219} {arXiv:2507.08219 [astro-ph.HE]}
  \BibitemShut {NoStop}%
\bibitem [{\citenamefont {Fumagalli}\ \emph {et~al.}(2024)\citenamefont
  {Fumagalli}, \citenamefont {Romero-Shaw}, \citenamefont {Gerosa},
  \citenamefont {De~Renzis}, \citenamefont {Kritos},\ and\ \citenamefont
  {Olejak}}]{Fumagalli:2024gko}%
  \BibitemOpen
  \bibfield  {author} {\bibinfo {author} {\bibfnamefont {Giulia}\ \bibnamefont
  {Fumagalli}}, \bibinfo {author} {\bibfnamefont {Isobel}\ \bibnamefont
  {Romero-Shaw}}, \bibinfo {author} {\bibfnamefont {Davide}\ \bibnamefont
  {Gerosa}}, \bibinfo {author} {\bibfnamefont {Viola}\ \bibnamefont
  {De~Renzis}}, \bibinfo {author} {\bibfnamefont {Konstantinos}\ \bibnamefont
  {Kritos}}, \ and\ \bibinfo {author} {\bibfnamefont {Aleksandra}\ \bibnamefont
  {Olejak}},\ }\bibfield  {title} {\enquote {\bibinfo {title} {{Residual
  eccentricity as a systematic uncertainty on the formation channels of binary
  black holes}},}\ }\href {\doibase 10.1103/PhysRevD.110.063012} {\bibfield
  {journal} {\bibinfo  {journal} {Phys. Rev. D}\ }\textbf {\bibinfo {volume}
  {110}},\ \bibinfo {pages} {063012} (\bibinfo {year} {2024})},\ \Eprint
  {http://arxiv.org/abs/2405.14945} {arXiv:2405.14945 [astro-ph.HE]}
  \BibitemShut {NoStop}%
\bibitem [{\citenamefont {Romero-Shaw}\ \emph {et~al.}(2020)\citenamefont
  {Romero-Shaw}, \citenamefont {Lasky}, \citenamefont {Thrane},\ and\
  \citenamefont {Bustillo}}]{Romero-Shaw:2020thy}%
  \BibitemOpen
  \bibfield  {author} {\bibinfo {author} {\bibfnamefont {Isobel~M.}\
  \bibnamefont {Romero-Shaw}}, \bibinfo {author} {\bibfnamefont {Paul~D.}\
  \bibnamefont {Lasky}}, \bibinfo {author} {\bibfnamefont {Eric}\ \bibnamefont
  {Thrane}}, \ and\ \bibinfo {author} {\bibfnamefont {Juan~Calderon}\
  \bibnamefont {Bustillo}},\ }\bibfield  {title} {\enquote {\bibinfo {title}
  {{GW190521: orbital eccentricity and signatures of dynamical formation in a
  binary black hole merger signal}},}\ }\href {\doibase
  10.3847/2041-8213/abbe26} {\bibfield  {journal} {\bibinfo  {journal}
  {Astrophys. J. Lett.}\ }\textbf {\bibinfo {volume} {903}},\ \bibinfo {pages}
  {L5} (\bibinfo {year} {2020})},\ \Eprint {http://arxiv.org/abs/2009.04771}
  {arXiv:2009.04771 [astro-ph.HE]} \BibitemShut {NoStop}%
\bibitem [{\citenamefont {Gayathri}\ \emph {et~al.}(2022)\citenamefont
  {Gayathri}, \citenamefont {Healy}, \citenamefont {Lange}, \citenamefont
  {O’Brien}, \citenamefont {Szczepańczyk}, \citenamefont {Bartos},
  \citenamefont {Campanelli}, \citenamefont {Klimenko}, \citenamefont
  {Lousto},\ and\ \citenamefont {O’Shaughnessy}}]{Gayathri2022}%
  \BibitemOpen
  \bibfield  {author} {\bibinfo {author} {\bibfnamefont {V.}~\bibnamefont
  {Gayathri}}, \bibinfo {author} {\bibfnamefont {J.}~\bibnamefont {Healy}},
  \bibinfo {author} {\bibfnamefont {J.}~\bibnamefont {Lange}}, \bibinfo
  {author} {\bibfnamefont {B.}~\bibnamefont {O’Brien}}, \bibinfo {author}
  {\bibfnamefont {M.}~\bibnamefont {Szczepańczyk}}, \bibinfo {author}
  {\bibfnamefont {Imre}\ \bibnamefont {Bartos}}, \bibinfo {author}
  {\bibfnamefont {M.}~\bibnamefont {Campanelli}}, \bibinfo {author}
  {\bibfnamefont {S.}~\bibnamefont {Klimenko}}, \bibinfo {author}
  {\bibfnamefont {C.~O.}\ \bibnamefont {Lousto}}, \ and\ \bibinfo {author}
  {\bibfnamefont {R.}~\bibnamefont {O’Shaughnessy}},\ }\bibfield  {title}
  {\enquote {\bibinfo {title} {Eccentricity estimate for black hole mergers
  with numerical relativity simulations},}\ }\href {\doibase
  10.1038/s41550-021-01568-w} {\bibfield  {journal} {\bibinfo  {journal}
  {Nature Astronomy}\ }\textbf {\bibinfo {volume} {6}},\ \bibinfo {pages}
  {344--349} (\bibinfo {year} {2022})}\BibitemShut {NoStop}%
\bibitem [{\citenamefont {Miller}\ \emph {et~al.}(2024)\citenamefont {Miller},
  \citenamefont {Isi}, \citenamefont {Chatziioannou}, \citenamefont {Varma},\
  and\ \citenamefont {Mandel}}]{Miller:2023ncs}%
  \BibitemOpen
  \bibfield  {author} {\bibinfo {author} {\bibfnamefont {Simona~J.}\
  \bibnamefont {Miller}}, \bibinfo {author} {\bibfnamefont {Maximiliano}\
  \bibnamefont {Isi}}, \bibinfo {author} {\bibfnamefont {Katerina}\
  \bibnamefont {Chatziioannou}}, \bibinfo {author} {\bibfnamefont {Vijay}\
  \bibnamefont {Varma}}, \ and\ \bibinfo {author} {\bibfnamefont {Ilya}\
  \bibnamefont {Mandel}},\ }\bibfield  {title} {\enquote {\bibinfo {title}
  {{GW190521: Tracing imprints of spin-precession on the most massive black
  hole binary}},}\ }\href {\doibase 10.1103/PhysRevD.109.024024} {\bibfield
  {journal} {\bibinfo  {journal} {Phys. Rev. D}\ }\textbf {\bibinfo {volume}
  {109}},\ \bibinfo {pages} {024024} (\bibinfo {year} {2024})},\ \Eprint
  {http://arxiv.org/abs/2310.01544} {arXiv:2310.01544 [astro-ph.HE]}
  \BibitemShut {NoStop}%
\bibitem [{\citenamefont {Calder{\'o}n~Bustillo}\ \emph
  {et~al.}(2021)\citenamefont {Calder{\'o}n~Bustillo}, \citenamefont
  {Sanchis-Gual}, \citenamefont {Torres-Forn{\'e}}, \citenamefont {Font},
  \citenamefont {Vajpeyi}, \citenamefont {Smith}, \citenamefont {Herdeiro},
  \citenamefont {Radu},\ and\ \citenamefont
  {Leong}}]{CalderonBustillo:2020fyi}%
  \BibitemOpen
  \bibfield  {author} {\bibinfo {author} {\bibfnamefont {Juan}\ \bibnamefont
  {Calder{\'o}n~Bustillo}}, \bibinfo {author} {\bibfnamefont {Nicolas}\
  \bibnamefont {Sanchis-Gual}}, \bibinfo {author} {\bibfnamefont {Alejandro}\
  \bibnamefont {Torres-Forn{\'e}}}, \bibinfo {author} {\bibfnamefont
  {Jos{\'e}~A.}\ \bibnamefont {Font}}, \bibinfo {author} {\bibfnamefont {Avi}\
  \bibnamefont {Vajpeyi}}, \bibinfo {author} {\bibfnamefont {Rory}\
  \bibnamefont {Smith}}, \bibinfo {author} {\bibfnamefont {Carlos}\
  \bibnamefont {Herdeiro}}, \bibinfo {author} {\bibfnamefont {Eugen}\
  \bibnamefont {Radu}}, \ and\ \bibinfo {author} {\bibfnamefont {Samson H.~W.}\
  \bibnamefont {Leong}},\ }\bibfield  {title} {\enquote {\bibinfo {title}
  {{GW190521 as a Merger of Proca Stars: A Potential New Vector Boson of
  $8.7\times 10^{-13}$ eV}},}\ }\href {\doibase 10.1103/PhysRevLett.126.081101}
  {\bibfield  {journal} {\bibinfo  {journal} {Phys. Rev. Lett.}\ }\textbf
  {\bibinfo {volume} {126}},\ \bibinfo {pages} {081101} (\bibinfo {year}
  {2021})},\ \Eprint {http://arxiv.org/abs/2009.05376} {arXiv:2009.05376
  [gr-qc]} \BibitemShut {NoStop}%
\bibitem [{\citenamefont {Cuceu}\ \emph {et~al.}(2025)\citenamefont {Cuceu},
  \citenamefont {Bizouard}, \citenamefont {Christensen},\ and\ \citenamefont
  {Sakellariadou}}]{Cuceu:2025fzi}%
  \BibitemOpen
  \bibfield  {author} {\bibinfo {author} {\bibfnamefont {Iuliu}\ \bibnamefont
  {Cuceu}}, \bibinfo {author} {\bibfnamefont {Marie~Anne}\ \bibnamefont
  {Bizouard}}, \bibinfo {author} {\bibfnamefont {Nelson}\ \bibnamefont
  {Christensen}}, \ and\ \bibinfo {author} {\bibfnamefont {Mairi}\ \bibnamefont
  {Sakellariadou}},\ }\bibfield  {title} {\enquote {\bibinfo {title}
  {{GW231123: Binary Black Hole Merger or Cosmic String?}}}\ }\href@noop {}
  {\bibfield  {journal} {\bibinfo  {journal} {arXiv:2507.20778}\ } (\bibinfo
  {year} {2025})}\BibitemShut {NoStop}%
\bibitem [{\citenamefont {Romero-Shaw}\ \emph {et~al.}(2023)\citenamefont
  {Romero-Shaw}, \citenamefont {Gerosa},\ and\ \citenamefont
  {Loutrel}}]{Romero-Shaw:2022fbf}%
  \BibitemOpen
  \bibfield  {author} {\bibinfo {author} {\bibfnamefont {Isobel~M.}\
  \bibnamefont {Romero-Shaw}}, \bibinfo {author} {\bibfnamefont {Davide}\
  \bibnamefont {Gerosa}}, \ and\ \bibinfo {author} {\bibfnamefont {Nicholas}\
  \bibnamefont {Loutrel}},\ }\bibfield  {title} {\enquote {\bibinfo {title}
  {{Eccentricity or spin precession? Distinguishing subdominant effects in
  gravitational-wave data}},}\ }\href {\doibase 10.1093/mnras/stad031}
  {\bibfield  {journal} {\bibinfo  {journal} {Mon. Not. Roy. Astron. Soc.}\
  }\textbf {\bibinfo {volume} {519}},\ \bibinfo {pages} {5352--5357} (\bibinfo
  {year} {2023})},\ \Eprint {http://arxiv.org/abs/2211.07528} {arXiv:2211.07528
  [astro-ph.HE]} \BibitemShut {NoStop}%
\bibitem [{\citenamefont {Xu}\ and\ \citenamefont {Hamilton}(2023)}]{Xu2023}%
  \BibitemOpen
  \bibfield  {author} {\bibinfo {author} {\bibfnamefont {Yumeng}\ \bibnamefont
  {Xu}}\ and\ \bibinfo {author} {\bibfnamefont {Eleanor}\ \bibnamefont
  {Hamilton}},\ }\bibfield  {title} {\enquote {\bibinfo {title} {Measurability
  of precession and eccentricity for heavy binary-black-hole mergers},}\ }\href
  {\doibase 10.1103/PhysRevD.107.103049} {\bibfield  {journal} {\bibinfo
  {journal} {Physical Review D}\ }\textbf {\bibinfo {volume} {107}},\ \bibinfo
  {pages} {103049} (\bibinfo {year} {2023})}\BibitemShut {NoStop}%
\bibitem [{\citenamefont {Divyajyoti}\ \emph {et~al.}(2025)\citenamefont
  {Divyajyoti} \emph {et~al.}}]{Divyajyoti:2025cwq}%
  \BibitemOpen
  \bibfield  {author} {\bibinfo {author} {\bibnamefont {Divyajyoti}} \emph
  {et~al.},\ }\bibfield  {title} {\enquote {\bibinfo {title} {{Biased parameter
  inference of eccentric, spin-precessing binary black holes}},}\ }\href@noop
  {} {\  (\bibinfo {year} {2025})},\ \Eprint {http://arxiv.org/abs/2510.04332}
  {arXiv:2510.04332 [gr-qc]} \BibitemShut {NoStop}%
\bibitem [{\citenamefont {Gupta}\ \emph {et~al.}(2025)\citenamefont {Gupta},
  \citenamefont {Narayan}, \citenamefont {London}, \citenamefont {Tiwari},\
  and\ \citenamefont {Sathyaprakash}}]{Gupta:2025paz}%
  \BibitemOpen
  \bibfield  {author} {\bibinfo {author} {\bibfnamefont {Ish}\ \bibnamefont
  {Gupta}}, \bibinfo {author} {\bibfnamefont {Purnima}\ \bibnamefont
  {Narayan}}, \bibinfo {author} {\bibfnamefont {Lionel}\ \bibnamefont
  {London}}, \bibinfo {author} {\bibfnamefont {Shubhanshu}\ \bibnamefont
  {Tiwari}}, \ and\ \bibinfo {author} {\bibfnamefont {Bangalore}\ \bibnamefont
  {Sathyaprakash}},\ }\bibfield  {title} {\enquote {\bibinfo {title} {{Testing
  general relativity with amplitudes of subdominant gravitational-wave
  modes}},}\ }\href@noop {} {\  (\bibinfo {year} {2025})},\ \Eprint
  {http://arxiv.org/abs/2511.11886} {arXiv:2511.11886 [gr-qc]} \BibitemShut
  {NoStop}%
\bibitem [{\citenamefont {O'Shea}\ and\ \citenamefont
  {Kumar}(2023)}]{OShea:2021faf}%
  \BibitemOpen
  \bibfield  {author} {\bibinfo {author} {\bibfnamefont {Eamonn}\ \bibnamefont
  {O'Shea}}\ and\ \bibinfo {author} {\bibfnamefont {Prayush}\ \bibnamefont
  {Kumar}},\ }\bibfield  {title} {\enquote {\bibinfo {title} {{Correlations in
  gravitational-wave reconstructions from eccentric binaries: A case study with
  GW151226 and GW170608}},}\ }\href {\doibase 10.1103/PhysRevD.108.104018}
  {\bibfield  {journal} {\bibinfo  {journal} {Phys. Rev. D}\ }\textbf {\bibinfo
  {volume} {108}},\ \bibinfo {pages} {104018} (\bibinfo {year} {2023})},\
  \Eprint {http://arxiv.org/abs/2107.07981} {arXiv:2107.07981 [astro-ph.HE]}
  \BibitemShut {NoStop}%
\bibitem [{\citenamefont {Favata}\ \emph {et~al.}(2022)\citenamefont {Favata},
  \citenamefont {Kim}, \citenamefont {Arun}, \citenamefont {Kim},\ and\
  \citenamefont {Lee}}]{Favata:2021vhw}%
  \BibitemOpen
  \bibfield  {author} {\bibinfo {author} {\bibfnamefont {Marc}\ \bibnamefont
  {Favata}}, \bibinfo {author} {\bibfnamefont {Chunglee}\ \bibnamefont {Kim}},
  \bibinfo {author} {\bibfnamefont {K.~G.}\ \bibnamefont {Arun}}, \bibinfo
  {author} {\bibfnamefont {JeongCho}\ \bibnamefont {Kim}}, \ and\ \bibinfo
  {author} {\bibfnamefont {Hyung~Won}\ \bibnamefont {Lee}},\ }\bibfield
  {title} {\enquote {\bibinfo {title} {{Constraining the orbital eccentricity
  of inspiralling compact binary systems with Advanced LIGO}},}\ }\href
  {\doibase 10.1103/PhysRevD.105.023003} {\bibfield  {journal} {\bibinfo
  {journal} {Phys. Rev. D}\ }\textbf {\bibinfo {volume} {105}},\ \bibinfo
  {pages} {023003} (\bibinfo {year} {2022})},\ \Eprint
  {http://arxiv.org/abs/2108.05861} {arXiv:2108.05861 [gr-qc]} \BibitemShut
  {NoStop}%
\bibitem [{\citenamefont {Divyajyoti}\ \emph {et~al.}(2024)\citenamefont
  {Divyajyoti}, \citenamefont {Kumar}, \citenamefont {Tibrewal}, \citenamefont
  {Romero-Shaw},\ and\ \citenamefont {Mishra}}]{Divyajyoti:2023rht}%
  \BibitemOpen
  \bibfield  {author} {\bibinfo {author} {\bibnamefont {Divyajyoti}}, \bibinfo
  {author} {\bibfnamefont {Sumit}\ \bibnamefont {Kumar}}, \bibinfo {author}
  {\bibfnamefont {Snehal}\ \bibnamefont {Tibrewal}}, \bibinfo {author}
  {\bibfnamefont {Isobel~M.}\ \bibnamefont {Romero-Shaw}}, \ and\ \bibinfo
  {author} {\bibfnamefont {Chandra~Kant}\ \bibnamefont {Mishra}},\ }\bibfield
  {title} {\enquote {\bibinfo {title} {{Blind spots and biases: The dangers of
  ignoring eccentricity in gravitational-wave signals from binary black
  holes}},}\ }\href {\doibase 10.1103/PhysRevD.109.043037} {\bibfield
  {journal} {\bibinfo  {journal} {Phys. Rev. D}\ }\textbf {\bibinfo {volume}
  {109}},\ \bibinfo {pages} {043037} (\bibinfo {year} {2024})},\ \Eprint
  {http://arxiv.org/abs/2309.16638} {arXiv:2309.16638 [gr-qc]} \BibitemShut
  {NoStop}%
\bibitem [{\citenamefont {Saini}\ \emph {et~al.}(2022)\citenamefont {Saini},
  \citenamefont {Favata},\ and\ \citenamefont {Arun}}]{Saini:2022igm}%
  \BibitemOpen
  \bibfield  {author} {\bibinfo {author} {\bibfnamefont {Pankaj}\ \bibnamefont
  {Saini}}, \bibinfo {author} {\bibfnamefont {Marc}\ \bibnamefont {Favata}}, \
  and\ \bibinfo {author} {\bibfnamefont {K.~G.}\ \bibnamefont {Arun}},\
  }\bibfield  {title} {\enquote {\bibinfo {title} {{Systematic bias on
  parametrized tests of general relativity due to neglect of orbital
  eccentricity}},}\ }\href {\doibase 10.1103/PhysRevD.106.084031} {\bibfield
  {journal} {\bibinfo  {journal} {Phys. Rev. D}\ }\textbf {\bibinfo {volume}
  {106}},\ \bibinfo {pages} {084031} (\bibinfo {year} {2022})},\ \Eprint
  {http://arxiv.org/abs/2203.04634} {arXiv:2203.04634 [gr-qc]} \BibitemShut
  {NoStop}%
\bibitem [{\citenamefont {Zeeshan}\ and\ \citenamefont
  {O'Shaughnessy}(2024)}]{Zeeshan:2024ovp}%
  \BibitemOpen
  \bibfield  {author} {\bibinfo {author} {\bibfnamefont {Muhammad}\
  \bibnamefont {Zeeshan}}\ and\ \bibinfo {author} {\bibfnamefont {Richard}\
  \bibnamefont {O'Shaughnessy}},\ }\bibfield  {title} {\enquote {\bibinfo
  {title} {{Eccentricity matters: Impact of eccentricity on inferred binary
  black hole populations}},}\ }\href {\doibase 10.1103/PhysRevD.110.063009}
  {\bibfield  {journal} {\bibinfo  {journal} {Phys. Rev. D}\ }\textbf {\bibinfo
  {volume} {110}},\ \bibinfo {pages} {063009} (\bibinfo {year} {2024})},\
  \Eprint {http://arxiv.org/abs/2404.08185} {arXiv:2404.08185 [gr-qc]}
  \BibitemShut {NoStop}%
\bibitem [{\citenamefont {Romero-Shaw}\ \emph {et~al.}(2021)\citenamefont
  {Romero-Shaw}, \citenamefont {Lasky},\ and\ \citenamefont
  {Thrane}}]{Romero-Shaw:2021ual}%
  \BibitemOpen
  \bibfield  {author} {\bibinfo {author} {\bibfnamefont {Isobel~M.}\
  \bibnamefont {Romero-Shaw}}, \bibinfo {author} {\bibfnamefont {Paul~D.}\
  \bibnamefont {Lasky}}, \ and\ \bibinfo {author} {\bibfnamefont {Eric}\
  \bibnamefont {Thrane}},\ }\bibfield  {title} {\enquote {\bibinfo {title}
  {{Signs of Eccentricity in Two Gravitational-wave Signals May Indicate a
  Subpopulation of Dynamically Assembled Binary Black Holes}},}\ }\href
  {\doibase 10.3847/2041-8213/ac3138} {\bibfield  {journal} {\bibinfo
  {journal} {Astrophys. J. Lett.}\ }\textbf {\bibinfo {volume} {921}},\
  \bibinfo {pages} {L31} (\bibinfo {year} {2021})},\ \Eprint
  {http://arxiv.org/abs/2108.01284} {arXiv:2108.01284 [astro-ph.HE]}
  \BibitemShut {NoStop}%
\bibitem [{\citenamefont {Iglesias}\ \emph {et~al.}(2024)\citenamefont
  {Iglesias} \emph {et~al.}}]{Iglesias:2022xfc}%
  \BibitemOpen
  \bibfield  {author} {\bibinfo {author} {\bibfnamefont {H.~L.}\ \bibnamefont
  {Iglesias}} \emph {et~al.},\ }\bibfield  {title} {\enquote {\bibinfo {title}
  {{Eccentricity Estimation for Five Binary Black Hole Mergers with
  Higher-order Gravitational-wave Modes}},}\ }\href {\doibase
  10.3847/1538-4357/ad5ff6} {\bibfield  {journal} {\bibinfo  {journal}
  {Astrophys. J.}\ }\textbf {\bibinfo {volume} {972}},\ \bibinfo {pages} {65}
  (\bibinfo {year} {2024})},\ \Eprint {http://arxiv.org/abs/2208.01766}
  {arXiv:2208.01766 [gr-qc]} \BibitemShut {NoStop}%
\bibitem [{\citenamefont {Gupte}\ \emph {et~al.}(2024)\citenamefont {Gupte}
  \emph {et~al.}}]{Gupte:2024jfe}%
  \BibitemOpen
  \bibfield  {author} {\bibinfo {author} {\bibfnamefont {Nihar}\ \bibnamefont
  {Gupte}} \emph {et~al.},\ }\bibfield  {title} {\enquote {\bibinfo {title}
  {{Evidence for eccentricity in the population of binary black holes observed
  by LIGO-Virgo-KAGRA}},}\ }\href@noop {} {\  (\bibinfo {year} {2024})},\
  \Eprint {http://arxiv.org/abs/2404.14286} {arXiv:2404.14286 [gr-qc]}
  \BibitemShut {NoStop}%
\bibitem [{\citenamefont {Planas}\ \emph {et~al.}(2025)\citenamefont {Planas},
  \citenamefont {Ramos-Buades}, \citenamefont {Garc{\'\i}a-Quir{\'o}s},
  \citenamefont {Estell{\'e}s}, \citenamefont {Husa},\ and\ \citenamefont
  {Haney}}]{Planas:2025jny}%
  \BibitemOpen
  \bibfield  {author} {\bibinfo {author} {\bibfnamefont {Maria de~Lluc}\
  \bibnamefont {Planas}}, \bibinfo {author} {\bibfnamefont {Antoni}\
  \bibnamefont {Ramos-Buades}}, \bibinfo {author} {\bibfnamefont {Cecilio}\
  \bibnamefont {Garc{\'\i}a-Quir{\'o}s}}, \bibinfo {author} {\bibfnamefont
  {H{\'e}ctor}\ \bibnamefont {Estell{\'e}s}}, \bibinfo {author} {\bibfnamefont
  {Sascha}\ \bibnamefont {Husa}}, \ and\ \bibinfo {author} {\bibfnamefont
  {Maria}\ \bibnamefont {Haney}},\ }\bibfield  {title} {\enquote {\bibinfo
  {title} {{Eccentric or circular? A reanalysis of binary black hole
  gravitational wave events for orbital eccentricity signatures}},}\
  }\href@noop {} {\  (\bibinfo {year} {2025})},\ \Eprint
  {http://arxiv.org/abs/2504.15833} {arXiv:2504.15833 [gr-qc]} \BibitemShut
  {NoStop}%
\bibitem [{\citenamefont {Romero-Shaw}\ \emph {et~al.}(2025)\citenamefont
  {Romero-Shaw}, \citenamefont {Stegmann}, \citenamefont {Tagawa},
  \citenamefont {Gerosa}, \citenamefont {Samsing}, \citenamefont {Gupte},\ and\
  \citenamefont {Green}}]{Romero-Shaw:2025vbc}%
  \BibitemOpen
  \bibfield  {author} {\bibinfo {author} {\bibfnamefont {Isobel}\ \bibnamefont
  {Romero-Shaw}}, \bibinfo {author} {\bibfnamefont {Jakob}\ \bibnamefont
  {Stegmann}}, \bibinfo {author} {\bibfnamefont {Hiromichi}\ \bibnamefont
  {Tagawa}}, \bibinfo {author} {\bibfnamefont {Davide}\ \bibnamefont {Gerosa}},
  \bibinfo {author} {\bibfnamefont {Johan}\ \bibnamefont {Samsing}}, \bibinfo
  {author} {\bibfnamefont {Nihar}\ \bibnamefont {Gupte}}, \ and\ \bibinfo
  {author} {\bibfnamefont {Stephen~R.}\ \bibnamefont {Green}},\ }\bibfield
  {title} {\enquote {\bibinfo {title} {{GW200208{\_}222617 as an eccentric
  black-hole binary merger: Properties and astrophysical implications}},}\
  }\href {\doibase 10.1103/jj7m-x66y} {\bibfield  {journal} {\bibinfo
  {journal} {Phys. Rev. D}\ }\textbf {\bibinfo {volume} {112}},\ \bibinfo
  {pages} {063052} (\bibinfo {year} {2025})},\ \Eprint
  {http://arxiv.org/abs/2506.17105} {arXiv:2506.17105 [astro-ph.HE]}
  \BibitemShut {NoStop}%
\bibitem [{\citenamefont {Klein}\ \emph {et~al.}(2018)\citenamefont {Klein},
  \citenamefont {Boetzel}, \citenamefont {Gopakumar}, \citenamefont {Jetzer},\
  and\ \citenamefont {de~Vittori}}]{Klein:2018ybm}%
  \BibitemOpen
  \bibfield  {author} {\bibinfo {author} {\bibfnamefont {Antoine}\ \bibnamefont
  {Klein}}, \bibinfo {author} {\bibfnamefont {Yannick}\ \bibnamefont
  {Boetzel}}, \bibinfo {author} {\bibfnamefont {Achamveedu}\ \bibnamefont
  {Gopakumar}}, \bibinfo {author} {\bibfnamefont {Philippe}\ \bibnamefont
  {Jetzer}}, \ and\ \bibinfo {author} {\bibfnamefont {Lorenzo}\ \bibnamefont
  {de~Vittori}},\ }\bibfield  {title} {\enquote {\bibinfo {title} {{Fourier
  domain gravitational waveforms for precessing eccentric binaries}},}\ }\href
  {\doibase 10.1103/PhysRevD.98.104043} {\bibfield  {journal} {\bibinfo
  {journal} {Phys. Rev. D}\ }\textbf {\bibinfo {volume} {98}},\ \bibinfo
  {pages} {104043} (\bibinfo {year} {2018})},\ \Eprint
  {http://arxiv.org/abs/1801.08542} {arXiv:1801.08542 [gr-qc]} \BibitemShut
  {NoStop}%
\bibitem [{\citenamefont {Gamba}\ \emph
  {et~al.}(2024{\natexlab{a}})\citenamefont {Gamba}, \citenamefont
  {Chiaramello},\ and\ \citenamefont {Neogi}}]{Gamba:2024cvy}%
  \BibitemOpen
  \bibfield  {author} {\bibinfo {author} {\bibfnamefont {Rossella}\
  \bibnamefont {Gamba}}, \bibinfo {author} {\bibfnamefont {Danilo}\
  \bibnamefont {Chiaramello}}, \ and\ \bibinfo {author} {\bibfnamefont {Sayan}\
  \bibnamefont {Neogi}},\ }\bibfield  {title} {\enquote {\bibinfo {title}
  {{Toward efficient effective-one-body models for generic, nonplanar
  orbits}},}\ }\href {\doibase 10.1103/PhysRevD.110.024031} {\bibfield
  {journal} {\bibinfo  {journal} {Phys. Rev. D}\ }\textbf {\bibinfo {volume}
  {110}},\ \bibinfo {pages} {024031} (\bibinfo {year} {2024}{\natexlab{a}})},\
  \Eprint {http://arxiv.org/abs/2404.15408} {arXiv:2404.15408 [gr-qc]}
  \BibitemShut {NoStop}%
\bibitem [{\citenamefont {Morras}\ \emph
  {et~al.}(2025{\natexlab{a}})\citenamefont {Morras}, \citenamefont {Pratten},\
  and\ \citenamefont {Schmidt}}]{Morras:2025xfu}%
  \BibitemOpen
  \bibfield  {author} {\bibinfo {author} {\bibfnamefont {Gonzalo}\ \bibnamefont
  {Morras}}, \bibinfo {author} {\bibfnamefont {Geraint}\ \bibnamefont
  {Pratten}}, \ and\ \bibinfo {author} {\bibfnamefont {Patricia}\ \bibnamefont
  {Schmidt}},\ }\bibfield  {title} {\enquote {\bibinfo {title} {{Orbital
  eccentricity in a neutron star - black hole binary}},}\ }\href@noop {} {\
  (\bibinfo {year} {2025}{\natexlab{a}})},\ \Eprint
  {http://arxiv.org/abs/2503.15393} {arXiv:2503.15393 [astro-ph.HE]}
  \BibitemShut {NoStop}%
\bibitem [{\citenamefont {Morras}\ \emph
  {et~al.}(2025{\natexlab{b}})\citenamefont {Morras}, \citenamefont {Pratten},\
  and\ \citenamefont {Schmidt}}]{Morras:2025nlp}%
  \BibitemOpen
  \bibfield  {author} {\bibinfo {author} {\bibfnamefont {Gonzalo}\ \bibnamefont
  {Morras}}, \bibinfo {author} {\bibfnamefont {Geraint}\ \bibnamefont
  {Pratten}}, \ and\ \bibinfo {author} {\bibfnamefont {Patricia}\ \bibnamefont
  {Schmidt}},\ }\bibfield  {title} {\enquote {\bibinfo {title} {{Improved
  post-Newtonian waveform model for inspiralling precessing-eccentric compact
  binaries}},}\ }\href {\doibase 10.1103/PhysRevD.111.084052} {\bibfield
  {journal} {\bibinfo  {journal} {Phys. Rev. D}\ }\textbf {\bibinfo {volume}
  {111}},\ \bibinfo {pages} {084052} (\bibinfo {year} {2025}{\natexlab{b}})},\
  \Eprint {http://arxiv.org/abs/2502.03929} {arXiv:2502.03929 [gr-qc]}
  \BibitemShut {NoStop}%
\bibitem [{\citenamefont {Abac}\ \emph
  {et~al.}(2025{\natexlab{d}})\citenamefont {Abac} \emph
  {et~al.}}]{LIGOScientific:2025hdt}%
  \BibitemOpen
  \bibfield  {author} {\bibinfo {author} {\bibfnamefont {A.~G.}\ \bibnamefont
  {Abac}} \emph {et~al.} (\bibinfo {collaboration} {LIGO Scientific, VIRGO,
  KAGRA}),\ }\bibfield  {title} {\enquote {\bibinfo {title} {{GWTC-4.0: An
  Introduction to Version 4.0 of the Gravitational-Wave Transient Catalog}},}\
  }\href@noop {} {\  (\bibinfo {year} {2025}{\natexlab{d}})},\ \Eprint
  {http://arxiv.org/abs/2508.18080} {arXiv:2508.18080 [gr-qc]} \BibitemShut
  {NoStop}%
\bibitem [{\citenamefont {{Pratten}}\ \emph {et~al.}(2021)\citenamefont
  {{Pratten}}, \citenamefont {{Garc{\'\i}a-Quir{\'o}s}}, \citenamefont
  {{Colleoni}}, \citenamefont {{Ramos-Buades}}, \citenamefont {{Estell{\'e}s}},
  \citenamefont {{Mateu-Lucena}}, \citenamefont {{Jaume}}, \citenamefont
  {{Haney}}, \citenamefont {{Keitel}}, \citenamefont {{Thompson}},\ and\
  \citenamefont {{Husa}}}]{2021PhRvD.103j4056P}%
  \BibitemOpen
  \bibfield  {author} {\bibinfo {author} {\bibfnamefont {Geraint}\ \bibnamefont
  {{Pratten}}}, \bibinfo {author} {\bibfnamefont {Cecilio}\ \bibnamefont
  {{Garc{\'\i}a-Quir{\'o}s}}}, \bibinfo {author} {\bibfnamefont {Marta}\
  \bibnamefont {{Colleoni}}}, \bibinfo {author} {\bibfnamefont {Antoni}\
  \bibnamefont {{Ramos-Buades}}}, \bibinfo {author} {\bibfnamefont
  {H{\'e}ctor}\ \bibnamefont {{Estell{\'e}s}}}, \bibinfo {author}
  {\bibfnamefont {Maite}\ \bibnamefont {{Mateu-Lucena}}}, \bibinfo {author}
  {\bibfnamefont {Rafel}\ \bibnamefont {{Jaume}}}, \bibinfo {author}
  {\bibfnamefont {Maria}\ \bibnamefont {{Haney}}}, \bibinfo {author}
  {\bibfnamefont {David}\ \bibnamefont {{Keitel}}}, \bibinfo {author}
  {\bibfnamefont {Jonathan~E.}\ \bibnamefont {{Thompson}}}, \ and\ \bibinfo
  {author} {\bibfnamefont {Sascha}\ \bibnamefont {{Husa}}},\ }\bibfield
  {title} {\enquote {\bibinfo {title} {{Computationally efficient models for
  the dominant and subdominant harmonic modes of precessing binary black
  holes}},}\ }\href {\doibase 10.1103/PhysRevD.103.104056} {\bibfield
  {journal} {\bibinfo  {journal} {\prd}\ }\textbf {\bibinfo {volume} {103}},\
  \bibinfo {eid} {104056} (\bibinfo {year} {2021})},\ \Eprint
  {http://arxiv.org/abs/2004.06503} {arXiv:2004.06503 [gr-qc]} \BibitemShut
  {NoStop}%
\bibitem [{\citenamefont {{Colleoni}}\ \emph {et~al.}(2025)\citenamefont
  {{Colleoni}}, \citenamefont {{Ramis Vidal}}, \citenamefont
  {{Garc{\'\i}a-Quir{\'o}s}}, \citenamefont {{Ak{\c{c}}ay}},\ and\
  \citenamefont {{Bera}}}]{2025PhRvD.111j4019C}%
  \BibitemOpen
  \bibfield  {author} {\bibinfo {author} {\bibfnamefont {Marta}\ \bibnamefont
  {{Colleoni}}}, \bibinfo {author} {\bibfnamefont {Felip~A.}\ \bibnamefont
  {{Ramis Vidal}}}, \bibinfo {author} {\bibfnamefont {Cecilio}\ \bibnamefont
  {{Garc{\'\i}a-Quir{\'o}s}}}, \bibinfo {author} {\bibfnamefont {Sarp}\
  \bibnamefont {{Ak{\c{c}}ay}}}, \ and\ \bibinfo {author} {\bibfnamefont
  {Sayantani}\ \bibnamefont {{Bera}}},\ }\bibfield  {title} {\enquote {\bibinfo
  {title} {{Fast frequency-domain gravitational waveforms for precessing
  binaries with a new twist}},}\ }\href {\doibase 10.1103/PhysRevD.111.104019}
  {\bibfield  {journal} {\bibinfo  {journal} {\prd}\ }\textbf {\bibinfo
  {volume} {111}},\ \bibinfo {eid} {104019} (\bibinfo {year} {2025})},\ \Eprint
  {http://arxiv.org/abs/2412.16721} {arXiv:2412.16721 [gr-qc]} \BibitemShut
  {NoStop}%
\bibitem [{\citenamefont {Hamilton}\ \emph {et~al.}(2021)\citenamefont
  {Hamilton}, \citenamefont {London}, \citenamefont {Thompson}, \citenamefont
  {Fauchon-Jones}, \citenamefont {Hannam}, \citenamefont {Kalaghatgi},
  \citenamefont {Khan}, \citenamefont {Pannarale},\ and\ \citenamefont
  {Vano-Vinuales}}]{PhysRevD.104.124027}%
  \BibitemOpen
  \bibfield  {author} {\bibinfo {author} {\bibfnamefont {Eleanor}\ \bibnamefont
  {Hamilton}}, \bibinfo {author} {\bibfnamefont {Lionel}\ \bibnamefont
  {London}}, \bibinfo {author} {\bibfnamefont {Jonathan~E.}\ \bibnamefont
  {Thompson}}, \bibinfo {author} {\bibfnamefont {Edward}\ \bibnamefont
  {Fauchon-Jones}}, \bibinfo {author} {\bibfnamefont {Mark}\ \bibnamefont
  {Hannam}}, \bibinfo {author} {\bibfnamefont {Chinmay}\ \bibnamefont
  {Kalaghatgi}}, \bibinfo {author} {\bibfnamefont {Sebastian}\ \bibnamefont
  {Khan}}, \bibinfo {author} {\bibfnamefont {Francesco}\ \bibnamefont
  {Pannarale}}, \ and\ \bibinfo {author} {\bibfnamefont {Alex}\ \bibnamefont
  {Vano-Vinuales}},\ }\bibfield  {title} {\enquote {\bibinfo {title} {Model of
  gravitational waves from precessing black-hole binaries through merger and
  ringdown},}\ }\href {\doibase 10.1103/PhysRevD.104.124027} {\bibfield
  {journal} {\bibinfo  {journal} {Phys. Rev. D}\ }\textbf {\bibinfo {volume}
  {104}},\ \bibinfo {pages} {124027} (\bibinfo {year} {2021})}\BibitemShut
  {NoStop}%
\bibitem [{\citenamefont {Thompson}\ \emph {et~al.}(2024)\citenamefont
  {Thompson}, \citenamefont {Hamilton}, \citenamefont {London}, \citenamefont
  {Ghosh}, \citenamefont {Kolitsidou}, \citenamefont {Hoy},\ and\ \citenamefont
  {Hannam}}]{PhysRevD.109.063012}%
  \BibitemOpen
  \bibfield  {author} {\bibinfo {author} {\bibfnamefont {Jonathan~E.}\
  \bibnamefont {Thompson}}, \bibinfo {author} {\bibfnamefont {Eleanor}\
  \bibnamefont {Hamilton}}, \bibinfo {author} {\bibfnamefont {Lionel}\
  \bibnamefont {London}}, \bibinfo {author} {\bibfnamefont {Shrobana}\
  \bibnamefont {Ghosh}}, \bibinfo {author} {\bibfnamefont {Panagiota}\
  \bibnamefont {Kolitsidou}}, \bibinfo {author} {\bibfnamefont {Charlie}\
  \bibnamefont {Hoy}}, \ and\ \bibinfo {author} {\bibfnamefont {Mark}\
  \bibnamefont {Hannam}},\ }\bibfield  {title} {\enquote {\bibinfo {title}
  {Phenomenological gravitational-wave model for precessing black-hole binaries
  with higher multipoles and asymmetries},}\ }\href {\doibase
  10.1103/PhysRevD.109.063012} {\bibfield  {journal} {\bibinfo  {journal}
  {Phys. Rev. D}\ }\textbf {\bibinfo {volume} {109}},\ \bibinfo {pages}
  {063012} (\bibinfo {year} {2024})}\BibitemShut {NoStop}%
\bibitem [{\citenamefont {{Hamilton}}\ \emph {et~al.}(2025)\citenamefont
  {{Hamilton}}, \citenamefont {{Colleoni}}, \citenamefont {{Thompson}},
  \citenamefont {{Hoy}}, \citenamefont {{Heffernan}}, \citenamefont
  {{Kinnear}}, \citenamefont {{Valencia}}, \citenamefont {{Ramis Vidal}},
  \citenamefont {{Garc{\'\i}a-Quir{\'o}s}}, \citenamefont {{Ghosh}},
  \citenamefont {{London}}, \citenamefont {{Hannam}},\ and\ \citenamefont
  {{Husa}}}]{2025arXiv250702604H}%
  \BibitemOpen
  \bibfield  {author} {\bibinfo {author} {\bibfnamefont {Eleanor}\ \bibnamefont
  {{Hamilton}}}, \bibinfo {author} {\bibfnamefont {Marta}\ \bibnamefont
  {{Colleoni}}}, \bibinfo {author} {\bibfnamefont {Jonathan~E.}\ \bibnamefont
  {{Thompson}}}, \bibinfo {author} {\bibfnamefont {Charlie}\ \bibnamefont
  {{Hoy}}}, \bibinfo {author} {\bibfnamefont {Anna}\ \bibnamefont
  {{Heffernan}}}, \bibinfo {author} {\bibfnamefont {Meryl}\ \bibnamefont
  {{Kinnear}}}, \bibinfo {author} {\bibfnamefont {Jorge}\ \bibnamefont
  {{Valencia}}}, \bibinfo {author} {\bibfnamefont {Felip~A}\ \bibnamefont
  {{Ramis Vidal}}}, \bibinfo {author} {\bibfnamefont {Cecilio}\ \bibnamefont
  {{Garc{\'\i}a-Quir{\'o}s}}}, \bibinfo {author} {\bibfnamefont {Shrobana}\
  \bibnamefont {{Ghosh}}}, \bibinfo {author} {\bibfnamefont {Lionel}\
  \bibnamefont {{London}}}, \bibinfo {author} {\bibfnamefont {Mark}\
  \bibnamefont {{Hannam}}}, \ and\ \bibinfo {author} {\bibfnamefont {Sascha}\
  \bibnamefont {{Husa}}},\ }\bibfield  {title} {\enquote {\bibinfo {title}
  {{PhenomXPNR: An improved gravitational wave model linking precessing
  inspirals and NR-calibrated merger-ringdown}},}\ }\href {\doibase
  10.48550/arXiv.2507.02604} {\bibfield  {journal} {\bibinfo  {journal} {arXiv
  e-prints}\ ,\ \bibinfo {eid} {arXiv:2507.02604}} (\bibinfo {year} {2025})},\
  \Eprint {http://arxiv.org/abs/2507.02604} {arXiv:2507.02604 [gr-qc]}
  \BibitemShut {NoStop}%
\bibitem [{\citenamefont {Estell\'es}\ \emph
  {et~al.}(2022{\natexlab{a}})\citenamefont {Estell\'es}, \citenamefont {Husa},
  \citenamefont {Colleoni}, \citenamefont {Keitel}, \citenamefont
  {Mateu-Lucena}, \citenamefont {Garc\'{\i}a-Quir\'os}, \citenamefont
  {Ramos-Buades},\ and\ \citenamefont {Borchers}}]{PhysRevD.105.084039}%
  \BibitemOpen
  \bibfield  {author} {\bibinfo {author} {\bibfnamefont {H\'ector}\
  \bibnamefont {Estell\'es}}, \bibinfo {author} {\bibfnamefont {Sascha}\
  \bibnamefont {Husa}}, \bibinfo {author} {\bibfnamefont {Marta}\ \bibnamefont
  {Colleoni}}, \bibinfo {author} {\bibfnamefont {David}\ \bibnamefont
  {Keitel}}, \bibinfo {author} {\bibfnamefont {Maite}\ \bibnamefont
  {Mateu-Lucena}}, \bibinfo {author} {\bibfnamefont {Cecilio}\ \bibnamefont
  {Garc\'{\i}a-Quir\'os}}, \bibinfo {author} {\bibfnamefont {Antoni}\
  \bibnamefont {Ramos-Buades}}, \ and\ \bibinfo {author} {\bibfnamefont
  {Angela}\ \bibnamefont {Borchers}},\ }\bibfield  {title} {\enquote {\bibinfo
  {title} {Time-domain phenomenological model of gravitational-wave subdominant
  harmonics for quasicircular nonprecessing binary black hole coalescences},}\
  }\href {\doibase 10.1103/PhysRevD.105.084039} {\bibfield  {journal} {\bibinfo
   {journal} {Phys. Rev. D}\ }\textbf {\bibinfo {volume} {105}},\ \bibinfo
  {pages} {084039} (\bibinfo {year} {2022}{\natexlab{a}})}\BibitemShut
  {NoStop}%
\bibitem [{\citenamefont {Estell\'es}\ \emph
  {et~al.}(2022{\natexlab{b}})\citenamefont {Estell\'es}, \citenamefont
  {Colleoni}, \citenamefont {Garc\'{\i}a-Quir\'os}, \citenamefont {Husa},
  \citenamefont {Keitel}, \citenamefont {Mateu-Lucena}, \citenamefont
  {Planas},\ and\ \citenamefont {Ramos-Buades}}]{PhysRevD.105.084040}%
  \BibitemOpen
  \bibfield  {author} {\bibinfo {author} {\bibfnamefont {H\'ector}\
  \bibnamefont {Estell\'es}}, \bibinfo {author} {\bibfnamefont {Marta}\
  \bibnamefont {Colleoni}}, \bibinfo {author} {\bibfnamefont {Cecilio}\
  \bibnamefont {Garc\'{\i}a-Quir\'os}}, \bibinfo {author} {\bibfnamefont
  {Sascha}\ \bibnamefont {Husa}}, \bibinfo {author} {\bibfnamefont {David}\
  \bibnamefont {Keitel}}, \bibinfo {author} {\bibfnamefont {Maite}\
  \bibnamefont {Mateu-Lucena}}, \bibinfo {author} {\bibfnamefont {Maria
  de~Lluc}\ \bibnamefont {Planas}}, \ and\ \bibinfo {author} {\bibfnamefont
  {Antoni}\ \bibnamefont {Ramos-Buades}},\ }\bibfield  {title} {\enquote
  {\bibinfo {title} {New twists in compact binary waveform modeling: A fast
  time-domain model for precession},}\ }\href {\doibase
  10.1103/PhysRevD.105.084040} {\bibfield  {journal} {\bibinfo  {journal}
  {Phys. Rev. D}\ }\textbf {\bibinfo {volume} {105}},\ \bibinfo {pages}
  {084040} (\bibinfo {year} {2022}{\natexlab{b}})}\BibitemShut {NoStop}%
\bibitem [{\citenamefont {Boh\'e}\ \emph {et~al.}(2017)\citenamefont {Boh\'e},
  \citenamefont {Shao}, \citenamefont {Taracchini}, \citenamefont {Buonanno},
  \citenamefont {Babak}, \citenamefont {Harry}, \citenamefont {Hinder},
  \citenamefont {Ossokine}, \citenamefont {P\"urrer}, \citenamefont {Raymond},
  \citenamefont {Chu}, \citenamefont {Fong}, \citenamefont {Kumar},
  \citenamefont {Pfeiffer}, \citenamefont {Boyle}, \citenamefont {Hemberger},
  \citenamefont {Kidder}, \citenamefont {Lovelace}, \citenamefont {Scheel},\
  and\ \citenamefont {Szil\'agyi}}]{PhysRevD.95.044028}%
  \BibitemOpen
  \bibfield  {author} {\bibinfo {author} {\bibfnamefont {Alejandro}\
  \bibnamefont {Boh\'e}}, \bibinfo {author} {\bibfnamefont {Lijing}\
  \bibnamefont {Shao}}, \bibinfo {author} {\bibfnamefont {Andrea}\ \bibnamefont
  {Taracchini}}, \bibinfo {author} {\bibfnamefont {Alessandra}\ \bibnamefont
  {Buonanno}}, \bibinfo {author} {\bibfnamefont {Stanislav}\ \bibnamefont
  {Babak}}, \bibinfo {author} {\bibfnamefont {Ian~W.}\ \bibnamefont {Harry}},
  \bibinfo {author} {\bibfnamefont {Ian}\ \bibnamefont {Hinder}}, \bibinfo
  {author} {\bibfnamefont {Serguei}\ \bibnamefont {Ossokine}}, \bibinfo
  {author} {\bibfnamefont {Michael}\ \bibnamefont {P\"urrer}}, \bibinfo
  {author} {\bibfnamefont {Vivien}\ \bibnamefont {Raymond}}, \bibinfo {author}
  {\bibfnamefont {Tony}\ \bibnamefont {Chu}}, \bibinfo {author} {\bibfnamefont
  {Heather}\ \bibnamefont {Fong}}, \bibinfo {author} {\bibfnamefont {Prayush}\
  \bibnamefont {Kumar}}, \bibinfo {author} {\bibfnamefont {Harald~P.}\
  \bibnamefont {Pfeiffer}}, \bibinfo {author} {\bibfnamefont {Michael}\
  \bibnamefont {Boyle}}, \bibinfo {author} {\bibfnamefont {Daniel~A.}\
  \bibnamefont {Hemberger}}, \bibinfo {author} {\bibfnamefont {Lawrence~E.}\
  \bibnamefont {Kidder}}, \bibinfo {author} {\bibfnamefont {Geoffrey}\
  \bibnamefont {Lovelace}}, \bibinfo {author} {\bibfnamefont {Mark~A.}\
  \bibnamefont {Scheel}}, \ and\ \bibinfo {author} {\bibfnamefont {B\'ela}\
  \bibnamefont {Szil\'agyi}},\ }\bibfield  {title} {\enquote {\bibinfo {title}
  {Improved effective-one-body model of spinning, nonprecessing binary black
  holes for the era of gravitational-wave astrophysics with advanced
  detectors},}\ }\href {\doibase 10.1103/PhysRevD.95.044028} {\bibfield
  {journal} {\bibinfo  {journal} {Phys. Rev. D}\ }\textbf {\bibinfo {volume}
  {95}},\ \bibinfo {pages} {044028} (\bibinfo {year} {2017})}\BibitemShut
  {NoStop}%
\bibitem [{\citenamefont {Cotesta}\ \emph {et~al.}(2018)\citenamefont
  {Cotesta}, \citenamefont {Buonanno}, \citenamefont {Boh\'e}, \citenamefont
  {Taracchini}, \citenamefont {Hinder},\ and\ \citenamefont
  {Ossokine}}]{PhysRevD.98.084028}%
  \BibitemOpen
  \bibfield  {author} {\bibinfo {author} {\bibfnamefont {Roberto}\ \bibnamefont
  {Cotesta}}, \bibinfo {author} {\bibfnamefont {Alessandra}\ \bibnamefont
  {Buonanno}}, \bibinfo {author} {\bibfnamefont {Alejandro}\ \bibnamefont
  {Boh\'e}}, \bibinfo {author} {\bibfnamefont {Andrea}\ \bibnamefont
  {Taracchini}}, \bibinfo {author} {\bibfnamefont {Ian}\ \bibnamefont
  {Hinder}}, \ and\ \bibinfo {author} {\bibfnamefont {Serguei}\ \bibnamefont
  {Ossokine}},\ }\bibfield  {title} {\enquote {\bibinfo {title} {Enriching the
  symphony of gravitational waves from binary black holes by tuning higher
  harmonics},}\ }\href {\doibase 10.1103/PhysRevD.98.084028} {\bibfield
  {journal} {\bibinfo  {journal} {Phys. Rev. D}\ }\textbf {\bibinfo {volume}
  {98}},\ \bibinfo {pages} {084028} (\bibinfo {year} {2018})}\BibitemShut
  {NoStop}%
\bibitem [{\citenamefont {Ossokine}\ \emph {et~al.}(2020)\citenamefont
  {Ossokine}, \citenamefont {Buonanno}, \citenamefont {Marsat}, \citenamefont
  {Cotesta}, \citenamefont {Babak}, \citenamefont {Dietrich}, \citenamefont
  {Haas}, \citenamefont {Hinder}, \citenamefont {Pfeiffer}, \citenamefont
  {P\"urrer}, \citenamefont {Woodford}, \citenamefont {Boyle}, \citenamefont
  {Kidder}, \citenamefont {Scheel},\ and\ \citenamefont
  {Szil\'agyi}}]{PhysRevD.102.044055}%
  \BibitemOpen
  \bibfield  {author} {\bibinfo {author} {\bibfnamefont {Serguei}\ \bibnamefont
  {Ossokine}}, \bibinfo {author} {\bibfnamefont {Alessandra}\ \bibnamefont
  {Buonanno}}, \bibinfo {author} {\bibfnamefont {Sylvain}\ \bibnamefont
  {Marsat}}, \bibinfo {author} {\bibfnamefont {Roberto}\ \bibnamefont
  {Cotesta}}, \bibinfo {author} {\bibfnamefont {Stanislav}\ \bibnamefont
  {Babak}}, \bibinfo {author} {\bibfnamefont {Tim}\ \bibnamefont {Dietrich}},
  \bibinfo {author} {\bibfnamefont {Roland}\ \bibnamefont {Haas}}, \bibinfo
  {author} {\bibfnamefont {Ian}\ \bibnamefont {Hinder}}, \bibinfo {author}
  {\bibfnamefont {Harald~P.}\ \bibnamefont {Pfeiffer}}, \bibinfo {author}
  {\bibfnamefont {Michael}\ \bibnamefont {P\"urrer}}, \bibinfo {author}
  {\bibfnamefont {Charles~J.}\ \bibnamefont {Woodford}}, \bibinfo {author}
  {\bibfnamefont {Michael}\ \bibnamefont {Boyle}}, \bibinfo {author}
  {\bibfnamefont {Lawrence~E.}\ \bibnamefont {Kidder}}, \bibinfo {author}
  {\bibfnamefont {Mark~A.}\ \bibnamefont {Scheel}}, \ and\ \bibinfo {author}
  {\bibfnamefont {B\'ela}\ \bibnamefont {Szil\'agyi}},\ }\bibfield  {title}
  {\enquote {\bibinfo {title} {Multipolar effective-one-body waveforms for
  precessing binary black holes: Construction and validation},}\ }\href
  {\doibase 10.1103/PhysRevD.102.044055} {\bibfield  {journal} {\bibinfo
  {journal} {Phys. Rev. D}\ }\textbf {\bibinfo {volume} {102}},\ \bibinfo
  {pages} {044055} (\bibinfo {year} {2020})}\BibitemShut {NoStop}%
\bibitem [{\citenamefont {Cotesta}\ \emph {et~al.}(2020)\citenamefont
  {Cotesta}, \citenamefont {Marsat},\ and\ \citenamefont
  {P\"urrer}}]{PhysRevD.101.124040}%
  \BibitemOpen
  \bibfield  {author} {\bibinfo {author} {\bibfnamefont {Roberto}\ \bibnamefont
  {Cotesta}}, \bibinfo {author} {\bibfnamefont {Sylvain}\ \bibnamefont
  {Marsat}}, \ and\ \bibinfo {author} {\bibfnamefont {Michael}\ \bibnamefont
  {P\"urrer}},\ }\bibfield  {title} {\enquote {\bibinfo {title}
  {Frequency-domain reduced-order model of aligned-spin effective-one-body
  waveforms with higher-order modes},}\ }\href {\doibase
  10.1103/PhysRevD.101.124040} {\bibfield  {journal} {\bibinfo  {journal}
  {Phys. Rev. D}\ }\textbf {\bibinfo {volume} {101}},\ \bibinfo {pages}
  {124040} (\bibinfo {year} {2020})}\BibitemShut {NoStop}%
\bibitem [{\citenamefont {Mihaylov}\ \emph {et~al.}(2021)\citenamefont
  {Mihaylov}, \citenamefont {Ossokine}, \citenamefont {Buonanno},\ and\
  \citenamefont {Ghosh}}]{PhysRevD.104.124087}%
  \BibitemOpen
  \bibfield  {author} {\bibinfo {author} {\bibfnamefont {Deyan~P.}\
  \bibnamefont {Mihaylov}}, \bibinfo {author} {\bibfnamefont {Serguei}\
  \bibnamefont {Ossokine}}, \bibinfo {author} {\bibfnamefont {Alessandra}\
  \bibnamefont {Buonanno}}, \ and\ \bibinfo {author} {\bibfnamefont {Abhirup}\
  \bibnamefont {Ghosh}},\ }\bibfield  {title} {\enquote {\bibinfo {title} {Fast
  post-adiabatic waveforms in the time domain: Applications to compact binary
  coalescences in ligo and virgo},}\ }\href {\doibase
  10.1103/PhysRevD.104.124087} {\bibfield  {journal} {\bibinfo  {journal}
  {Phys. Rev. D}\ }\textbf {\bibinfo {volume} {104}},\ \bibinfo {pages}
  {124087} (\bibinfo {year} {2021})}\BibitemShut {NoStop}%
\bibitem [{\citenamefont {{Pompili}}\ \emph {et~al.}(2023)\citenamefont
  {{Pompili}}, \citenamefont {{Buonanno}}, \citenamefont {{Estell{\'e}s}},
  \citenamefont {{Khalil}}, \citenamefont {{van de Meent}}, \citenamefont
  {{Mihaylov}}, \citenamefont {{Ossokine}}, \citenamefont {{P{\"u}rrer}},
  \citenamefont {{Ramos-Buades}}, \citenamefont {{Mehta}}, \citenamefont
  {{Cotesta}}, \citenamefont {{Marsat}}, \citenamefont {{Boyle}}, \citenamefont
  {{Kidder}}, \citenamefont {{Pfeiffer}}, \citenamefont {{Scheel}},
  \citenamefont {{R{\"u}ter}}, \citenamefont {{Vu}}, \citenamefont {{Dudi}},
  \citenamefont {{Ma}}, \citenamefont {{Mitman}}, \citenamefont {{Melchor}},
  \citenamefont {{Thomas}},\ and\ \citenamefont
  {{Sanchez}}}]{2023PhRvD.108l4035P}%
  \BibitemOpen
  \bibfield  {author} {\bibinfo {author} {\bibfnamefont {Lorenzo}\ \bibnamefont
  {{Pompili}}}, \bibinfo {author} {\bibfnamefont {Alessandra}\ \bibnamefont
  {{Buonanno}}}, \bibinfo {author} {\bibfnamefont {H{\'e}ctor}\ \bibnamefont
  {{Estell{\'e}s}}}, \bibinfo {author} {\bibfnamefont {Mohammed}\ \bibnamefont
  {{Khalil}}}, \bibinfo {author} {\bibfnamefont {Maarten}\ \bibnamefont {{van
  de Meent}}}, \bibinfo {author} {\bibfnamefont {Deyan~P.}\ \bibnamefont
  {{Mihaylov}}}, \bibinfo {author} {\bibfnamefont {Serguei}\ \bibnamefont
  {{Ossokine}}}, \bibinfo {author} {\bibfnamefont {Michael}\ \bibnamefont
  {{P{\"u}rrer}}}, \bibinfo {author} {\bibfnamefont {Antoni}\ \bibnamefont
  {{Ramos-Buades}}}, \bibinfo {author} {\bibfnamefont {Ajit~Kumar}\
  \bibnamefont {{Mehta}}}, \bibinfo {author} {\bibfnamefont {Roberto}\
  \bibnamefont {{Cotesta}}}, \bibinfo {author} {\bibfnamefont {Sylvain}\
  \bibnamefont {{Marsat}}}, \bibinfo {author} {\bibfnamefont {Michael}\
  \bibnamefont {{Boyle}}}, \bibinfo {author} {\bibfnamefont {Lawrence~E.}\
  \bibnamefont {{Kidder}}}, \bibinfo {author} {\bibfnamefont {Harald~P.}\
  \bibnamefont {{Pfeiffer}}}, \bibinfo {author} {\bibfnamefont {Mark~A.}\
  \bibnamefont {{Scheel}}}, \bibinfo {author} {\bibfnamefont {Hannes~R.}\
  \bibnamefont {{R{\"u}ter}}}, \bibinfo {author} {\bibfnamefont {Nils}\
  \bibnamefont {{Vu}}}, \bibinfo {author} {\bibfnamefont {Reetika}\
  \bibnamefont {{Dudi}}}, \bibinfo {author} {\bibfnamefont {Sizheng}\
  \bibnamefont {{Ma}}}, \bibinfo {author} {\bibfnamefont {Keefe}\ \bibnamefont
  {{Mitman}}}, \bibinfo {author} {\bibfnamefont {Denyz}\ \bibnamefont
  {{Melchor}}}, \bibinfo {author} {\bibfnamefont {Sierra}\ \bibnamefont
  {{Thomas}}}, \ and\ \bibinfo {author} {\bibfnamefont {Jennifer}\ \bibnamefont
  {{Sanchez}}},\ }\bibfield  {title} {\enquote {\bibinfo {title} {{Laying the
  foundation of the effective-one-body waveform models SEOBNRv5: Improved
  accuracy and efficiency for spinning nonprecessing binary black holes}},}\
  }\href {\doibase 10.1103/PhysRevD.108.124035} {\bibfield  {journal} {\bibinfo
   {journal} {\prd}\ }\textbf {\bibinfo {volume} {108}},\ \bibinfo {eid}
  {124035} (\bibinfo {year} {2023})},\ \Eprint
  {http://arxiv.org/abs/2303.18039} {arXiv:2303.18039 [gr-qc]} \BibitemShut
  {NoStop}%
\bibitem [{\citenamefont {{Nagar}}\ \emph {et~al.}(2020)\citenamefont
  {{Nagar}}, \citenamefont {{Riemenschneider}}, \citenamefont {{Pratten}},
  \citenamefont {{Rettegno}},\ and\ \citenamefont
  {{Messina}}}]{2020PhRvD.102b4077N}%
  \BibitemOpen
  \bibfield  {author} {\bibinfo {author} {\bibfnamefont {Alessandro}\
  \bibnamefont {{Nagar}}}, \bibinfo {author} {\bibfnamefont {Gunnar}\
  \bibnamefont {{Riemenschneider}}}, \bibinfo {author} {\bibfnamefont
  {Geraint}\ \bibnamefont {{Pratten}}}, \bibinfo {author} {\bibfnamefont
  {Piero}\ \bibnamefont {{Rettegno}}}, \ and\ \bibinfo {author} {\bibfnamefont
  {Francesco}\ \bibnamefont {{Messina}}},\ }\bibfield  {title} {\enquote
  {\bibinfo {title} {{Multipolar effective one body waveform model for
  spin-aligned black hole binaries}},}\ }\href {\doibase
  10.1103/PhysRevD.102.024077} {\bibfield  {journal} {\bibinfo  {journal}
  {\prd}\ }\textbf {\bibinfo {volume} {102}},\ \bibinfo {eid} {024077}
  (\bibinfo {year} {2020})},\ \Eprint {http://arxiv.org/abs/2001.09082}
  {arXiv:2001.09082 [gr-qc]} \BibitemShut {NoStop}%
\bibitem [{\citenamefont {Riemenschneider}\ \emph {et~al.}(2021)\citenamefont
  {Riemenschneider}, \citenamefont {Rettegno}, \citenamefont {Breschi},
  \citenamefont {Albertini}, \citenamefont {Gamba}, \citenamefont {Bernuzzi},\
  and\ \citenamefont {Nagar}}]{PhysRevD.104.104045}%
  \BibitemOpen
  \bibfield  {author} {\bibinfo {author} {\bibfnamefont {Gunnar}\ \bibnamefont
  {Riemenschneider}}, \bibinfo {author} {\bibfnamefont {Piero}\ \bibnamefont
  {Rettegno}}, \bibinfo {author} {\bibfnamefont {Matteo}\ \bibnamefont
  {Breschi}}, \bibinfo {author} {\bibfnamefont {Angelica}\ \bibnamefont
  {Albertini}}, \bibinfo {author} {\bibfnamefont {Rossella}\ \bibnamefont
  {Gamba}}, \bibinfo {author} {\bibfnamefont {Sebastiano}\ \bibnamefont
  {Bernuzzi}}, \ and\ \bibinfo {author} {\bibfnamefont {Alessandro}\
  \bibnamefont {Nagar}},\ }\bibfield  {title} {\enquote {\bibinfo {title}
  {Assessment of consistent next-to-quasicircular corrections and postadiabatic
  approximation in effective-one-body multipolar waveforms for binary black
  hole coalescences},}\ }\href {\doibase 10.1103/PhysRevD.104.104045}
  {\bibfield  {journal} {\bibinfo  {journal} {Phys. Rev. D}\ }\textbf {\bibinfo
  {volume} {104}},\ \bibinfo {pages} {104045} (\bibinfo {year}
  {2021})}\BibitemShut {NoStop}%
\bibitem [{\citenamefont {Akcay}\ \emph {et~al.}(2021)\citenamefont {Akcay},
  \citenamefont {Gamba},\ and\ \citenamefont {Bernuzzi}}]{PhysRevD.103.024014}%
  \BibitemOpen
  \bibfield  {author} {\bibinfo {author} {\bibfnamefont {Sarp}\ \bibnamefont
  {Akcay}}, \bibinfo {author} {\bibfnamefont {Rossella}\ \bibnamefont {Gamba}},
  \ and\ \bibinfo {author} {\bibfnamefont {Sebastiano}\ \bibnamefont
  {Bernuzzi}},\ }\bibfield  {title} {\enquote {\bibinfo {title} {Hybrid
  post-newtonian effective-one-body scheme for spin-precessing compact-binary
  waveforms up to merger},}\ }\href {\doibase 10.1103/PhysRevD.103.024014}
  {\bibfield  {journal} {\bibinfo  {journal} {Phys. Rev. D}\ }\textbf {\bibinfo
  {volume} {103}},\ \bibinfo {pages} {024014} (\bibinfo {year}
  {2021})}\BibitemShut {NoStop}%
\bibitem [{\citenamefont {Gamba}\ \emph {et~al.}(2022)\citenamefont {Gamba},
  \citenamefont {Ak\ifmmode~\mbox{\c{c}}\else \c{c}\fi{}ay}, \citenamefont
  {Bernuzzi},\ and\ \citenamefont {Williams}}]{PhysRevD.106.024020}%
  \BibitemOpen
  \bibfield  {author} {\bibinfo {author} {\bibfnamefont {Rossella}\
  \bibnamefont {Gamba}}, \bibinfo {author} {\bibfnamefont {Sarp}\ \bibnamefont
  {Ak\ifmmode~\mbox{\c{c}}\else \c{c}\fi{}ay}}, \bibinfo {author}
  {\bibfnamefont {Sebastiano}\ \bibnamefont {Bernuzzi}}, \ and\ \bibinfo
  {author} {\bibfnamefont {Jake}\ \bibnamefont {Williams}},\ }\bibfield
  {title} {\enquote {\bibinfo {title} {Effective-one-body waveforms for
  precessing coalescing compact binaries with post-newtonian twist},}\ }\href
  {\doibase 10.1103/PhysRevD.106.024020} {\bibfield  {journal} {\bibinfo
  {journal} {Phys. Rev. D}\ }\textbf {\bibinfo {volume} {106}},\ \bibinfo
  {pages} {024020} (\bibinfo {year} {2022})}\BibitemShut {NoStop}%
\bibitem [{\citenamefont {Gamba}\ \emph
  {et~al.}(2024{\natexlab{b}})\citenamefont {Gamba}, \citenamefont
  {Chiaramello},\ and\ \citenamefont {Neogi}}]{PhysRevD.110.024031}%
  \BibitemOpen
  \bibfield  {author} {\bibinfo {author} {\bibfnamefont {Rossella}\
  \bibnamefont {Gamba}}, \bibinfo {author} {\bibfnamefont {Danilo}\
  \bibnamefont {Chiaramello}}, \ and\ \bibinfo {author} {\bibfnamefont {Sayan}\
  \bibnamefont {Neogi}},\ }\bibfield  {title} {\enquote {\bibinfo {title}
  {Toward efficient effective-one-body models for generic, nonplanar orbits},}\
  }\href {\doibase 10.1103/PhysRevD.110.024031} {\bibfield  {journal} {\bibinfo
   {journal} {Phys. Rev. D}\ }\textbf {\bibinfo {volume} {110}},\ \bibinfo
  {pages} {024031} (\bibinfo {year} {2024}{\natexlab{b}})}\BibitemShut
  {NoStop}%
\bibitem [{\citenamefont {{Varma}}\ \emph
  {et~al.}(2019{\natexlab{a}})\citenamefont {{Varma}}, \citenamefont {{Field}},
  \citenamefont {{Scheel}}, \citenamefont {{Blackman}}, \citenamefont
  {{Gerosa}}, \citenamefont {{Stein}}, \citenamefont {{Kidder}},\ and\
  \citenamefont {{Pfeiffer}}}]{2019PhRvR...1c3015V}%
  \BibitemOpen
  \bibfield  {author} {\bibinfo {author} {\bibfnamefont {Vijay}\ \bibnamefont
  {{Varma}}}, \bibinfo {author} {\bibfnamefont {Scott~E.}\ \bibnamefont
  {{Field}}}, \bibinfo {author} {\bibfnamefont {Mark~A.}\ \bibnamefont
  {{Scheel}}}, \bibinfo {author} {\bibfnamefont {Jonathan}\ \bibnamefont
  {{Blackman}}}, \bibinfo {author} {\bibfnamefont {Davide}\ \bibnamefont
  {{Gerosa}}}, \bibinfo {author} {\bibfnamefont {Leo~C.}\ \bibnamefont
  {{Stein}}}, \bibinfo {author} {\bibfnamefont {Lawrence~E.}\ \bibnamefont
  {{Kidder}}}, \ and\ \bibinfo {author} {\bibfnamefont {Harald~P.}\
  \bibnamefont {{Pfeiffer}}},\ }\bibfield  {title} {\enquote {\bibinfo {title}
  {{Surrogate models for precessing binary black hole simulations with unequal
  masses}},}\ }\href {\doibase 10.1103/PhysRevResearch.1.033015} {\bibfield
  {journal} {\bibinfo  {journal} {Physical Review Research}\ }\textbf {\bibinfo
  {volume} {1}},\ \bibinfo {eid} {033015} (\bibinfo {year}
  {2019}{\natexlab{a}})},\ \Eprint {http://arxiv.org/abs/1905.09300}
  {arXiv:1905.09300 [gr-qc]} \BibitemShut {NoStop}%
\bibitem [{\citenamefont {{Varma}}\ \emph
  {et~al.}(2019{\natexlab{b}})\citenamefont {{Varma}}, \citenamefont {{Field}},
  \citenamefont {{Scheel}}, \citenamefont {{Blackman}}, \citenamefont
  {{Kidder}},\ and\ \citenamefont {{Pfeiffer}}}]{2019PhRvD..99f4045V}%
  \BibitemOpen
  \bibfield  {author} {\bibinfo {author} {\bibfnamefont {Vijay}\ \bibnamefont
  {{Varma}}}, \bibinfo {author} {\bibfnamefont {Scott~E.}\ \bibnamefont
  {{Field}}}, \bibinfo {author} {\bibfnamefont {Mark~A.}\ \bibnamefont
  {{Scheel}}}, \bibinfo {author} {\bibfnamefont {Jonathan}\ \bibnamefont
  {{Blackman}}}, \bibinfo {author} {\bibfnamefont {Lawrence~E.}\ \bibnamefont
  {{Kidder}}}, \ and\ \bibinfo {author} {\bibfnamefont {Harald~P.}\
  \bibnamefont {{Pfeiffer}}},\ }\bibfield  {title} {\enquote {\bibinfo {title}
  {{Surrogate model of hybridized numerical relativity binary black hole
  waveforms}},}\ }\href {\doibase 10.1103/PhysRevD.99.064045} {\bibfield
  {journal} {\bibinfo  {journal} {\prd}\ }\textbf {\bibinfo {volume} {99}},\
  \bibinfo {eid} {064045} (\bibinfo {year} {2019}{\natexlab{b}})},\ \Eprint
  {http://arxiv.org/abs/1812.07865} {arXiv:1812.07865 [gr-qc]} \BibitemShut
  {NoStop}%
\bibitem [{\citenamefont {{de Lluc Planas}}\ \emph {et~al.}(2025)\citenamefont
  {{de Lluc Planas}}, \citenamefont {{Ramos-Buades}}, \citenamefont
  {{Garc{\'\i}a-Quir{\'o}s}}, \citenamefont {{Estell{\'e}s}}, \citenamefont
  {{Husa}},\ and\ \citenamefont {{Haney}}}]{2025arXiv250313062D}%
  \BibitemOpen
  \bibfield  {author} {\bibinfo {author} {\bibfnamefont {Maria}\ \bibnamefont
  {{de Lluc Planas}}}, \bibinfo {author} {\bibfnamefont {Antoni}\ \bibnamefont
  {{Ramos-Buades}}}, \bibinfo {author} {\bibfnamefont {Cecilio}\ \bibnamefont
  {{Garc{\'\i}a-Quir{\'o}s}}}, \bibinfo {author} {\bibfnamefont {H{\'e}ctor}\
  \bibnamefont {{Estell{\'e}s}}}, \bibinfo {author} {\bibfnamefont {Sascha}\
  \bibnamefont {{Husa}}}, \ and\ \bibinfo {author} {\bibfnamefont {Maria}\
  \bibnamefont {{Haney}}},\ }\bibfield  {title} {\enquote {\bibinfo {title}
  {{Time-domain phenomenological multipolar waveforms for aligned-spin binary
  black holes in elliptical orbits}},}\ }\href {\doibase
  10.48550/arXiv.2503.13062} {\bibfield  {journal} {\bibinfo  {journal} {arXiv
  e-prints}\ ,\ \bibinfo {eid} {arXiv:2503.13062}} (\bibinfo {year} {2025})},\
  \Eprint {http://arxiv.org/abs/2503.13062} {arXiv:2503.13062 [gr-qc]}
  \BibitemShut {NoStop}%
\bibitem [{\citenamefont {Gamboa}\ \emph {et~al.}(2024)\citenamefont {Gamboa},
  \citenamefont {Buonanno}, \citenamefont {Enficiaud}, \citenamefont {Khalil},
  \citenamefont {Ramos-Buades}, \citenamefont {Pompili}, \citenamefont
  {Estellés}, \citenamefont {Boyle}, \citenamefont {Kidder}, \citenamefont
  {Pfeiffer}, \citenamefont {Rüter},\ and\ \citenamefont
  {Scheel}}]{Gamboa2024}%
  \BibitemOpen
  \bibfield  {author} {\bibinfo {author} {\bibfnamefont {Aldo}\ \bibnamefont
  {Gamboa}}, \bibinfo {author} {\bibfnamefont {Alessandra}\ \bibnamefont
  {Buonanno}}, \bibinfo {author} {\bibfnamefont {Raffi}\ \bibnamefont
  {Enficiaud}}, \bibinfo {author} {\bibfnamefont {Mohammed}\ \bibnamefont
  {Khalil}}, \bibinfo {author} {\bibfnamefont {Antoni}\ \bibnamefont
  {Ramos-Buades}}, \bibinfo {author} {\bibfnamefont {Lorenzo}\ \bibnamefont
  {Pompili}}, \bibinfo {author} {\bibfnamefont {Héctor}\ \bibnamefont
  {Estellés}}, \bibinfo {author} {\bibfnamefont {Michael}\ \bibnamefont
  {Boyle}}, \bibinfo {author} {\bibfnamefont {Lawrence~E.}\ \bibnamefont
  {Kidder}}, \bibinfo {author} {\bibfnamefont {Harald~P.}\ \bibnamefont
  {Pfeiffer}}, \bibinfo {author} {\bibfnamefont {Hannes~R.}\ \bibnamefont
  {Rüter}}, \ and\ \bibinfo {author} {\bibfnamefont {Mark~A.}\ \bibnamefont
  {Scheel}},\ }\bibfield  {title} {\enquote {\bibinfo {title} {Accurate
  waveforms for eccentric, aligned-spin binary black holes: The multipolar
  effective-one-body model seobnrv5ehm},}\ }\href@noop {} {\  (\bibinfo {year}
  {2024})}\BibitemShut {NoStop}%
\bibitem [{\citenamefont {{Ramos-Buades}}\ \emph {et~al.}(2022)\citenamefont
  {{Ramos-Buades}}, \citenamefont {{Buonanno}}, \citenamefont {{Khalil}},\ and\
  \citenamefont {{Ossokine}}}]{2022PhRvD.105d4035R}%
  \BibitemOpen
  \bibfield  {author} {\bibinfo {author} {\bibfnamefont {Antoni}\ \bibnamefont
  {{Ramos-Buades}}}, \bibinfo {author} {\bibfnamefont {Alessandra}\
  \bibnamefont {{Buonanno}}}, \bibinfo {author} {\bibfnamefont {Mohammed}\
  \bibnamefont {{Khalil}}}, \ and\ \bibinfo {author} {\bibfnamefont {Serguei}\
  \bibnamefont {{Ossokine}}},\ }\bibfield  {title} {\enquote {\bibinfo {title}
  {{Effective-one-body multipolar waveforms for eccentric binary black holes
  with nonprecessing spins}},}\ }\href {\doibase 10.1103/PhysRevD.105.044035}
  {\bibfield  {journal} {\bibinfo  {journal} {\prd}\ }\textbf {\bibinfo
  {volume} {105}},\ \bibinfo {eid} {044035} (\bibinfo {year} {2022})},\ \Eprint
  {http://arxiv.org/abs/2112.06952} {arXiv:2112.06952 [gr-qc]} \BibitemShut
  {NoStop}%
\bibitem [{\citenamefont {{Chiaramello}}\ and\ \citenamefont
  {{Nagar}}(2020)}]{2020PhRvD.101j1501C}%
  \BibitemOpen
  \bibfield  {author} {\bibinfo {author} {\bibfnamefont {Danilo}\ \bibnamefont
  {{Chiaramello}}}\ and\ \bibinfo {author} {\bibfnamefont {Alessandro}\
  \bibnamefont {{Nagar}}},\ }\bibfield  {title} {\enquote {\bibinfo {title}
  {{Faithful analytical effective-one-body waveform model for spin-aligned,
  moderately eccentric, coalescing black hole binaries}},}\ }\href {\doibase
  10.1103/PhysRevD.101.101501} {\bibfield  {journal} {\bibinfo  {journal}
  {\prd}\ }\textbf {\bibinfo {volume} {101}},\ \bibinfo {eid} {101501}
  (\bibinfo {year} {2020})},\ \Eprint {http://arxiv.org/abs/2001.11736}
  {arXiv:2001.11736 [gr-qc]} \BibitemShut {NoStop}%
\bibitem [{\citenamefont {{Islam}}\ \emph {et~al.}(2021)\citenamefont
  {{Islam}}, \citenamefont {{Varma}}, \citenamefont {{Lodman}}, \citenamefont
  {{Field}}, \citenamefont {{Khanna}}, \citenamefont {{Scheel}}, \citenamefont
  {{Pfeiffer}}, \citenamefont {{Gerosa}},\ and\ \citenamefont
  {{Kidder}}}]{2021PhRvD.103f4022I}%
  \BibitemOpen
  \bibfield  {author} {\bibinfo {author} {\bibfnamefont {Tousif}\ \bibnamefont
  {{Islam}}}, \bibinfo {author} {\bibfnamefont {Vijay}\ \bibnamefont
  {{Varma}}}, \bibinfo {author} {\bibfnamefont {Jackie}\ \bibnamefont
  {{Lodman}}}, \bibinfo {author} {\bibfnamefont {Scott~E.}\ \bibnamefont
  {{Field}}}, \bibinfo {author} {\bibfnamefont {Gaurav}\ \bibnamefont
  {{Khanna}}}, \bibinfo {author} {\bibfnamefont {Mark~A.}\ \bibnamefont
  {{Scheel}}}, \bibinfo {author} {\bibfnamefont {Harald~P.}\ \bibnamefont
  {{Pfeiffer}}}, \bibinfo {author} {\bibfnamefont {Davide}\ \bibnamefont
  {{Gerosa}}}, \ and\ \bibinfo {author} {\bibfnamefont {Lawrence~E.}\
  \bibnamefont {{Kidder}}},\ }\bibfield  {title} {\enquote {\bibinfo {title}
  {{Eccentric binary black hole surrogate models for the gravitational waveform
  and remnant properties: Comparable mass, nonspinning case}},}\ }\href
  {\doibase 10.1103/PhysRevD.103.064022} {\bibfield  {journal} {\bibinfo
  {journal} {\prd}\ }\textbf {\bibinfo {volume} {103}},\ \bibinfo {eid}
  {064022} (\bibinfo {year} {2021})},\ \Eprint
  {http://arxiv.org/abs/2101.11798} {arXiv:2101.11798 [gr-qc]} \BibitemShut
  {NoStop}%
\bibitem [{\citenamefont {Ramos-Buades}\ \emph {et~al.}(2023)\citenamefont
  {Ramos-Buades}, \citenamefont {Buonanno}, \citenamefont {Estell{\'e}s},
  \citenamefont {Khalil}, \citenamefont {Mihaylov}, \citenamefont {Ossokine},
  \citenamefont {Pompili},\ and\ \citenamefont
  {Shiferaw}}]{Ramos-Buades:2023ehm}%
  \BibitemOpen
  \bibfield  {author} {\bibinfo {author} {\bibfnamefont {Antoni}\ \bibnamefont
  {Ramos-Buades}}, \bibinfo {author} {\bibfnamefont {Alessandra}\ \bibnamefont
  {Buonanno}}, \bibinfo {author} {\bibfnamefont {H{\'e}ctor}\ \bibnamefont
  {Estell{\'e}s}}, \bibinfo {author} {\bibfnamefont {Mohammed}\ \bibnamefont
  {Khalil}}, \bibinfo {author} {\bibfnamefont {Deyan~P.}\ \bibnamefont
  {Mihaylov}}, \bibinfo {author} {\bibfnamefont {Serguei}\ \bibnamefont
  {Ossokine}}, \bibinfo {author} {\bibfnamefont {Lorenzo}\ \bibnamefont
  {Pompili}}, \ and\ \bibinfo {author} {\bibfnamefont {Mahlet}\ \bibnamefont
  {Shiferaw}},\ }\bibfield  {title} {\enquote {\bibinfo {title} {{Next
  generation of accurate and efficient multipolar precessing-spin
  effective-one-body waveforms for binary black holes}},}\ }\href {\doibase
  10.1103/PhysRevD.108.124037} {\bibfield  {journal} {\bibinfo  {journal}
  {Phys. Rev. D}\ }\textbf {\bibinfo {volume} {108}},\ \bibinfo {pages}
  {124037} (\bibinfo {year} {2023})},\ \Eprint
  {http://arxiv.org/abs/2303.18046} {arXiv:2303.18046 [gr-qc]} \BibitemShut
  {NoStop}%
\bibitem [{\citenamefont {{Owen}}\ and\ \citenamefont
  {{Sathyaprakash}}(1999)}]{1999PhRvD..60b2002O}%
  \BibitemOpen
  \bibfield  {author} {\bibinfo {author} {\bibfnamefont {Benjamin~J.}\
  \bibnamefont {{Owen}}}\ and\ \bibinfo {author} {\bibfnamefont {B.~S.}\
  \bibnamefont {{Sathyaprakash}}},\ }\bibfield  {title} {\enquote {\bibinfo
  {title} {{Matched filtering of gravitational waves from inspiraling compact
  binaries: Computational cost and template placement}},}\ }\href {\doibase
  10.1103/PhysRevD.60.022002} {\bibfield  {journal} {\bibinfo  {journal}
  {\prd}\ }\textbf {\bibinfo {volume} {60}},\ \bibinfo {eid} {022002} (\bibinfo
  {year} {1999})},\ \Eprint {http://arxiv.org/abs/gr-qc/9808076}
  {arXiv:gr-qc/9808076 [gr-qc]} \BibitemShut {NoStop}%
\bibitem [{\citenamefont {Boyle}\ \emph {et~al.}(2007)\citenamefont {Boyle},
  \citenamefont {Brown}, \citenamefont {Kidder}, \citenamefont {Mroue},
  \citenamefont {Pfeiffer}, \citenamefont {Scheel}, \citenamefont {Cook},\ and\
  \citenamefont {Teukolsky}}]{Boyle:2007ft}%
  \BibitemOpen
  \bibfield  {author} {\bibinfo {author} {\bibfnamefont {Michael}\ \bibnamefont
  {Boyle}}, \bibinfo {author} {\bibfnamefont {Duncan~A.}\ \bibnamefont
  {Brown}}, \bibinfo {author} {\bibfnamefont {Lawrence~E.}\ \bibnamefont
  {Kidder}}, \bibinfo {author} {\bibfnamefont {Abdul~H.}\ \bibnamefont
  {Mroue}}, \bibinfo {author} {\bibfnamefont {Harald~P.}\ \bibnamefont
  {Pfeiffer}}, \bibinfo {author} {\bibfnamefont {Mark~A.}\ \bibnamefont
  {Scheel}}, \bibinfo {author} {\bibfnamefont {Gregory~B.}\ \bibnamefont
  {Cook}}, \ and\ \bibinfo {author} {\bibfnamefont {Saul~A.}\ \bibnamefont
  {Teukolsky}},\ }\bibfield  {title} {\enquote {\bibinfo {title}
  {{High-accuracy comparison of numerical relativity simulations with
  post-Newtonian expansions}},}\ }\href {\doibase 10.1103/PhysRevD.76.124038}
  {\bibfield  {journal} {\bibinfo  {journal} {Phys. Rev. D}\ }\textbf {\bibinfo
  {volume} {76}},\ \bibinfo {pages} {124038} (\bibinfo {year} {2007})},\
  \Eprint {http://arxiv.org/abs/0710.0158} {arXiv:0710.0158 [gr-qc]}
  \BibitemShut {NoStop}%
\bibitem [{\citenamefont {Hannam}\ \emph {et~al.}(2008)\citenamefont {Hannam},
  \citenamefont {Husa}, \citenamefont {Sperhake}, \citenamefont {Bruegmann},\
  and\ \citenamefont {Gonzalez}}]{Hannam:2007ik}%
  \BibitemOpen
  \bibfield  {author} {\bibinfo {author} {\bibfnamefont {Mark}\ \bibnamefont
  {Hannam}}, \bibinfo {author} {\bibfnamefont {Sascha}\ \bibnamefont {Husa}},
  \bibinfo {author} {\bibfnamefont {Ulrich}\ \bibnamefont {Sperhake}}, \bibinfo
  {author} {\bibfnamefont {Bernd}\ \bibnamefont {Bruegmann}}, \ and\ \bibinfo
  {author} {\bibfnamefont {Jose~A.}\ \bibnamefont {Gonzalez}},\ }\bibfield
  {title} {\enquote {\bibinfo {title} {{Where post-Newtonian and
  numerical-relativity waveforms meet}},}\ }\href {\doibase
  10.1103/PhysRevD.77.044020} {\bibfield  {journal} {\bibinfo  {journal} {Phys.
  Rev. D}\ }\textbf {\bibinfo {volume} {77}},\ \bibinfo {pages} {044020}
  (\bibinfo {year} {2008})},\ \Eprint {http://arxiv.org/abs/0706.1305}
  {arXiv:0706.1305 [gr-qc]} \BibitemShut {NoStop}%
\bibitem [{\citenamefont {Buonanno}\ \emph {et~al.}(2009)\citenamefont
  {Buonanno}, \citenamefont {Iyer}, \citenamefont {Ochsner}, \citenamefont
  {Pan},\ and\ \citenamefont {Sathyaprakash}}]{Buonanno:2009zt}%
  \BibitemOpen
  \bibfield  {author} {\bibinfo {author} {\bibfnamefont {Alessandra}\
  \bibnamefont {Buonanno}}, \bibinfo {author} {\bibfnamefont {Bala}\
  \bibnamefont {Iyer}}, \bibinfo {author} {\bibfnamefont {Evan}\ \bibnamefont
  {Ochsner}}, \bibinfo {author} {\bibfnamefont {Yi}~\bibnamefont {Pan}}, \ and\
  \bibinfo {author} {\bibfnamefont {B.~S.}\ \bibnamefont {Sathyaprakash}},\
  }\bibfield  {title} {\enquote {\bibinfo {title} {{Comparison of
  post-Newtonian templates for compact binary inspiral signals in
  gravitational-wave detectors}},}\ }\href {\doibase
  10.1103/PhysRevD.80.084043} {\bibfield  {journal} {\bibinfo  {journal} {Phys.
  Rev. D}\ }\textbf {\bibinfo {volume} {80}},\ \bibinfo {pages} {084043}
  (\bibinfo {year} {2009})},\ \Eprint {http://arxiv.org/abs/0907.0700}
  {arXiv:0907.0700 [gr-qc]} \BibitemShut {NoStop}%
\bibitem [{\citenamefont {Ajith}\ \emph {et~al.}(2007)\citenamefont {Ajith}
  \emph {et~al.}}]{Ajith:2007qp}%
  \BibitemOpen
  \bibfield  {author} {\bibinfo {author} {\bibfnamefont {Parameswaran}\
  \bibnamefont {Ajith}} \emph {et~al.},\ }\bibfield  {title} {\enquote
  {\bibinfo {title} {{Phenomenological template family for black-hole
  coalescence waveforms}},}\ }\href {\doibase 10.1088/0264-9381/24/19/S31}
  {\bibfield  {journal} {\bibinfo  {journal} {Class. Quant. Grav.}\ }\textbf
  {\bibinfo {volume} {24}},\ \bibinfo {pages} {S689--S700} (\bibinfo {year}
  {2007})},\ \Eprint {http://arxiv.org/abs/0704.3764} {arXiv:0704.3764 [gr-qc]}
  \BibitemShut {NoStop}%
\bibitem [{\citenamefont {Schutz}(2011)}]{Schutz:2011tw}%
  \BibitemOpen
  \bibfield  {author} {\bibinfo {author} {\bibfnamefont {Bernard~F.}\
  \bibnamefont {Schutz}},\ }\bibfield  {title} {\enquote {\bibinfo {title}
  {{Networks of gravitational wave detectors and three figures of merit}},}\
  }\href {\doibase 10.1088/0264-9381/28/12/125023} {\bibfield  {journal}
  {\bibinfo  {journal} {Class. Quant. Grav.}\ }\textbf {\bibinfo {volume}
  {28}},\ \bibinfo {pages} {125023} (\bibinfo {year} {2011})},\ \Eprint
  {http://arxiv.org/abs/1102.5421} {arXiv:1102.5421 [astro-ph.IM]} \BibitemShut
  {NoStop}%
\bibitem [{\citenamefont {Cutler}\ and\ \citenamefont
  {Flanagan}(1994)}]{Cutler:1994ys}%
  \BibitemOpen
  \bibfield  {author} {\bibinfo {author} {\bibfnamefont {Curt}\ \bibnamefont
  {Cutler}}\ and\ \bibinfo {author} {\bibfnamefont {Eanna~E.}\ \bibnamefont
  {Flanagan}},\ }\bibfield  {title} {\enquote {\bibinfo {title} {{Gravitational
  waves from merging compact binaries: How accurately can one extract the
  binary's parameters from the inspiral wave form?}}}\ }\href {\doibase
  10.1103/PhysRevD.49.2658} {\bibfield  {journal} {\bibinfo  {journal} {Phys.
  Rev. D}\ }\textbf {\bibinfo {volume} {49}},\ \bibinfo {pages} {2658--2697}
  (\bibinfo {year} {1994})},\ \Eprint {http://arxiv.org/abs/gr-qc/9402014}
  {arXiv:gr-qc/9402014} \BibitemShut {NoStop}%
\bibitem [{\citenamefont {Pankow}\ \emph {et~al.}(2015)\citenamefont {Pankow},
  \citenamefont {Brady}, \citenamefont {Ochsner},\ and\ \citenamefont
  {O'Shaughnessy}}]{Pankow:2015cra}%
  \BibitemOpen
  \bibfield  {author} {\bibinfo {author} {\bibfnamefont {C.}~\bibnamefont
  {Pankow}}, \bibinfo {author} {\bibfnamefont {P.}~\bibnamefont {Brady}},
  \bibinfo {author} {\bibfnamefont {E.}~\bibnamefont {Ochsner}}, \ and\
  \bibinfo {author} {\bibfnamefont {R.}~\bibnamefont {O'Shaughnessy}},\
  }\bibfield  {title} {\enquote {\bibinfo {title} {{Novel scheme for rapid
  parallel parameter estimation of gravitational waves from compact binary
  coalescences}},}\ }\href {\doibase 10.1103/PhysRevD.92.023002} {\bibfield
  {journal} {\bibinfo  {journal} {Phys. Rev. D}\ }\textbf {\bibinfo {volume}
  {92}},\ \bibinfo {pages} {023002} (\bibinfo {year} {2015})},\ \Eprint
  {http://arxiv.org/abs/1502.04370} {arXiv:1502.04370 [gr-qc]} \BibitemShut
  {NoStop}%
\bibitem [{\citenamefont {Lange}\ \emph {et~al.}(2018)\citenamefont {Lange},
  \citenamefont {O'Shaughnessy},\ and\ \citenamefont {Rizzo}}]{Lange:2018pyp}%
  \BibitemOpen
  \bibfield  {author} {\bibinfo {author} {\bibfnamefont {Jacob}\ \bibnamefont
  {Lange}}, \bibinfo {author} {\bibfnamefont {Richard}\ \bibnamefont
  {O'Shaughnessy}}, \ and\ \bibinfo {author} {\bibfnamefont {Monica}\
  \bibnamefont {Rizzo}},\ }\bibfield  {title} {\enquote {\bibinfo {title}
  {{Rapid and accurate parameter inference for coalescing, precessing compact
  binaries}},}\ }\href@noop {} {\  (\bibinfo {year} {2018})},\ \Eprint
  {http://arxiv.org/abs/1805.10457} {arXiv:1805.10457 [gr-qc]} \BibitemShut
  {NoStop}%
\bibitem [{\citenamefont {Wofford}\ \emph {et~al.}(2022)\citenamefont {Wofford}
  \emph {et~al.}}]{Wofford:2022ykb}%
  \BibitemOpen
  \bibfield  {author} {\bibinfo {author} {\bibfnamefont {J.}~\bibnamefont
  {Wofford}} \emph {et~al.},\ }\bibfield  {title} {\enquote {\bibinfo {title}
  {{Expanding RIFT: Improving performance for GW parameter inference}},}\
  }\href@noop {} {\  (\bibinfo {year} {2022})},\ \Eprint
  {http://arxiv.org/abs/2210.07912} {arXiv:2210.07912 [gr-qc]} \BibitemShut
  {NoStop}%
\bibitem [{\citenamefont {Wagner}\ \emph {et~al.}(2025)\citenamefont {Wagner},
  \citenamefont {O'Shaughnessy}, \citenamefont {Yelikar}, \citenamefont
  {Manning}, \citenamefont {Fernando}, \citenamefont {Lange}, \citenamefont
  {Tiwari}, \citenamefont {Fernando},\ and\ \citenamefont
  {Williams}}]{Wagner:2025bih}%
  \BibitemOpen
  \bibfield  {author} {\bibinfo {author} {\bibfnamefont {Katelyn~J.}\
  \bibnamefont {Wagner}}, \bibinfo {author} {\bibfnamefont {R.}~\bibnamefont
  {O'Shaughnessy}}, \bibinfo {author} {\bibfnamefont {A.}~\bibnamefont
  {Yelikar}}, \bibinfo {author} {\bibfnamefont {N.}~\bibnamefont {Manning}},
  \bibinfo {author} {\bibfnamefont {D.}~\bibnamefont {Fernando}}, \bibinfo
  {author} {\bibfnamefont {J.}~\bibnamefont {Lange}}, \bibinfo {author}
  {\bibfnamefont {V.}~\bibnamefont {Tiwari}}, \bibinfo {author} {\bibfnamefont
  {A.}~\bibnamefont {Fernando}}, \ and\ \bibinfo {author} {\bibfnamefont
  {D.}~\bibnamefont {Williams}},\ }\bibfield  {title} {\enquote {\bibinfo
  {title} {{Narrowing RIFT: Focused simulation-based-inference for interpreting
  exceptional GW sources}},}\ }\href@noop {} {\  (\bibinfo {year} {2025})},\
  \Eprint {http://arxiv.org/abs/2505.11655} {arXiv:2505.11655 [astro-ph.IM]}
  \BibitemShut {NoStop}%
\bibitem [{\citenamefont {Lange}\ \emph {et~al.}(2017)\citenamefont {Lange},
  \citenamefont {O'Shaughnessy}, \citenamefont {Boyle}, \citenamefont
  {Calder\'on~Bustillo}, \citenamefont {Campanelli}, \citenamefont {Chu},
  \citenamefont {Clark}, \citenamefont {Demos}, \citenamefont {Fong},
  \citenamefont {Healy}, \citenamefont {Hemberger}, \citenamefont {Hinder},
  \citenamefont {Jani}, \citenamefont {Khamesra}, \citenamefont {Kidder},
  \citenamefont {Kumar}, \citenamefont {Laguna}, \citenamefont {Lousto},
  \citenamefont {Lovelace}, \citenamefont {Ossokine}, \citenamefont {Pfeiffer},
  \citenamefont {Scheel}, \citenamefont {Shoemaker}, \citenamefont {Szilagyi},
  \citenamefont {Teukolsky},\ and\ \citenamefont
  {Zlochower}}]{PhysRevD.96.104041}%
  \BibitemOpen
  \bibfield  {author} {\bibinfo {author} {\bibfnamefont {J.}~\bibnamefont
  {Lange}}, \bibinfo {author} {\bibfnamefont {R.}~\bibnamefont
  {O'Shaughnessy}}, \bibinfo {author} {\bibfnamefont {M.}~\bibnamefont
  {Boyle}}, \bibinfo {author} {\bibfnamefont {J.}~\bibnamefont
  {Calder\'on~Bustillo}}, \bibinfo {author} {\bibfnamefont {M.}~\bibnamefont
  {Campanelli}}, \bibinfo {author} {\bibfnamefont {T.}~\bibnamefont {Chu}},
  \bibinfo {author} {\bibfnamefont {J.~A.}\ \bibnamefont {Clark}}, \bibinfo
  {author} {\bibfnamefont {N.}~\bibnamefont {Demos}}, \bibinfo {author}
  {\bibfnamefont {H.}~\bibnamefont {Fong}}, \bibinfo {author} {\bibfnamefont
  {J.}~\bibnamefont {Healy}}, \bibinfo {author} {\bibfnamefont {D.~A.}\
  \bibnamefont {Hemberger}}, \bibinfo {author} {\bibfnamefont {I.}~\bibnamefont
  {Hinder}}, \bibinfo {author} {\bibfnamefont {K.}~\bibnamefont {Jani}},
  \bibinfo {author} {\bibfnamefont {B.}~\bibnamefont {Khamesra}}, \bibinfo
  {author} {\bibfnamefont {L.~E.}\ \bibnamefont {Kidder}}, \bibinfo {author}
  {\bibfnamefont {P.}~\bibnamefont {Kumar}}, \bibinfo {author} {\bibfnamefont
  {P.}~\bibnamefont {Laguna}}, \bibinfo {author} {\bibfnamefont {C.~O.}\
  \bibnamefont {Lousto}}, \bibinfo {author} {\bibfnamefont {G.}~\bibnamefont
  {Lovelace}}, \bibinfo {author} {\bibfnamefont {S.}~\bibnamefont {Ossokine}},
  \bibinfo {author} {\bibfnamefont {H.}~\bibnamefont {Pfeiffer}}, \bibinfo
  {author} {\bibfnamefont {M.~A.}\ \bibnamefont {Scheel}}, \bibinfo {author}
  {\bibfnamefont {D.~M.}\ \bibnamefont {Shoemaker}}, \bibinfo {author}
  {\bibfnamefont {B.}~\bibnamefont {Szilagyi}}, \bibinfo {author}
  {\bibfnamefont {S.}~\bibnamefont {Teukolsky}}, \ and\ \bibinfo {author}
  {\bibfnamefont {Y.}~\bibnamefont {Zlochower}},\ }\bibfield  {title} {\enquote
  {\bibinfo {title} {Parameter estimation method that directly compares
  gravitational wave observations to numerical relativity},}\ }\href {\doibase
  10.1103/PhysRevD.96.104041} {\bibfield  {journal} {\bibinfo  {journal} {Phys.
  Rev. D}\ }\textbf {\bibinfo {volume} {96}},\ \bibinfo {pages} {104041}
  (\bibinfo {year} {2017})}\BibitemShut {NoStop}%
\bibitem [{LIG(2022)}]{LIGO-T2200043}%
  \BibitemOpen
  \href {https://dcc.ligo.org/LIGO-T2200043/public} {\emph {\bibinfo {title}
  {{Noise Curves for use in simulations pre-O4}}}},\ \bibinfo {type} {Tech.
  Rep.}\ \bibinfo {number} {LIGO-T2200043-v3}\ (\bibinfo  {institution}
  {LIGO},\ \bibinfo {year} {2022})\BibitemShut {NoStop}%
\bibitem [{\citenamefont {Dickey}(1971)}]{Dickey:1971}%
  \BibitemOpen
  \bibfield  {author} {\bibinfo {author} {\bibfnamefont {James~M.}\
  \bibnamefont {Dickey}},\ }\bibfield  {title} {\enquote {\bibinfo {title}
  {{The weighted likelihood ratio, linear hypotheses on normal location
  parameters}},}\ }\href@noop {} {\bibfield  {journal} {\bibinfo  {journal}
  {Ann. Math. Statist.}\ }\textbf {\bibinfo {volume} {42}},\ \bibinfo {pages}
  {204--223} (\bibinfo {year} {1971})}\BibitemShut {NoStop}%
\bibitem [{\citenamefont {Verdinelli}\ and\ \citenamefont
  {Wasserman}(1995)}]{Verdinelli:1995}%
  \BibitemOpen
  \bibfield  {author} {\bibinfo {author} {\bibfnamefont {Isabella}\
  \bibnamefont {Verdinelli}}\ and\ \bibinfo {author} {\bibfnamefont {Larry}\
  \bibnamefont {Wasserman}},\ }\bibfield  {title} {\enquote {\bibinfo {title}
  {{Bayes factors, nuisance parameters, and imprecise tests}},}\ }\href@noop {}
  {\bibfield  {journal} {\bibinfo  {journal} {J. Am. Stat. Assoc.}\ }\textbf
  {\bibinfo {volume} {90}},\ \bibinfo {pages} {614--618} (\bibinfo {year}
  {1995})}\BibitemShut {NoStop}%
\bibitem [{\citenamefont {Ajith}\ \emph {et~al.}(2011)\citenamefont {Ajith}
  \emph {et~al.}}]{Ajith:2009bn}%
  \BibitemOpen
  \bibfield  {author} {\bibinfo {author} {\bibfnamefont {P.}~\bibnamefont
  {Ajith}} \emph {et~al.},\ }\bibfield  {title} {\enquote {\bibinfo {title}
  {{Inspiral-merger-ringdown waveforms for black-hole binaries with
  non-precessing spins}},}\ }\href {\doibase 10.1103/PhysRevLett.106.241101}
  {\bibfield  {journal} {\bibinfo  {journal} {Phys. Rev. Lett.}\ }\textbf
  {\bibinfo {volume} {106}},\ \bibinfo {pages} {241101} (\bibinfo {year}
  {2011})},\ \Eprint {http://arxiv.org/abs/0909.2867} {arXiv:0909.2867 [gr-qc]}
  \BibitemShut {NoStop}%
\bibitem [{\citenamefont {Santamaria}\ \emph {et~al.}(2010)\citenamefont
  {Santamaria} \emph {et~al.}}]{Santamaria:2010yb}%
  \BibitemOpen
  \bibfield  {author} {\bibinfo {author} {\bibfnamefont {L.}~\bibnamefont
  {Santamaria}} \emph {et~al.},\ }\bibfield  {title} {\enquote {\bibinfo
  {title} {{Matching post-Newtonian and numerical relativity waveforms:
  systematic errors and a new phenomenological model for non-precessing black
  hole binaries}},}\ }\href {\doibase 10.1103/PhysRevD.82.064016} {\bibfield
  {journal} {\bibinfo  {journal} {Phys. Rev. D}\ }\textbf {\bibinfo {volume}
  {82}},\ \bibinfo {pages} {064016} (\bibinfo {year} {2010})},\ \Eprint
  {http://arxiv.org/abs/1005.3306} {arXiv:1005.3306 [gr-qc]} \BibitemShut
  {NoStop}%
\bibitem [{\citenamefont {Gerosa}\ \emph {et~al.}(2015)\citenamefont {Gerosa},
  \citenamefont {Kesden}, \citenamefont {Sperhake}, \citenamefont {Berti},\
  and\ \citenamefont {O'Shaughnessy}}]{Gerosa:2015tea}%
  \BibitemOpen
  \bibfield  {author} {\bibinfo {author} {\bibfnamefont {Davide}\ \bibnamefont
  {Gerosa}}, \bibinfo {author} {\bibfnamefont {Michael}\ \bibnamefont
  {Kesden}}, \bibinfo {author} {\bibfnamefont {Ulrich}\ \bibnamefont
  {Sperhake}}, \bibinfo {author} {\bibfnamefont {Emanuele}\ \bibnamefont
  {Berti}}, \ and\ \bibinfo {author} {\bibfnamefont {Richard}\ \bibnamefont
  {O'Shaughnessy}},\ }\bibfield  {title} {\enquote {\bibinfo {title}
  {{Multi-timescale analysis of phase transitions in precessing black-hole
  binaries}},}\ }\href {\doibase 10.1103/PhysRevD.92.064016} {\bibfield
  {journal} {\bibinfo  {journal} {Phys. Rev. D}\ }\textbf {\bibinfo {volume}
  {92}},\ \bibinfo {pages} {064016} (\bibinfo {year} {2015})},\ \Eprint
  {http://arxiv.org/abs/1506.03492} {arXiv:1506.03492 [gr-qc]} \BibitemShut
  {NoStop}%
\bibitem [{\citenamefont {Schmidt}\ \emph {et~al.}(2015)\citenamefont
  {Schmidt}, \citenamefont {Ohme},\ and\ \citenamefont
  {Hannam}}]{Schmidt:2014iyl}%
  \BibitemOpen
  \bibfield  {author} {\bibinfo {author} {\bibfnamefont {Patricia}\
  \bibnamefont {Schmidt}}, \bibinfo {author} {\bibfnamefont {Frank}\
  \bibnamefont {Ohme}}, \ and\ \bibinfo {author} {\bibfnamefont {Mark}\
  \bibnamefont {Hannam}},\ }\bibfield  {title} {\enquote {\bibinfo {title}
  {{Towards models of gravitational waveforms from generic binaries II:
  Modelling precession effects with a single effective precession
  parameter}},}\ }\href {\doibase 10.1103/PhysRevD.91.024043} {\bibfield
  {journal} {\bibinfo  {journal} {Phys. Rev. D}\ }\textbf {\bibinfo {volume}
  {91}},\ \bibinfo {pages} {024043} (\bibinfo {year} {2015})},\ \Eprint
  {http://arxiv.org/abs/1408.1810} {arXiv:1408.1810 [gr-qc]} \BibitemShut
  {NoStop}%
\bibitem [{\citenamefont {Veitch}\ \emph {et~al.}(2015)\citenamefont {Veitch}
  \emph {et~al.}}]{Veitch:2014wba}%
  \BibitemOpen
  \bibfield  {author} {\bibinfo {author} {\bibfnamefont {J.}~\bibnamefont
  {Veitch}} \emph {et~al.},\ }\bibfield  {title} {\enquote {\bibinfo {title}
  {{Parameter estimation for compact binaries with ground-based
  gravitational-wave observations using the LALInference software library}},}\
  }\href {\doibase 10.1103/PhysRevD.91.042003} {\bibfield  {journal} {\bibinfo
  {journal} {Phys. Rev. D}\ }\textbf {\bibinfo {volume} {91}},\ \bibinfo
  {pages} {042003} (\bibinfo {year} {2015})},\ \Eprint
  {http://arxiv.org/abs/1409.7215} {arXiv:1409.7215 [gr-qc]} \BibitemShut
  {NoStop}%
\bibitem [{\citenamefont {Thrane}\ and\ \citenamefont
  {Talbot}(2019)}]{Thrane:2018qnx}%
  \BibitemOpen
  \bibfield  {author} {\bibinfo {author} {\bibfnamefont {Eric}\ \bibnamefont
  {Thrane}}\ and\ \bibinfo {author} {\bibfnamefont {Colm}\ \bibnamefont
  {Talbot}},\ }\bibfield  {title} {\enquote {\bibinfo {title} {{An introduction
  to Bayesian inference in gravitational-wave astronomy: parameter estimation,
  model selection, and hierarchical models}},}\ }\href {\doibase
  10.1017/pasa.2019.2} {\bibfield  {journal} {\bibinfo  {journal} {Publ.
  Astron. Soc. Austral.}\ }\textbf {\bibinfo {volume} {36}},\ \bibinfo {pages}
  {e010} (\bibinfo {year} {2019})},\ \bibinfo {note} {[Erratum:
  Publ.Astron.Soc.Austral. 37, e036 (2020)]},\ \Eprint
  {http://arxiv.org/abs/1809.02293} {arXiv:1809.02293 [astro-ph.IM]}
  \BibitemShut {NoStop}%
\bibitem [{\citenamefont {Abac}\ \emph
  {et~al.}(2025{\natexlab{e}})\citenamefont {Abac} \emph
  {et~al.}}]{LIGOScientific:2025yae}%
  \BibitemOpen
  \bibfield  {author} {\bibinfo {author} {\bibfnamefont {A.~G.}\ \bibnamefont
  {Abac}} \emph {et~al.} (\bibinfo {collaboration} {LIGO Scientific, VIRGO,
  KAGRA}),\ }\bibfield  {title} {\enquote {\bibinfo {title} {{GWTC-4.0: Methods
  for Identifying and Characterizing Gravitational-wave Transients}},}\
  }\href@noop {} {\  (\bibinfo {year} {2025}{\natexlab{e}})},\ \Eprint
  {http://arxiv.org/abs/2508.18081} {arXiv:2508.18081 [gr-qc]} \BibitemShut
  {NoStop}%
\end{thebibliography}
